\documentclass[10pt,epsfig]{article}

\usepackage{enumerate}
\usepackage{float}
\usepackage{subfig}
\usepackage{color}
\usepackage[table]{xcolor}
\usepackage{slashed}
\usepackage{multirow}
\usepackage{cite}

\let\counterwithin\relax
\usepackage{chngcntr}
\usepackage{amssymb, amsmath,mathrsfs}

\usepackage{graphics}
\usepackage{graphicx}
\usepackage{epsf}
\usepackage{epsfig}
\usepackage{float}
\usepackage{makecell}
\usepackage{multirow}
\usepackage{color}
\usepackage{xcolor}
\usepackage{simplewick}

\usepackage[utf8]{inputenc}
\usepackage{amsmath}
\usepackage{amssymb}
\usepackage{subfig}
\usepackage[normalem]{ulem}

\usepackage{mathtools}
\usepackage{amsmath}
\usepackage{diagbox}

\newcommand\undermat[2]{
	\makebox[0.5pt][l]{$\smash{\underbrace{\phantom{%
					\begin{matrix}#2\end{matrix}}}_{ \let\scriptstyle\textstyle\text{\large $#1$}}}$}#2}
\newcommand\overmat[2]{
	\makebox[-1pt][l]{$\smash{\overbrace{\phantom{%
					\begin{matrix}#2\end{matrix}}}^{ \let\scriptstyle\textstyle\text{\large $#1$}}}$}#2}    
\usepackage{tikz}
\usepackage[vcentermath]{youngtab}
\usepackage{slashed} 
\usepackage[font=small]{caption} 
\usepackage{times} 

\usepackage{bm}       
\usepackage{bbm} 

\usepackage{relsize}  

\usepackage{autobreak}  

\usepackage{makeidx} 
\usepackage{bbding} 
\usepackage{listings} 
\usepackage{ytableau}

\usepackage[colorlinks=true,
linkcolor=blue,
urlcolor=red,
citecolor=red]{hyperref}

\long\def\rpl#1!!#2!!{\textcolor{red}{#1} \textcolor{blue}{#2}}

\usepackage[top=1in, left=0.95in, bottom=1.1in, right=0.95in]{geometry}
\def\baselinestretch{1.27}
\usepackage[toc]{appendix}

\newcommand{\beq}{\begin {equation}}  
\newcommand{\eeq}{\end   {equation}} 
\newcommand{\bea}{\begin {eqnarray}} 
\newcommand{\eea}{\end   {eqnarray}}  
\newcommand{\baa}{\begin {array}   } 
\newcommand{\eaa}{\end   {array}   }     
\newcommand{\bit}{\begin {itemize} }
\newcommand{\eit}{\end   {itemize} }
\newcommand{\be }{\begin {equation}} 
\newcommand{\ee }{\end   {equation}}
\newcommand{\nn }{\nonumber        }

\newcommand{\mc}[1]{\mathcal{#1}}

\newcommand{\vev}[1]{ \left\langle {#1}  \right\rangle }

\newcommand{\eq}[1]{\begin{equation}\begin{split} #1 \end{split}\end{equation}}

\newcommand{\comment}[1]{}

\newcolumntype{M}[1]{>{\centering\arraybackslash}m{#1}}
\newcolumntype{N}{@{}m{0pt}@{}}

\allowdisplaybreaks
\begin{document}

\begin{center}

{\Large \textbf  {On-shell Operator Construction in the Effective Field Theory of Gravity}}\\[10mm]

Hao-Lin Li$^{a, c}$\footnote{haolin.li@uclouvain.be}, 
Zhe Ren$^{a, b}$\footnote{renzhe@itp.ac.cn}, 
Ming-Lei Xiao$^{d, e}$\footnote{minglei.xiao@northwestern.edu}, 
Jiang-Hao Yu$^{a, b, f, g, h}$\footnote{jhyu@itp.ac.cn}, 
Yu-Hui Zheng$^{a, b}$\footnote{zhengyuhui@itp.ac.cn}\\[10mm]

\noindent 
$^a${\em \small CAS Key Laboratory of Theoretical Physics, Institute of Theoretical Physics, Chinese Academy of Sciences,    \\ Beijing 100190, P. R. China}  \\
$^b${\em \small School of Physical Sciences, University of Chinese Academy of Sciences,   Beijing 100049, P.R. China}   \\
$^c${\em \small Centre for Cosmology, Particle Physics and Phenomenology (CP3), Universite Catholique de Louvain,\\
Chem. du Cyclotron 2, 1348, Louvain-la-neuve, Belgium}\\
$^d${\em \small Department of Physics and Astronomy, Northwestern University, Evanston, Illinois 60208, USA}\\
$^e${\em \small High Energy Physics Division, Argonne National Laboratory, Lemont, Illinois 60439, USA}\\
$^f${\em \small Center for High Energy Physics, Peking University, Beijing 100871, China} \\
$^g${\em \small School of Fundamental Physics and Mathematical Sciences, Hangzhou Institute for Advanced Study, UCAS, Hangzhou 310024, China} \\
$^h${\em \small International Centre for Theoretical Physics Asia-Pacific, Beijing/Hangzhou, China}\\[10mm]

\date{\today}   
          
\end{center}

\begin{abstract} 

We construct the on-shell amplitude basis and the corresponding effective operators for generic modified gravity theory, such as pure gravity with higher derivatives, scalar-tensor gravity, Einstein-Yang-Mills, etc. Taking the Weyl tensor as the building block, we utilize the Young tensor technique to obtain independent operators, without equation of motion and total derivative redundancies. We update our algorithm and vastly increase the speed for finding the monomial basis (m-basis) of effective operators expressed in terms of Weyl tensors with Lorentz indices, the familiar form for the General Relativity community. Finally, we obtain the complete and independent amplitude and operator basis for GRSMEFT and GRLEFT up to mass dimension 10.

\end{abstract}

\newpage

\setcounter{tocdepth}{4}
\setcounter{secnumdepth}{4}

\tableofcontents

\setcounter{footnote}{0}

\def\baselinestretch{1.5}
\counterwithin{equation}{section}

\newpage

\section{Introduction}

The modern era of gravitational theory was initiated by Einstein's General Relativity (GR), which provided the geometric interpretation of the classical gravity. However, puzzles already existed when the quantum theory of physics was also discovered at around the same time, which pointed out that some compromise should be made for the gravitational theory to be compatible with a quantum theory of matter. The quantum field theory of gravity is expected to be non-renormalizable since the UV divergences arise at the loop order, requiring an infinite number of higher curvature counterterms. From modern point of view~\cite{Weinberg:1978kz,Donoghue:1994dn,Burgess:2003jk,Donoghue:2012zc}, the non-renormalizability of the Einstein-Hilbert Lagrangian is understood to mean that the theory is an effective field theory (EFT), in which loops are made finite by counterterms provided by the higher curvature terms in the Lagrangian order-by-order. It suggests that the classical theory of gravity should be regarded as a low-energy effective theory of some UV completion, often referred to as quantum gravity. Although many impressive efforts have been made to construct such a UV completion based on the self-consistency requirement, no direct experiments can yet be conducted to test those theories since the typical scale is at the Planck mass $M_{pl}$, parametrically larger than the energy scale of current and near-future experiments. In this circumstance, the effective field theory of gravity~\cite{Donoghue:1994dn,Burgess:2003jk,Donoghue:2012zc} can be used to describe the quantum effects of gravity at low energies without knowing the complete theory.

At the same time, the accumulating data from gravitational observations, especially through the new probe of gravitational waves, are pushing our understanding of gravity beyond Einstein's general relativity. Modified gravity~\cite{Clifton:2011jh,Nojiri:2017ncd}, which can derive from some specific theory of quantum gravity or just serve as a phenomenological model, is an active and diverse area of research in theoretical physics and cosmology, which aims to address various cosmological and gravitational puzzles, such as the accelerated expansion of the universe, the nature of dark matter and dark energy, and the behavior of gravity in extreme environments. 
From modern point of view, any modified gravity should be understood as effective field theory of gravity, which allows us to study gravity in a more tractable and model-independent way at low-energy region. 
The EFT approach in general starts from constructing a Lagrangian that includes all possible operators consistent with the symmetries of the theory and the power counting rules. In the case of gravity, the most important symmetry is the diffeomorphism, which asserts that the only degree of freedom for gravity should come from the Riemann curvature tensor of the spacetime manifold. By including higher-dimensional operators involving the Riemann curvature tensor, the gravity EFT captures the effects of higher-energy physics on the gravitational interactions in an organized and controlled manner. By matching the EFT predictions with experimental or observational data, we can constrain the values of the coefficients associated with the higher-dimensional operators. This can provide insights into the underlying fundamental theory of gravity or uncover new phenomena that could be observed in experiments.

The higher-dimensional operators are usually suppressed by the cutoff scale of the EFT $\Lambda_{\rm UV}$. In the case of gravity EFT, it is typically taken to be the Planck scale $M_{\rm pl}$, and thus the low energy effects of these higher dimensional operators are highly suppressed at energies much lower than the Planck scale. However, to address puzzles in the early universe the higher dimensional operators in modified gravity could play the important roles, such as Starobinsky theory~\cite{Starobinsky:1980te} and $f(R)$ gravity, scalar-tensor gravity~\cite{Fujii:2003pa}, Horava-Lifshitz gravity~\cite{Horava:2009uw}, etc. In the Higgs inflation~\cite{Bezrukov:2007ep}, the Higgs boson could play the role of inflaton and thus the non-minimal coupling between the Higgs boson and Ricci scalar affects the Higgs potential significantly during inflation. Furthermore, there are also higher dimensional theories of gravity in which the cutoff scale $\Lambda_{\rm UV}$ is significantly lower than the Planck scale, such as the large extra dimensions~\cite{Arkani-Hamed:1998jmv}, warped dimensions~\cite{Randall:1999ee}, DGP localized gravity~\cite{Dvali:2000hr}, low scale string theory~\cite{Lust:2009kp}, etc. In these cases, higher dimensional operators between graviton field and standard model fields is not so suppressed and thus their effects might be testified at future TeV colliders. To systematically investigate the sub-Planckian low energy effects, it is necessary to write down the gravity EFT operators. 
At the same time, the gravity scattering amplitude plays more and more important roles in the gravity EFT amplitude from double copy~\cite{Broedel:2012rc,Bern:2022wqg}, the positivity bounds of gravity EFTs~\cite{Adams:2006sv,Cheung:2016yqr,Bellazzini:2019xts,deRham:2021bll}, corrections on the gravitational potential and related effects~\cite{Brandhuber:2019qpg,AccettulliHuber:2020oou} and calculations on gravitational wave for testing the gravity EFTs~\cite{Endlich:2017tqa,Buonanno:2022pgc}, etc. All of these call for the complete and independent amplitude and operator bases for gravity EFTs to serve as a key input for the future gravity EFT amplitude calculation. 
However, this task is highly non-trivial because of the redundancies of total derivative and field redefinition.

In recent years, techniques of on-shell scattering amplitudes have inspired a lot of new ideas in the area of effective field theory. 
In particular, it is found to be extremely convenient to treat effective operators as on-shell amplitudes \cite{Shadmi:2018xan,Ma:2019gtx,Durieux:2020gip,Li:2020gnx,Li:2020zfq,Li:2022tec}, so that the equation of motion (EOM) redundancy is automatically removed and the integration-by-part (IBP) redundancy is translated to the requirement of momentum conservation. In this prescription, also known as the amplitude/operator correspondence, the operators yielding form factors that vanish on-shell should always be converted to a form with different constituting fields via field redefinition (usually equivalent to applying the classical equations of motion). 
In the case of gravity EFT, according to the on-shell prescription, we eliminate all the Ricci tensor $R_{\mu\nu}$ and Ricci scalar $R$ in the Lagrangian by field redefinitions because they do not generate on-shell gravitons. 
One example is when the non-minimal gravitational coupling $Rf(\phi)$ is present in the so-called Jordan frame, one could go to the Einstein frame by a Weyl transformation where the non-minimal coupling is removed and the scalar potential is modified. From the on-shell perspective, the operators corresponding to on-shell amplitudes are written in the Einstein frame, thus no non-minimal couplings involving $R$ should be present.
In fact, the building block that does generate on-shell gravitons is the Weyl tensor $C_{\mu\nu\rho\sigma}$, and the gravitational non-minimal on-shell couplings \cite{Levi:2015msa}, which can be written in terms of $C_{\mu\nu\rho\sigma}$, cannot be removed by field redefinitions and must be retained in the operator basis. We emphasize that the operator basis from on-shell method only describes local observables, hence the topological terms such as Gauss-Bonnet which are total derivative operators are not included in our basis.

Consequently, we generalize the amplitude/operator correspondence to involve the massless spin-2 particles, such that 
the operator bases can be obtained through constructing the corresponding amplitude bases, which are constructed using the Young tensor method~\cite{Li:2022tec} that has been applied to various EFTs~\cite{Li:2020gnx,Li:2020xlh,Li:2020tsi,Li:2021tsq}. In the Young tensor method, the independent Lorentz structures of on-shell amplitudes are identified as the semi-standard Young tableaux of the primary Young diagram of the $SL(2,\mathbb{C}) \times SU(N)$ group~\cite{Henning:2019enq,Henning:2019mcv,Li:2020gnx}, and the independent gauge structures are constructed by applying the modified Littlewood-Richardson rule at the Young tableau level. We also utilize the symmetric group to take account of the flavor structure of operators in a systematic way when the operators involve repeated fields~\cite{Fonseca:2019yya,Li:2020xlh}.

In this work, we consider all kinds of interactions in gravity EFTs, including pure gravity couplings with higher derivatives, gravity coupled to a scalar field and Goldstone field, gravity coupled to a gauge field in Yang-Mills theory and gravity coupled to a fermion and apply the above method to obtain both the operator bases and the amplitude bases in these EFTs up to mass dimension 10. Furthermore, we consider two specific gravity EFTs: the standard model effective field theory (SMEFT) with graviton field extension, and the low energy effective field theory (LEFT) with graviton field extension, called GRSMEFT and GRLEFT, respectively. The GRSMEFT operators have been counted~\cite{Ruhdorfer:2019qmk} using the Hilbert series method, and listed up to dimension 8 with one-flavor fermion using the traditional approach. In this work, we utilize the Young tensor method to list both the operator bases and the amplitude bases in the GRSMEFT and the GRLEFT up to mass dimension 10.

The paper is organized as follows. In section~\ref{sec:spinor}, firstly 
we introduce the tetrad formalism to define spinors in curved spacetime, so that the building blocks of gravity EFTs can be presented as spinors. Then we generalize the amplitude-operator correspondence to involve massless spin-2 fields and prove the validity of the generalization. At the end of the section, we use examples to show the conversions between operators presented by spinor indices and operators presented by Lorentz indices. In section~\ref{sec:grEFTs}, we list the operator bases and amplitude bases of pure gravity couplings and gravity coupled to a Goldstone scalar, a gauge boson, a fermion up to dimension 10. In section~\ref{sec:GRSMEFTopes} and section~\ref{sec:GRLEFTopes}, we list the operator bases and amplitude bases in the GRSMEFT and the GRLEFT up to dimension 10 respectively.

\section{Gravity in Spinor-helicity Formalism}\label{sec:spinor}

\subsection{Notation and Building Blocks}

\subsubsection{The Tetrad (Vielbein) Formalism and Spinors in Curved Spacetime}

In curved spacetime, the tangent space at each point is a Minkowski space, so we can introduce a local Lorentz frame at each point $x$, with basis vectors $e_\mu(x)$ constituting a tetrad of the 4-dimensional spacetime, or vielbein in other dimensions. For a coordinate patch $\{x^{\mu'}\}$ of the spacetime manifold ($\mu'$ labels the spacetime coordinates), we have the local basis transformation $e_{\mu}(x) = e_{\mu}{}^{\mu'}(x) \dfrac{\partial}{\partial x^{\mu'}}$.
Requiring the basis vectors to be orthonormal gives,
    \begin{eqnarray}
    	e_{\mu'}{}^{\mu}(x) e_{\nu' \mu}(x) = g_{\mu'\nu'}(x),
    \end{eqnarray}
    or equivalently
    \begin{eqnarray}
    	e_{\mu'}{}^{\mu}(x) e^{\mu' \nu}(x)=\eta^{\mu\nu},
    \end{eqnarray}
    where $g_{\mu'\nu'}$ is the metric tensor for the coordinate patch, and 
    \begin{eqnarray}
    	\eta^{\mu\nu} = \eta_{\mu\nu} = \begin{pmatrix}
    		1 & 0 & 0 & 0 \\
    		0 & -1 & 0 & 0 \\
    		0 & 0 & -1 & 0 \\
    		0 & 0 & 0 & -1
    	\end{pmatrix}.
    \end{eqnarray}
    The tetrad is related to the metric by
    \begin{eqnarray}
        g_{\mu' \nu'}=e_{\mu'}{}^{\mu} e_{\nu'}{}^{\nu} \eta_{\mu\nu}.
    \end{eqnarray}
    Using $e$ we can change indices from spacetime indices to tangent space indices or vice versa,
    \begin{eqnarray}\label{eq:changeindices}
    	V^{\mu}=e_{\mu'}{}^{\mu} V^{\mu'}, \quad V^{\mu'} = e^{\mu'}{}_{\mu} V^{\mu}.
    \end{eqnarray}
    
    In order to define the spinors in a curved spacetime, first we introduce the Van der Waerden symbols ($\sigma$ matrices) which give a mapping from spin space to Minkowski space as follows,
    \eq{\label{eq:VanderWaarden}
    	(\sigma^{0})_{\alpha\dot{\alpha}}=\left(\begin{array}{cc}
    		1 & 0 \\
    		0 & 1
    	\end{array}\right), &\quad 
        (\sigma^{1})_{\alpha\dot{\alpha}}=\left(\begin{array}{cc}
        	0 & 1 \\
        	1 & 0
        \end{array}\right), \\
        (\sigma^{2})_{\alpha\dot{\alpha}}=\left(\begin{array}{cc}
        	0 & -i \\
        	i & 0
        \end{array}\right), &\quad 
        (\sigma^{3})_{\alpha\dot{\alpha}}=\left(\begin{array}{cc}
        	1 & 0 \\
        	0 & -1
        \end{array}\right).
    }
    The Van der Waerden symbols fulfill the following algebraic properties:
    \begin{eqnarray}
    	\sigma^{\mu}_{\alpha\dot{\alpha}} \sigma_{\mu \beta\dot{\beta}} = 2\epsilon_{\alpha\beta} \epsilon_{\dot{\alpha}\dot{\beta}}, \quad \textrm{Tr}[\sigma^{\mu}\bar{\sigma}^{\nu}]=\sigma^{\mu}_{\alpha\dot{\alpha}} \bar{\sigma}^{\nu\dot{\alpha}\alpha}=2\eta^{\mu\nu},
    \end{eqnarray}
    where 
    \begin{eqnarray}
    	\epsilon_{\alpha\beta}=-\epsilon^{\alpha\beta}=\epsilon_{\dot{\alpha}\dot{\beta}}=-\epsilon^{\dot{\alpha}\dot{\beta}}=\left(\begin{array}{cc}
    		0 & -1 \\
    		1 & 0
    	\end{array}\right), \quad \bar{\sigma}=\left(\sigma^0,-\vec{\sigma}\right),
    \end{eqnarray}
    so that $\epsilon_{\alpha\beta} \epsilon^{\beta\gamma} = \delta^{\gamma}_{\alpha}$. $\epsilon_{\alpha\beta}$, $\epsilon_{\dot{\alpha}\dot{\beta}}$ can be used to raise and lower spinor indices in the following way:
    \begin{eqnarray}
    	\epsilon_{\alpha\beta} \xi^{\beta} = \xi_{\alpha}, \quad \xi^{\alpha}=\epsilon^{\alpha\beta} \xi_{\beta}.
    \end{eqnarray}
    In curved spacetime, the Minkowski space is replaced by the tangent space at each point. We can choose a smooth orthonormal (restricted) tetrad $e_{\mu'}{}^{\mu}(x)$ and define the $\sigma$'s in the local frame of the tangent space as eq.~(\ref{eq:VanderWaarden}). Since the components of a tensor in the local frame and the spacetime frame are connected by eq.~(\ref{eq:changeindices}),
    we can define the mapping from the tensors in spacetime frame to spinors in the local frame as
    \begin{eqnarray}\label{eq:sigmainst}
    	\sigma^{\mu'}_{\alpha\dot{\alpha}}(x)=e^{\mu'}{}_{\mu}(x) \sigma^{\mu}_{\alpha\dot{\alpha}},
    \end{eqnarray}
    fulfilling
    \begin{eqnarray}
    	\sigma^{\mu'}_{\alpha\dot{\alpha}} \sigma_{\mu' \beta\dot{\beta}} = 2\epsilon_{\alpha\beta} \epsilon_{\dot{\alpha}\dot{\beta}}, \quad \sigma^{\mu'}_{\alpha\dot{\alpha}} \sigma^{\nu'\alpha\dot{\alpha}}=2g^{\mu'\nu'}.
    \end{eqnarray}
    For example, the gauge field in the chiral basis $F_{\rm L/R} = \frac12(F \pm i\tilde{F})$ can be presented as
    \eq{
		F_{{\rm L}\alpha\beta}&=\frac{i}{2}F_{{\rm L}\mu'\nu'}\sigma^{\mu'\nu'}_{\alpha\beta}= \frac{i}{2}F_{{\rm L}\mu'\nu'}e^{\mu'}{}_{\mu}e^{\nu'}{}_{\nu}\sigma^{\mu\nu}_{\alpha\beta} = \frac{i}{2}F_{{\rm L}\mu\nu}\sigma^{\mu\nu}_{\alpha\beta},\\  \quad F_{{\rm R} \dot\alpha\dot\beta}&=-\frac{i}{2}F_{{\rm R}\mu'\nu'}\bar\sigma^{\mu'\nu'}_{\dot\alpha\dot\beta}= -\frac{i}{2}F_{{\rm R}\mu'\nu'}e^{\mu'}{}_{\mu}e^{\nu'}{}_{\nu}\bar\sigma^{\mu\nu}_{\dot\alpha\dot\beta} =-\frac{i}{2}F_{{\rm R}\mu\nu}\bar\sigma^{\mu\nu}_{\dot\alpha\dot\beta},
	}
	where
\begin{align}
	\sigma^{\mu\nu}_{\alpha\beta}=&\frac{i}{2}(\sigma^{\mu}\bar{\sigma}^{\nu}-\sigma^{\nu}\bar{\sigma}^{\mu})_{\alpha\beta},\\
	\bar{\sigma}^{\mu\nu}_{\dot{\alpha}\dot{\beta}}=&\frac{i}{2}(\bar{\sigma}^{\mu}\sigma^{\nu}-\bar{\sigma}^{\nu}\sigma^{\mu})_{\dot{\alpha}\dot{\beta}}.
\end{align}
    
The covariant derivative satisfies\footnote{We emphasize that the covariant derivative $D$ is different from the connection $\nabla$ since $D$ counts the Lorentz and gauge structure of a tensor in the tangent space while $\nabla$ does not. For example, $D_{\mu'} e_{\nu'}{}^{\mu} = \partial_{\mu'} e_{\nu'}{}^{\mu}-\Gamma^{\lambda'}{}_{\mu' \nu'} e_{\lambda'}{}^{\mu}+\omega_{\mu'}{}^{\mu}{}_{\nu} e_{\nu'}{}^{\nu}$ while $\nabla_{\mu'} e_{\nu'}{}^{\mu} = \partial_{\mu'} e_{\nu'}{}^{\mu}-\Gamma^{\lambda'}{}_{\mu' \nu'} e_{\lambda'}{}^{\mu}$.}
    \begin{eqnarray}
    	D_{\lambda'} g_{\mu'\nu'}=0, \quad D_{\mu'} e_{\nu'}{}^{\mu} =0,
    \end{eqnarray}
    and corrects local Lorentz and spinor indices with the spin connection $\omega_{\mu'}{}^{\mu\nu}$, where
    \begin{eqnarray}
        &\omega_{\mu'}{}^{\mu\nu}=e_{\nu'}{}^{\mu} \partial_{\mu'} e^{\nu' \nu} + e_{\nu'}{}^{\mu} \Gamma^{\nu'}{}_{\mu'\lambda'} e^{\lambda' \nu}, \\
        &\Gamma^{\lambda'}{}_{\mu' \nu'}=\frac{1}{2} g^{\lambda' \sigma'}\left(\partial_{\mu'} g_{\nu' \sigma'}+\partial_{\nu'} g_{\sigma' \mu'}-\partial_{\sigma'} g_{\mu' \nu'}\right).
    \end{eqnarray}
    In general cases, the covariant derivative acting on a field $f$ takes the form
    \begin{eqnarray}
    	D_{\mu'} f^x=\left(\delta^x{}_y \partial_{\mu'}- \frac{i}{2}\omega_{\mu'}{}^{\mu\nu}\left(X_{[\mu\nu]}\right)^x{}_y \right) f^y,
    \end{eqnarray}
    For example, for the spinor representation $X_{[\mu\nu]} = -\dfrac{\sigma_{\mu\nu}}{2}$, while for the vector representation $(X_{[\mu\nu]})^{\rho}{}_{\kappa}=i(\delta_{\mu}{}^{\rho} \eta_{\nu\kappa} - \delta_{\nu}{}^{\rho}\eta_{\mu\kappa})$. The components of the covariant derivative in the tangent space are just
    \begin{eqnarray}
        D_{\mu} = e^{\mu'}{}_{\mu} D_{\mu'}, \quad D_{\alpha\dot{\alpha}} = D_{\mu} \sigma^{\mu}_{\alpha\dot{\alpha}},
    \end{eqnarray}
    and satisfies
    \begin{equation}\label{eq:CDC1}
        [D_{\mu}, D_{\nu}] = -iX_{[\rho\kappa]} R_{\mu\nu}{}^{\rho\kappa},
    \end{equation}
    where $R_{\mu\nu}{}^{\rho\kappa}$ is the Riemann tensor in the tangent space and the gauge structure of the covariant derivative is not considered. In the following content, all operators and relations are written down in a local frame of the tangent space.
    
\subsubsection{Building Blocks of Gravity EFTs}

In general relativity, the Riemann tensor $R_{\mu\nu\rho\sigma}$ can be decomposed into irreducible representations of the Lorentz group as
\begin{eqnarray}
		R_{\mu\nu\rho\sigma} \sim (1,1) \oplus (2,0) \oplus (0,2) \oplus (0,0).
\end{eqnarray}
The $(1,1)$ and the $(0,0)$ representations can be recognized as the traceless part of Ricci tensor $R_{\mu\nu}$ and the Ricci scalar $R$ respectively, and the component transforming as $(2,0) \oplus (0,2)$ is the Weyl tensor $C_{\mu\nu\rho\sigma}$.
\begin{eqnarray}
        	C_{\mu\nu\rho\sigma} \equiv R_{\mu\nu\rho\sigma} - \left(\eta_{\mu\left[\rho\right.} R_{\left.\sigma\right]\nu} - \eta_{\nu\left[\rho\right.} R_{\left.\sigma\right]\mu}\right) + \dfrac{1}{3} \eta_{\mu\left[\rho\right.} \eta_{\left.\sigma\right]\nu} R.
        \end{eqnarray}
        
In gravity EFTs, any effective operator involving $R_{\mu\nu}$ or $R$ can be eliminated by a field redefinition of the metric~\cite{Ruhdorfer:2019qmk}, which is equivalent to substituting the EOM of GR, that is, the Einstein's equations, at the leading order of the field redefinition. For example, consider the following effective action
\begin{eqnarray}\label{eq:fredeg}
    S_{\text{eff}}=\int d^4x \sqrt{-g} \left[-\frac{M_{\text{pl}}^2}{2} R+\mc{L}_{\text{matter}} +a R^2 +bR_{\mu\nu}R^{\mu\nu}+c\phi^2 R +\dots\right],
\end{eqnarray}
where $\phi$ is a scalar field and the term $c\phi^2R$ is a non-minimal coupling of gravity and the scalar field. After a field redefinition of the metric
\begin{eqnarray}\label{eq:fredrule}
    g_{\mu\nu} \to g_{\mu\nu} - \frac{2}{M_{\text{pl}}^2} \Delta g_{\mu\nu}, \quad \Delta g_{\mu\nu}=\left(a+\frac{b}{2}\right)Rg_{\mu\nu}-bR_{\mu\nu}+c\phi^2g_{\mu\nu},
\end{eqnarray}
the effective action eq.~(\ref{eq:fredeg}) becomes
\begin{eqnarray}
\begin{aligned}\label{eq:fredeg2}
    S_{\text{eff}} \to &\int d^4x \sqrt{-g} \left[-\frac{M_{\text{pl}}^2}{2} R+\mc{L}_{\text{matter}} -\frac{1}{M_{\text{pl}}^2}\left(a+\frac{b}{2}\right) TR+\frac{b}{M_{\text{pl}}^2}T^{\mu\nu}R_{\mu\nu}-\frac{c}{M_{\text{pl}}^2}T\phi^2\right. \\
    &\left.+\mc{O}\left(\frac{1}{M_{\text{pl}}^4}\right)+\dots\right],
\end{aligned}
\end{eqnarray}
where $T^{\mu\nu}\equiv\frac{2}{\sqrt{-g}}\frac{\delta (\sqrt{-g}\mc{L}_{\text{matter}})}{\delta g_{\mu\nu}}$ and $T\equiv T^{\mu\nu}g_{\mu\nu}$. Comparing eq.~(\ref{eq:fredeg2}) with eq.~(\ref{eq:fredeg}), one can explicitly see that the dimension-4 effective operators involving $R_{\mu\nu}$ or $R$ in eq.~(\ref{eq:fredeg}) disappear while higher-dimensional operators emerge in eq.~(\ref{eq:fredeg2}), and the order $\mc{O}\left(\frac{1}{M_{\text{pl}}^2}\right)$ terms in eq.~(\ref{eq:fredeg2}) can be obtained by substituting the following trace-reversed Einstein's equations into the dimension-4 effective operators once in eq.~(\ref{eq:fredeg}),
\begin{eqnarray}
    R_{\mu\nu} = \frac{1}{M_{\text{pl}}^2}(T_{\mu\nu}-\frac{1}{2}Tg_{\mu\nu}).
\end{eqnarray}
Similarly, other field redefinitions of metric can be performed to further eliminate the operators involving $R_{\mu\nu}$ or $R$ in eq.~(\ref{eq:fredeg2}) until all of them are eliminated. It should be noted that the field redefinition $\Delta g_{\mu\nu} \propto \phi^2 g_{\mu\nu}$ is a Weyl transformation that remove the non-minimal coupling $c\phi^2R$, so the effective action eq.~(\ref{eq:fredeg2}) is in the Einstein frame, and so are the effective operators in the following content.

Since the Weyl tensor and the Riemann tensor only differ by terms that involve $R_{\mu\nu}$ or $R$, we can use the Weyl tensor $C_{\mu\nu\rho\sigma}$ instead of the Riemann tensor $R_{\mu\nu\rho\sigma}$ to construct EFT operators. The following contracted Bianchi identities,
        \begin{eqnarray}
        	D^{\mu} R_{\mu\nu\rho\sigma}=2D_{[\rho} R_{\sigma] \nu}, \quad D^{\mu} R_{\mu\nu} = \dfrac{1}{2} D_{\nu} R,
        \end{eqnarray}
        imply the EOM of the Weyl tensor
        \begin{eqnarray}\label{eq:EOMWeyl}
        	D^{\mu} C_{\mu \nu \rho \sigma}=D_{[\rho} R_{\sigma] \nu}+\frac{1}{6} \eta_{\nu[\rho} D_{\sigma]} R,
        \end{eqnarray}
where the terms on the right-hand side only involve the Ricci tensor $R_{\mu\nu}$ and Ricci scalar $R$, and thus can be replaced via the EOM. Furthermore, the Bianchi identity
        \begin{equation}
            D_{\kappa} R_{\mu \nu \rho \sigma}+D_{\rho} R_{\mu \nu \sigma \kappa}+D_{\sigma} R_{\mu \nu \kappa \rho}=0
        \end{equation}
implies that similar relation holds for the Weyl tensor
        \begin{equation}\label{eq:BianchiWeyl}
            D_{\kappa} C_{\mu \nu \rho \sigma}+D_{\rho} C_{\mu \nu \sigma \kappa}+D_{\sigma} C_{\mu \nu \kappa \rho}=0
        \end{equation}
up to terms that can be removed by the EOM. Eq.~(\ref{eq:BianchiWeyl}), together with the EOM of the Weyl tensor eq.~(\ref{eq:EOMWeyl}), implies that $D^2 C_{\mu \nu \rho \sigma}$ is redundant as well. Similarly for $\tilde{C}^{\mu\nu\rho\sigma}= \dfrac{1}{2} \epsilon^{\mu\nu\lambda\xi} C_{\lambda\xi}{}^{\rho\sigma}$, one can straightforwardly find that $D^{\mu} \tilde{C}_{\mu \nu \rho \sigma}$ and $D^2 \tilde{C}_{\mu \nu \rho \sigma}$ are redundant from eq.~(\ref{eq:BianchiWeyl}).
        
        In analogy to the gauge field, the Weyl tensor $C_{\mu\nu\rho\sigma}$ can be decomposed into its irreducible parts of the Lorentz group,
\begin{eqnarray}
    C_{\rm L/\rm R}{}_{\mu\nu\rho\sigma} = \dfrac{1}{2} \left(C_{\mu\nu\rho\sigma} \mp i\tilde{C}_{\mu\nu\rho\sigma}\right),
\end{eqnarray}
where $\tilde{C}^{\mu\nu\rho\sigma}= \dfrac{1}{2} \epsilon^{\mu\nu\lambda\xi} C_{\lambda\xi}{}^{\rho\sigma}$, $\epsilon^{0123}=1$. Furthermore, the left-handed and right-handed Weyl tensors can be defined in the local spinor space through the Van der Waerden symbols introduced in eq.~(\ref{eq:VanderWaarden}),
\begin{eqnarray}
    C_{\rm L}{}_{\alpha\beta\gamma\delta} = -\dfrac{1}{4} C_{\rm L}{}_{\mu\nu\rho\lambda} \sigma^{\mu\nu}_{\alpha\beta} \sigma^{\rho\lambda}_{\gamma\delta}, \quad C_{\rm R}{}_{\dot{\alpha}\dot{\beta}\dot{\gamma}\dot{\delta}} = -\dfrac{1}{4} C_{\rm R}{}_{\mu\nu\rho\lambda} \bar{\sigma}^{\mu\nu}_{\dot{\alpha}\dot{\beta}} \bar{\sigma}^{\rho\lambda}_{\dot{\gamma}\dot{\delta}}.
\end{eqnarray}
It is straightforward to verify that all the spinor indices in $C_{\rm L}$ and $C_{\rm R}$ are totally symmetric.

To conclude, the Lorentz structure of the building blocks in general gravity EFTs are presented as fields and covariant derivatives denoted by the irreducible representations under the Lorentz group $SL(2,\mathbb{C}) = SU(2)_{l}\times SU(2)_{r}$ in the tangent space, that is,
\eq{&\phi \in (0,0), \quad \psi_{\alpha }  \in (1/2,0), \quad \psi^{\dagger}_{\dot\alpha } \in (0,1/2), \\
&F_{{\rm L}\alpha\beta} = \frac{i}{2}F_{\mu\nu}\sigma^{\mu\nu}_{\alpha\beta}\in (1,0),  \quad F_{{\rm R} \dot\alpha\dot\beta} = -\frac{i}{2}F_{\mu\nu}\bar\sigma^{\mu\nu}_{\dot\alpha\dot\beta}\in(0,1), \\
&C_{\rm L}{}_{\alpha\beta\gamma\delta} = -\dfrac{1}{4} C_{\mu\nu\rho\lambda} \sigma^{\mu\nu}_{\alpha\beta} \sigma^{\rho\lambda}_{\gamma\delta} \in (2,0), \quad C_{\rm R}{}_{\dot{\alpha}\dot{\beta}\dot{\gamma}\dot{\delta}} = -\dfrac{1}{4} C_{\mu\nu\rho\lambda} \bar{\sigma}^{\mu\nu}_{\dot{\alpha}\dot{\beta}} \bar{\sigma}^{\rho\lambda}_{\dot{\gamma}\dot{\delta}} \in (0,2), \\
&D_{\alpha\dot\alpha} = D_{\mu}\sigma^{\mu}_{\alpha\dot\alpha} \in (1/2,1/2).}

\subsection{Amplitude-operator Correspondence and Young Tableau Method}
The amplitude-operator correspondence for massless fields with helicity $|h| \leq 1$ was introduced in Ref.~\cite{Li:2020xlh} and generalized to cases involving massive scalars and fermions in Ref.~\cite{Li:2020tsi}. In order to apply the correspondence to general gravity EFTs, it should be further generalized to involve fields with helicity $|h| = 2$. Intuitively, the generalized correspondence can be presented as follows,
\eq{\label{eq:dictionary}
	\begin{array}{lcl}
		D^{r_i-2}C_{{\tiny\rm L/R}\,i}		&	\Leftrightarrow	&		\lambda_i^{r_i\pm2}\tilde\lambda_i^{r_i\mp2},		\\
		D^{r_i-1}F_{{\tiny\rm L/R}\,i}		&	\Leftrightarrow	&				\lambda_i^{r_i\pm1}\tilde\lambda_i^{r_i\mp1},  \\
		D^{r_i-1/2}\psi^{(\dagger)}_i	&	\Leftrightarrow	&	\lambda_i^{r_i\pm1/2}\tilde\lambda_i^{r_i\mp1/2},	\\
		D^{r_i}\phi_i				&	\Leftrightarrow	&	\lambda_i^{r_i}\tilde\lambda_i^{r_i},
	\end{array}
}
where $i$ labels the $i$th field in an operator, $i \in \{1,2,\cdots,N\}$, and $N$ denotes the number of fields in the operator. $r_i$ is a positive integer or half-integer depending on whether the field is bosonic or fermionic. However, as we illustrated for $|h| \leq 1$ fields in Ref.~\cite{Li:2020xlh}, the correspondence for $C_{{\tiny\rm L/R}\,i}$ is exact if and only if all left(right)-handed spinor indices in the building block $D^{r_i-2}C_{{\tiny\rm L/R}\,i}$ are totally symmetric.

To prove that, let us consider all the left-handed spinor indices in $D^{r_i-2}C_{\rm L}{}_i$. The symmetry of the right-handed spinor indices spinor indices in $D^{r_i-2}C_{\rm L}{}_i$ and the symmetry of the spinor  in $D^{r_i-2}C_{\rm R}{}_i$ can be proved similarly. Firstly it should be noted that the spinor indices of $C_{{\tiny\rm L}\,i}$ are totally symmetric since $C_{{\tiny\rm L}} \in (2,0)$. Secondly, if two left-handed spinor indices of the covariant derivatives in $D^{r_i-2}C_{{\tiny\rm L}\,i}$ are antisymmetric, then
\eq{\label{eq:antisymdev}
D_{[\alpha\dot\alpha}D_{\beta]\dot\beta} = D_{\mu}D_{\nu}\sigma^{\mu}_{[\alpha\dot\alpha}\sigma^{\nu}_{\beta]\dot\beta} = - D^2 \epsilon_{\alpha\beta}\epsilon_{\dot\alpha\dot\beta} + \frac{i}{2}[D_{\mu},D_{\nu}]\epsilon_{\alpha\beta}\bar\sigma^{\mu\nu}_{\dot\alpha\dot\beta}.
}
The first term on the right side in eq.~(\ref{eq:antisymdev}) acting on $C_{\rm L}$ gives $D^2 C_{\rm L}$, and the second term is just the covariant derivative commutator eq.~(\ref{eq:CDC1}), which after the replacement will transform the operator to other type, so the left side in eq.~(\ref{eq:antisymdev}) can be equivalently set to zero if one enumerates operators type by type. Thirdly, if one left-handed spinor index of the covariant derivative and one left-handed spinor index of the $C_{\rm L}$ are antisymmetric, then
\begin{eqnarray}
        	D_{\left[\alpha\dot{\alpha}\right.} C_{\rm L}{}_{\left.\beta\right]\gamma\delta\epsilon} &=& -\dfrac{1}{4} D_{\mu} C_{\nu\rho\sigma\lambda} \sigma^{\mu}_{\left[\alpha\dot{\alpha}\right.} \sigma^{\nu\rho}_{\left.\beta\right]\gamma} \sigma^{\sigma\lambda}_{\delta\epsilon} \nn \\
        	&=& \dfrac{i}{8} \epsilon_{\alpha\beta} D_{\mu} C_{\nu\rho\sigma\lambda} \left[2\eta^{\rho\mu} (\sigma^{\nu})_{\gamma\dot{\alpha}} + i \epsilon^{\nu\rho\mu\kappa} (\sigma_{\kappa})_{\gamma\dot{\alpha}}\right] (\sigma^{\sigma\lambda})_{\delta\epsilon} \nn \\
        	&=& -\dfrac{i}{4} \epsilon_{\alpha\beta} (\sigma^{\nu})_{\gamma\dot{\alpha}} (\sigma^{\sigma\lambda})_{\delta\epsilon} D^{\mu} C_{\mu\nu\sigma\lambda},
        \end{eqnarray}
        which is the EOM of the Weyl tensor and thus can also be set to zero. 
        
        To conclude, only the totally symmetric part of $D^{r_i-2}C_{{\tiny\rm L/R}\,i}$ needs to be considered in operator construction perspective, so the amplitude-operator correspondence can be naturally generalized for $|h|=2$ fields as in eq.~(\ref{eq:dictionary}). With such an amplitude-operator correspondence, a set of non-redundant operators can be constructed by finding an on-shell amplitude basis utilizing the Young tableau method introduced in Ref.~\cite{Li:2020gnx}.
        
        Here we briefly introduce the Young tableau method to obtain on-shell amplitudes basis. The on-shell amplitudes of $N$ particles involving $2n$ $\lambda$s and $2\tilde{n}$ $\tilde\lambda$s span a irreducible representation of $SU(N)$ group denoted by the following primary Young diagram,\\
        \begin{eqnarray}\label{eq:primary_YD}
	Y_{N,n,\tilde{n}} \quad = \quad \arraycolsep=0.3pt\def\arraystretch{1}
	\rotatebox[]{90}{\text{$N-2$}} \left\{
	\begin{array}{cccccc}
		\yng(1,1) &\ \ldots{}&\ \yng(1,1)& \overmat{n}{\yng(1,1)&\ \ldots{}\  &\yng(1,1)} \\
		\vdotswithin{}& & \vdotswithin{}&&&\\
		\undermat{\tilde{n}}{\yng(1,1)\ &\ldots{}&\ \yng(1,1)} &&&
	\end{array}
	\right.
	\\
	\nonumber 
\end{eqnarray}
        The independent on-shell amplitudes can be recognized as the basis vectors of the irreducible representation, and the basis vectors are given by semi-standard Young tableaux (SSYTs) of the Young diagram. Here the amplitudes being independent means that the redundancies of total momentum conservation and the Schouten identity are removed, and the independence is guaranteed by the Fock's condition of Young tableaux. The SSYTs can be obtained from filling the Young diagram with numbers $i\in\{1,\cdots,N\}$, and the multiplicity of the number $i$, $\#i=\tilde{n}-2h_i$, where $h_i$ denotes the helicity of the $i$th particle, and the label $i$ of the corresponding particle are determined by setting the helicity sequence $\{h_i\}$ in an ascending order. The SSYTs can be translated to amplitudes column by column using the following rules
\begin{eqnarray}
    \langle ij\rangle \sim \young(i,j)\quad ,\qquad 
\mathcal{E}^{k_1...k_{N-2}ij}[ij] \sim   \left. \begin{array}{c}
\ytableausetup{centertableaux, boxsize=2em} \begin{ytableau} k_1 \\ k_2 \end{ytableau}
 \\
\vdotswithin{}\\
\begin{ytableau}
\scriptstyle k_{N-2}
\end{ytableau}
\end{array}\right\} \rotatebox[]{270}{\text{$N-2$}} \quad,
\end{eqnarray}
where $\mc{E}$ is the Levi-Civita tensor of the $SU(N)$ group.

The gauge structures of on-shell amplitudes are the Levi-Civita tensors contracted with fundamental representation indices of fields if all the fields are expressed with fundamental representation indices only. In fact, a field as a irreducible representation of the gauge group can always be presented in a form with fundamental representations indices with certain permutation symmetries. Here we lists some simple examples to show that field of non-fundamental representaion can be expressed in terms of tensor with fundamental indices only, which also correspond to Young tableaux denoting the permutation symmetries of their indices,
\eq{\label{eq:egrep}
	&\epsilon_{acd}\lambda^A{}_b^d G^A = G_{abc} \sim \young(ab,c)\,,\\
	&\epsilon_{abc}Q^{\dagger,{c}} = Q^\dagger_{ab} \sim \young(a,b)\,,\\
	&\epsilon_{jk}\tau^I{}_i^k W^I = W_{ij}\sim \young(ij) \,,\\
	&\epsilon_{ij}H^{\dagger, j} = H_i \sim \young(i) \,.
}
The independent gauge basis are obtained through constructing singlet representations Young tableaux from the gauge Young tableaux of all fields for each gauge group using the modified Littlewood-Richardson rule\cite{Li:2020gnx}. The direct product of the gauge basis and the Lorentz basis gives the y-basis of on-shell amplitudes.

\subsection{Conversion}
In Ref.~\cite{Li:2020gnx}, in order to get monomial operators, we started from the amplitude y-basis, which is directly translated to a basis of monomial operators with spinor indexed building blocks such as $D_{\alpha\dot\alpha}$ and $F_{{\rm L}\alpha\beta}$, and convert them into the form with Lorentz indexed building blocks $D_\mu$ and $F_{{\rm L}\mu\nu}$. Nevertheless, it is usually not efficient, especially when the conversion between the two forms are time consuming, or when there are way more monomials than necessary. In the gravity EFT, the conversion inevitably involves a very long chain of $\sigma$ or $\gamma$ matrices, which is indeed time consuming.
In Ref.~\cite{Low:2022iim}, the chiral perturbation theory was studied where the monomial basis (m-basis) is obtained in a reversed process, which we adopt here.
We directly construct a set of Lorentz invariant operators $\{\mc{O}'_i\}$ by traversing different contractions of the Lorentz indices, which form an over-complete m-basis. It is a lot easier to dressing all the Lorentz indices in operators with $\sigma$ and extract the on-shell amplitudes they generate. Afterwards, with the reduction rules introduced in Ref.~\cite{Li:2020gnx}, one can reduce the amplitudes onto the y-basis, and then independent operators among $\{\mc{O}'_i\}$ can be selected by requiring their coordinates under the y-basis to be numerically independent vectors. 
We'll see two examples to illustrate the process.

The first example is the type $D^2 L L^{\dagger} Q Q^{\dagger}$, it is straightforward to write down the following over-complete operators based on the properties of $\gamma$ matrices,
\begin{align}\label{eq:overcompletebasis1}
	\begin{array}{c|c}
		D^2 L L^{\dagger} Q Q^{\dagger} & \\
		\hline
		\mathcal{O}'_1 & (D_{\mu} L_{pi} D^{\mu} Q_{raj}) (L^{\dagger}_s{}^i Q^{\dagger}_t{}^{aj}) \\
		\mathcal{O}'_2 & (D_{\mu} L_{pi} D^{\mu} Q_{raj}) (L^{\dagger}_s{}^j Q^{\dagger}_t{}^{ai}) \\
		\mathcal{O}'_3 & (D_{\mu} L_{pi} Q_{raj}) (D^{\mu} L^{\dagger}_s{}^i Q^{\dagger}_t{}^{aj}) \\
		\mathcal{O}'_4 & (D_{\mu} L_{pi} Q_{raj}) (D^{\mu} L^{\dagger}_s{}^j Q^{\dagger}_t{}^{ai}) \\
		\mathcal{O}'_5 & (D_{\mu} L_{pi} D_{\nu} Q_{raj}) (L^{\dagger}_s{}^i \bar{\sigma}^{\mu\nu} Q^{\dagger}_t{}^{aj}) \\
		\mathcal{O}'_6 & (D_{\mu} L_{pi} D_{\nu} Q_{raj}) (L^{\dagger}_s{}^j \bar{\sigma}^{\mu\nu} Q^{\dagger}_t{}^{ai}) \\
		\mathcal{O}'_7 & (D^{\nu} L_{pi} \sigma_{\mu} L^{\dagger}_s{}^i) (D^{\mu} Q_{raj} \sigma_{\nu} Q^{\dagger}_t{}^{aj}) \\
		\mathcal{O}'_8 & (D^{\nu} L_{pi} \sigma_{\mu} Q^{\dagger}_t{}^{aj}) (D^{\mu} Q_{raj} \sigma_{\nu} L^{\dagger}_s{}^i) \\
		\mathcal{O}'_9 & (D^{\mu} L_{pi} \sigma_{\mu} L^{\dagger}_s{}^i) (D^{\nu} Q_{raj} \sigma_{\nu} Q^{\dagger}_t{}^{aj}). \\
	\end{array}
\end{align}
The on-shell amplitudes generated by the operators in eq.~(\ref{eq:overcompletebasis1}) can be expanded on the amplitude y-basis:
\begin{align}
	\begin{array}{c|c}
		\text{Y-basis} & \mathcal{A}\left(L_i(1),Q_{aj}(2), L^{\dagger k}(3), Q^{\dagger bl}(4)  \right) \\ 
		\hline
		\mathcal{A}_1^{(y)} & \delta^i_k\delta^j_l\delta^a_b \langle 12\rangle\langle 34\rangle[34]^2 \\
		\mathcal{A}_2^{(y)} & \delta^i_k\delta^j_l\delta^a_b \langle 13\rangle\langle 24\rangle[34]^2 \\
		\mathcal{A}_3^{(y)} & \delta^i_l\delta^j_k\delta^a_b \langle 12\rangle\langle 34\rangle[34]^2 \\
		\mathcal{A}_4^{(y)} & \delta^i_l\delta^j_k\delta^a_b \langle 13\rangle\langle 24\rangle[34]^2 \\
	\end{array}
\end{align}
and the the coefficient matrix is obtained as
\begin{eqnarray}\label{eq:coematrix1}
	\left(
	\begin{array}{cccc}
		\color{red}{\frac12} & \color{red}{0} & \color{red}{0} & \color{red}{0} \\
		\color{red}{0} & \color{red}{0} & \color{red}{\frac12} & \color{red}{0} \\
		\color{red}{0} & \color{red}{-\frac{1}{2}} & \color{red}{0} & \color{red}{0} \\
		\color{red}{0} & \color{red}{0} & \color{red}{0} & \color{red}{-\frac{1}{2}} \\
		-\frac{i}{2} & i & 0 & 0 \\
		0 & 0 & -\frac{i}{2} & i \\
		-1 & 1 & 0 & 0 \\
		0 & 1 & 0 & 0 \\
		0 & 0 & 0 & 0 \\
	\end{array}
	\right),
\end{eqnarray}
where the coefficients in the last row being all 0 is the consequence of $\mathcal{O}'_9$ corresponding to the EOM redundancy. From eq.~(\ref{eq:coematrix1}) one can choose 4 independent rows of the matrix as a complete basis of $D^2 L L^{\dagger} Q Q^{\dagger}$, for example, the 4 rows which are marked red in eq.~(\ref{eq:coematrix1}). We can also give priority to $\mc{O}'_7$, $\mc{O}'_8$ if necessary. We don't even need to list all the alternative operators $\mc{O}'_i$ at the beginning. Our package checks the independence of each candidate during the construction until obtaining the complete base. The method avoids dealing with a long chain of $\sigma$ matrices and is especially time saving for types including gravitons as will be demonstrated in the next example.
The second example is type $C_{\rm L}^5$. To focus on the independence of the Lorentz structure, it is straightforward to write down the following contractions.
\comment{\begin{align}\label{eq:overcompletebasis}
    \begin{array}{c|c}
        C_{\rm L}^2F_{\rm L}^2 &  \\
        \hline
        \mc{O}'_1 & C_{\rm L1}{}_{\mu\nu\rho\sigma}C_{\rm L2}^{\mu\nu\rho\sigma}F_{\rm L3}{}_{\lambda\eta}F_{\rm L4}^{\lambda\eta} \\
        \mc{O}'_2 & C_{\rm L1}{}_{\mu\nu\rho\sigma}C_{\rm L2}^{\mu\nu\rho\lambda}F_{\rm L3}{}_{\lambda\eta}F_{\rm L4}^{\sigma\eta} \\
        \mc{O}'_3 & C_{\rm L1}{}_{\mu\nu\rho\sigma}C_{\rm L2}^{\mu\nu\lambda\eta}F_{\rm L3}{}_{\rho\sigma}F_{\rm L4}^{\lambda\eta} \\
        \mc{O}'_4 & C_{\rm L1}{}_{\mu\nu\rho\sigma}C_{\rm L2}^{\mu\nu}{}_{\lambda\eta}F_{\rm L3}^{\rho\lambda}F_{\rm L4}^{\sigma\eta} \\
        \vdots & \vdots
    \end{array}
\end{align}
}
\begin{align}\label{eq:overcompletebasis}
    \begin{array}{c|c}
        C_{\rm L}^5 &  \\
        \hline
        \mc{O}'_1 & C_{\rm L1}{}_{\mu\nu\lambda\rho}C_{\rm L2}{}_{\eta\xi}{}^{\lambda\rho}C_{\rm L3}^{\mu\nu\eta\xi}C_{\rm L4}{}_{\tau\upsilon\phi\chi}C_{\rm L5}^{\tau\upsilon\phi\chi} \\
        \mc{O}'_2 & C_{\rm L1}{}_{\mu\nu\lambda\rho}C_{\rm L2}^{\mu\nu\lambda\rho}C_{\rm L3}{}_{\eta\xi\tau\upsilon}C_{\rm L4}{}_{\phi\chi}{}^{\tau\upsilon}C_{\rm L5}^{\eta\xi\phi\chi} \\
        \mc{O}'_3 & C_{\rm L1}{}_{\mu\nu\lambda\rho}C_{\rm L2}{}_{\eta\xi}{}^{\lambda\rho}C_{\rm L3}{}_{\tau\upsilon}{}^{\eta\xi}C_{\rm L4}{}_{\phi\chi}{}^{\mu\nu}C_{\rm L5}^{\phi\chi\tau\upsilon} \\
        \mc{O}'_4 & C_{\rm L1}{}_{\mu\nu\lambda\rho}C_{\rm L2}{}_{\eta\xi\tau\upsilon}C_{\rm L3}^{\mu\nu\lambda\rho}C_{\rm L4}{}_{\phi\chi}{}^{\tau\upsilon}C_{\rm L5}^{\phi\chi\eta\xi} \\
        \mc{O}'_5 & C_{\rm L1}{}_{\mu\nu\lambda\rho}C_{\rm L2}{}_{\eta\xi\tau\upsilon}C_{\rm L3}^{\tau\upsilon\lambda\rho}C_{\rm L4}{}_{\phi\chi}{}^{\mu\nu}C_{\rm L5}^{\phi\chi\eta\xi} \\
        \mc{O}'_6 & C_{\rm L1}{}_{\mu\nu\lambda\rho}C_{\rm L2}{}_{\eta\xi}{}^{\lambda\rho}C_{\rm L3}{}_{\tau\upsilon}{}^{\eta\xi}C_{\rm L4}{}_{\phi\chi}{}^{\tau\upsilon}C_{\rm L5}^{\phi\chi\mu\nu} \\
        \mc{O}'_7 & C_{\rm L1}{}_{\mu\nu\lambda\rho}C_{\rm L2}{}_{\eta\xi\tau\upsilon}C_{\rm L3}^{\tau\upsilon\lambda\rho}C_{\rm L4}{}_{\phi\chi}{}^{\eta\xi}C_{\rm L5}^{\phi\chi\mu\nu} \\
        \mc{O}'_8 & C_{\rm L1}{}_{\mu\nu\lambda\rho}C_{\rm L2}{}_{\eta\xi\tau\upsilon}C_{\rm L3}^{\eta\xi\tau\upsilon}C_{\rm L4}{}_{\phi\chi}{}^{\lambda\rho}C_{\rm L5}^{\phi\chi\mu\nu} \\
        \mc{O}'_9 & C_{\rm L1}{}_{\mu\nu\lambda\rho}C_{\rm L2}{}_{\eta\xi}{}^{\lambda\rho}C_{\rm L3}{}_{\tau\upsilon\phi\chi}C_{\rm L4}^{\mu\nu\eta\xi}C_{\rm L5}^{\tau\upsilon\phi\chi} \\
        \mc{O}'_{10} & C_{\rm L1}{}_{\mu\nu\lambda\rho}C_{\rm L2}{}_{\eta\xi}{}^{\lambda\rho}C_{\rm L3}{}_{\tau\upsilon\phi\chi}C_{\rm L4}^{\mu\nu\phi\chi}C_{\rm L5}^{\tau\upsilon\eta\xi} \\
        \mc{O}'_{11} & C_{\rm L1}{}_{\mu\nu\lambda\rho}C_{\rm L2}{}_{\eta\xi\tau\upsilon}C_{\rm L3}{}_{\phi\chi}{}^{\lambda\rho}C_{\rm L4}^{\mu\nu\tau\upsilon}C_{\rm L5}^{\phi\chi\eta\xi} \\
        \mc{O}'_{12} & C_{\rm L1}{}_{\mu\nu\lambda\rho}C_{\rm L2}{}_{\eta\xi\tau\upsilon}C_{\rm L3}{}_{\phi\chi}{}^{\tau\upsilon}C_{\rm L4}^{\mu\nu\lambda\rho}C_{\rm L5}^{\phi\chi\eta\xi} \\
        \mc{O}'_{13} & C_{\rm L1}{}_{\mu\nu\lambda\rho}C_{\rm L2}{}_{\eta\xi}{}^{\lambda\rho}C_{\rm L3}{}_{\phi\chi\tau\upsilon}C_{\rm L4}^{\phi\chi\eta\xi}C_{\rm L5}^{\tau\upsilon\mu\nu} \\
        \mc{O}'_{14} & C_{\rm L1}{}_{\mu\nu\lambda\rho}C_{\rm L2}{}_{\eta\xi\tau\upsilon}C_{\rm L3}{}_{\phi\chi}{}^{\lambda\rho}C_{\rm L4}^{\eta\xi\tau\upsilon}C_{\rm L5}^{\phi\chi\mu\nu} \\
        \mc{O}'_{15} & C_{\rm L1}{}_{\mu\nu\lambda\rho}C_{\rm L2}{}_{\eta\xi\tau\upsilon}C_{\rm L3}{}_{\phi\chi}{}^{\tau\upsilon}C_{\rm L4}^{\eta\xi\lambda\rho}C_{\rm L5}^{\phi\chi\mu\nu} \\
        \mc{O}'_{16} & C_{\rm L1}{}_{\mu\nu\lambda\rho}C_{\rm L2}{}_{\eta\xi\tau\upsilon}C_{\rm L3}{}_{\phi\chi}{}^{\lambda\rho}C_{\rm L4}^{\mu\nu\phi\chi}C_{\rm L5}^{\eta\xi\tau\upsilon} \\
        \mc{O}'_{17} & C_{\rm L1}{}_{\mu\nu\lambda\rho}C_{\rm L2}{}_{\eta\xi\tau\upsilon}C_{\rm L3}{}_{\phi\chi}{}^{\tau\upsilon}C_{\rm L4}^{\eta\xi\phi\chi}C_{\rm L5}^{\mu\nu\lambda\rho} \\
        \mc{O}'_{18} & C_{\rm L1}{}_{\eta\xi\tau\upsilon}C_{\rm L2}{}_{\phi\chi}{}^{\tau\upsilon}C_{\rm L3}{}_{\mu\nu\lambda\rho}C_{\rm L4}^{\mu\nu\lambda\rho}C_{\rm L5}^{\eta\xi\phi\chi} \\
        \vdots & \vdots
    \end{array}
\end{align}
Again, each operator in eq.~(\ref{eq:overcompletebasis}) can be expanded on the amplitude y-basis
\comment{\begin{align}
	\begin{array}{c|c}
		\text{Y-basis} & \mathcal{A}\left(C_{\rm L1}, C_{\rm L2}, F_{\rm L3}, F_{\rm L4} \right) \\ 
		\hline
		\mathcal{A}_1^{(y)} & \vev{12}^4\vev{34}^2 \\
		\mathcal{A}_2^{(y)} & \vev{12}^3\vev{13}\vev{24}\vev{34}\\
		\mathcal{A}_3^{(y)} & \vev{12}^2\vev{13}^2\vev{24}^2
	\end{array}
\end{align}
}
\begin{align}
	\begin{array}{c|c}
		\text{Y-basis} & \mathcal{A}\left(C_{\rm L1}, C_{\rm L2}, C_{\rm L3}, C_{\rm L4}, C_{\rm L5} \right) \\ 
		\hline
		\mathcal{A}_1^{(y)} & \vev{12}^4\vev{34}^2\vev{35}^2\vev{45}^2 \\
		\mathcal{A}_2^{(y)} & \vev{13}^4\vev{24}^2\vev{25}^2\vev{45}^2 \\
		\mathcal{A}_3^{(y)} & \vev{13}^3\vev{14}\vev{24}^2\vev{25}^2\vev{35}\vev{45} \\
		\mathcal{A}_4^{(y)} & \vev{13}^2\vev{14}^2\vev{24}^2\vev{25}^2\vev{35}^2 \\
		\mathcal{A}_5^{(y)} & \vev{12}^3\vev{14}\vev{24}\vev{34}\vev{35}^3\vev{45} \\
		\mathcal{A}_6^{(y)} & \vev{12}^3\vev{13}\vev{24}\vev{34}\vev{35}^2\vev{45}^2 \\
		\mathcal{A}_7^{(y)} & \vev{12}^3\vev{13}\vev{23}\vev{34}\vev{35}\vev{45}^3 \\
		\mathcal{A}_8^{(y)} & \vev{12}\vev{13}^3\vev{23}\vev{24}\vev{25}\vev{45}^3 \\
		\mathcal{A}_9^{(y)} & \vev{12}\vev{13}^3\vev{24}^2\vev{25}\vev{35}\vev{45}^2 \\
		\mathcal{A}_{10}^{(y)} & \vev{12}\vev{13}\vev{14}^2\vev{24}^2\vev{25}\vev{35}^3 \\
		\mathcal{A}_{11}^{(y)} & \vev{12}\vev{13}^2\vev{14}\vev{24}^2\vev{25}\vev{35}^2\vev{45} \\
		\mathcal{A}_{12}^{(y)} & \vev{12}^2\vev{13}^2\vev{23}^2\vev{45}^4 \\
		\mathcal{A}_{13}^{(y)} & \vev{12}^2\vev{14}^2\vev{24}^2\vev{35}^4 \\
		\mathcal{A}_{14}^{(y)} & \vev{12}^2\vev{13}^2\vev{23}\vev{24}\vev{35}\vev{45}^3 \\
		\mathcal{A}_{15}^{(y)} & \vev{12}^2\vev{13}\vev{14}\vev{24}^2\vev{35}^3\vev{45} \\
		\mathcal{A}_{15}^{(y)} & \vev{12}^2\vev{13}^2\vev{24}^2\vev{35}^2\vev{45}^2 
	\end{array}.
\end{align}
The coefficient matrix is obtained as
\comment{\begin{eqnarray}\label{eq:coematrix}
	\left(
	\begin{array}{ccc}
		\color{red}{\frac18} & \color{red}{0} & \color{red}{0} \\
		\color{red}{0} & \color{red}{\frac{1}{16}} & \color{red}{0} \\
		\color{red}{0} & \color{red}{0} & \color{red}{\frac18} \\
		0 & -\frac{1}{16} & \frac{1}{16} \\
		\multicolumn{3}{c}{\cdots}
	\end{array}
	\right),
\end{eqnarray}
}
\begin{eqnarray}\label{eq:coematrix}
	\frac{1}{32}\left(
	\begin{array}{cccccccccccccccc}
		\color{red}{0} & \color{red}{0} & \color{red}{0} & \color{red}{0} & \color{red}{0} & \color{red}{0} & \color{red}{0} & \color{red}{0} & \color{red}{0} & \color{red}{0} & \color{red}{0} & \color{red}{1} & \color{red}{0} & \color{red}{0} & \color{red}{0} & \color{red}{0} \\
		\color{red}{1} & \color{red}{0} & \color{red}{0} & \color{red}{0} & \color{red}{0} & \color{red}{0} & \color{red}{0} & \color{red}{0} & \color{red}{0} & \color{red}{0} & \color{red}{0} & \color{red}{0} & \color{red}{0} & \color{red}{0} & \color{red}{0} & \color{red}{0} \\
		\color{red}{1} & \color{red}{0} & \color{red}{0} & \color{red}{0} & \color{red}{0} & \color{red}{-2} & \color{red}{0} & \color{red}{0} & \color{red}{0} & \color{red}{0} & \color{red}{0} & \color{red}{0} & \color{red}{0} & \color{red}{0} & \color{red}{0} & \color{red}{1} \\
		\color{red}{0} & \color{red}{1} & \color{red}{0} & \color{red}{0} & \color{red}{0} & \color{red}{0} & \color{red}{0} & \color{red}{0} & \color{red}{0} & \color{red}{0} & \color{red}{0} & \color{red}{0} & \color{red}{0} & \color{red}{0} & \color{red}{0} & \color{red}{0} \\
		\color{red}{0} & \color{red}{1} & \color{red}{0} & \color{red}{0} & \color{red}{0} & \color{red}{0} & \color{red}{0} & \color{red}{2} & \color{red}{-2} & \color{red}{0} & \color{red}{0} & \color{red}{1} & \color{red}{0} & \color{red}{-2} & \color{red}{0} & \color{red}{1} \\
		\color{red}{1} & \color{red}{0} & \color{red}{0} & \color{red}{0} & \color{red}{0} & \color{red}{-2} & \color{red}{2} & \color{red}{0} & \color{red}{0} & \color{red}{0} & \color{red}{0} & \color{red}{1} & \color{red}{0} & \color{red}{-2} & \color{red}{0} & \color{red}{1} \\
		\color{red}{0} & \color{red}{1} & \color{red}{0} & \color{red}{0} & \color{red}{0} & \color{red}{0} & \color{red}{0} & \color{red}{0} & \color{red}{-2} & \color{red}{0} & \color{red}{0} & \color{red}{0} & \color{red}{0} & \color{red}{0} & \color{red}{0} & \color{red}{1} \\
		\color{red}{1} & \color{red}{1} & \color{red}{0} & \color{red}{0} & \color{red}{0} & \color{red}{-4} & \color{red}{2} & \color{red}{2} & \color{red}{-4} & \color{red}{0} & \color{red}{0} & \color{red}{1} & \color{red}{0} & \color{red}{-6} & \color{red}{0} & \color{red}{6} \\
		\color{red}{0} & \color{red}{0} & \color{red}{0} & \color{red}{0} & \color{red}{0} & \color{red}{0} & \color{red}{0} & \color{red}{0} & \color{red}{0} & \color{red}{0} & \color{red}{0} & \color{red}{0} & \color{red}{1} & \color{red}{0} & \color{red}{0} & \color{red}{0} \\
		\color{red}{1} & \color{red}{0} & \color{red}{0} & \color{red}{0} & \color{red}{2} & \color{red}{-2} & \color{red}{0} & \color{red}{0} & \color{red}{0} & \color{red}{0} & \color{red}{0} & \color{red}{0} & \color{red}{1} & \color{red}{0} & \color{red}{-2} & \color{red}{1} \\
		\color{red}{0} & \color{red}{0} & \color{red}{0} & \color{red}{1} & \color{red}{0} & \color{red}{0} & \color{red}{0} & \color{red}{0} & \color{red}{0} & \color{red}{0} & \color{red}{0} & \color{red}{0} & \color{red}{0} & \color{red}{0} & \color{red}{0} & \color{red}{0} \\
		\color{red}{1} & \color{red}{0} & \color{red}{0} & \color{red}{1} & \color{red}{2} & \color{red}{-4} & \color{red}{0} & \color{red}{0} & \color{red}{0} & \color{red}{-2} & \color{red}{2} & \color{red}{0} & \color{red}{1} & \color{red}{0} & \color{red}{-4} & \color{red}{3} \\
		\color{red}{0} & \color{red}{0} & \color{red}{0} & \color{red}{0} & \color{red}{0} & \color{red}{0} & \color{red}{0} & \color{red}{0} & \color{red}{0} & \color{red}{0} & \color{red}{0} & \color{red}{0} & \color{red}{1} & \color{red}{0} & \color{red}{-2} & \color{red}{1} \\
		\color{red}{0} & \color{red}{0} & \color{red}{0} & \color{red}{1} & \color{red}{0} & \color{red}{0} & \color{red}{0} & \color{red}{0} & \color{red}{0} & \color{red}{0} & \color{red}{-2} & \color{red}{0} & \color{red}{0} & \color{red}{0} & \color{red}{0} & \color{red}{1} \\
		\color{red}{0} & \color{red}{0} & \color{red}{0} & \color{red}{1} & \color{red}{0} & \color{red}{0} & \color{red}{0} & \color{red}{0} & \color{red}{0} & \color{red}{-2} & \color{red}{0} & \color{red}{0} & \color{red}{1} & \color{red}{0} & \color{red}{0} & \color{red}{0} \\
		\color{red}{0} & \color{red}{1} & \color{red}{-2} & \color{red}{1} & \color{red}{0} & \color{red}{0} & \color{red}{0} & \color{red}{2} & \color{red}{-4} & \color{red}{0} & \color{red}{2} & \color{red}{1} & \color{red}{0} & \color{red}{-4} & \color{red}{0} & \color{red}{3} \\
		0 & 1 & -2 & 1 & 0 & 0 & 0 & 0 & -2 & -2 & 4 & 0 & 1 & 0 & -2 & 1 \\
		1 & 0 & 0 & 0 & 2 & -6 & 2 & 0 & 0 & 0 & 0 & 1 & 1 & -4 & -4 & 6 \\
		\multicolumn{16}{c}{\cdots}
	\end{array}
	\right),
\end{eqnarray}
From eq.~(\ref{eq:coematrix}) one can choose 16 independent rows of the matrix as a complete Lorentz basis of $C_{\rm L}^5$, for example, the 16 rows which are marked red in eq.~(\ref{eq:coematrix}). Although the list of Lorentz invariant operators in eq.~\eqref{eq:overcompletebasis} goes on beyond the 18 explicitly written ones, we may stop here because we have already found 16 independent structures, and any other structures can be expressed as linear combinations of the basis just by solving the linear equation of the coordinates. 

Note that the 16 operators are independent under the assumption that the five Weyl tensors are distinguishable by an extra label $\{1,2,3,4,5\}$ as in eq.~\eqref{eq:overcompletebasis}. In gravity EFTs, all the Weyl tensors are indistinguishable, and we should consider the totally symmetric amplitudes that satisfy the Bose symmetry of the gravitons. This is done by directly symmetrizing the monomial y-basis amplitude with the corresponding particle labels. After symmetrization, the resulting amplitudes can still be projected onto the y-basis and obtained the corresponding coordinates forming the coefficient matrix similar to eq.\eqref{eq:coematrix}, from which one can select the independent components and record the y-basis amplitudes they symmetrize from as the so-called f-basis as introduced in Ref.~\cite{Li:2022tec}. These monomial f-basis amplitudes after the selection are reported as the amplitude basis in the remaining sections of our paper. Specifically, for the aforementioned $C_{\rm L}^5$, all the amplitude are symmetrized to a single form $\vev{12}^4\vev{34}^2\vev{35}^2\vev{45}^2+ \text{symm}(12345)$ resulting only one f-basis. Such a technique can also be used to find the f-basis in the presence of repeated gauge field strength tensors and generalize to the cases including repeated fermion with generation larger than one, where non-trivial flavor symmetries can appear and are encoded in the symmetry properties of the Wilson coefficients~\cite{Li:2020xlh,Li:2022tec}. Obviously, this technique can also apply to the operator form and obtain the f-basis monomial for each operator type/class and is exactly the meaning of the ``operator" reported in tables in the rest of our paper, and the relevant concepts are first illustrated in Ref.~\cite{Li:2022tec}.

\section{On-shell Amplitude/Operators in Generic Gravity EFTs}\label{sec:grEFTs}
In this section, we list the Lorentz factors in Gravity EFTs up to dimension 10. We divide them into the following classes. Note that the operators on the right column are not in one-to-one correspondence to the amplitudes on the middle column since some of the corresponding amplitudes are too long to exhibit. Thus we put the independent monomials in this table, and the true physical amplitudes should be understood as the total (anti-)symmetrization of particle number labels of the identical bosons(fermions).
For example, the dim-9 pure gravity operator $C_{\rm L}^4$, in the column of amplitudes, it gives a form of monomial $\langle 12\rangle^4\langle 34\rangle^4$, the physical amplitude should be understood as:
\begin{eqnarray}
   {\cal A}_{phys}^{C_{\rm L}^4} =  \langle 12\rangle^4\langle 34\rangle^4 + \text{symm } (1234).
\end{eqnarray}
Similarly,  as an example, in the Einstein-Yang-Mills theory, for the operator type $C_{\rm L}F_{\rm L}^2F_{\rm R}^2$, the last amplitude should be symmetrized to:
\begin{eqnarray}
    \delta^{A_2A_4}\delta^{A_3A_5}\vev{12}^2\vev{13}^3[45]^2 \to  \left(\delta^{A_2A_4}\delta^{A_3A_5}\vev{12}^2\vev{13}^3[45]^2 + \text{symm } (23) \right)+ \text{symm } (45).
\end{eqnarray}

In the presence of fermion, the Wilson coefficients with flavor indices may present in the amplitude form, they are denoted in the following format\footnote{In principle, the Wilson coefficients also present in the amplitude for the operators without fermions, but in those cases, the corresponding Wilson coefficients are just one-dimension parameters, so we neglect them for notational convenience.}:
\begin{eqnarray}
    {\cal C}^{PT_1,PT_2,\dots}_{\{f_i\},\{f_j\},\dots},
\end{eqnarray}
where $PT_1,PT_2,\dots$ represent the shape of the Young diagrams corresponding to the irreducible representations of the permutation group for the repeated fermion fields \footnote{in this section we only consider two kinds of neutral fermion $\psi$ and $\psi^\dagger$ in the Gravity-Fermion theory.}, and $\{f_i\},\{f_j\},\dots$ are sets of flavor indices separated by commas for each group of repeated fermion fields. For example, for the operator class ${\cal C}_{\rm L}^2\psi^2\psi^{\dagger 2}$ of the Gravity-Fermion theory, the amplitude contains the Wilson coefficients $C^{[2],[2]}_{f_3f_4,f_5,f_6}$, it tells that $f_3,f_4$ are flavor indices for identical particles annihilated by the repeated fields $\psi$, and $f_5,f_6$ are flavor indices for identical particles annihilated by the repeated fields $\psi^\dagger$. The superscript $[2],[2]$, denotes the partition and the shape of the Young diagrams denoting the irreducible representations of the Wilson coefficient under the symmetric group regarding to the permutation of the flavor indices $f_3,f_4$ and $f_5,f_6$ respectively. Here $[2]$ represents $\tiny\yng(2)$, which indicates that $C^{[2],[2]}_{f_3f_4,f_5,f_6}$ are symmetric under the permutation of the either $f_3,f_4$ or $f_5,f_6$. In more general cases, the Wilson coefficients tensor is said to have permutation symmetry of certain Young diagrams, meaning it is unchanged under the acting of the corresponding normalized Young symmetrizer of the standard Young tableau. For example, for the operator of the class $C_{\rm L}\psi^3\psi^\dagger D$, the amplitude of which contains the Wilson coefficients ${\cal C}^{[2,1]}_{f_2f_3f_4,f_5}$, and this coefficients has the following properties:
\begin{eqnarray}
    {\cal Y}\left(\tiny{\young(23,4)}\right)\circ{\cal C}^{[2,1]}_{f_2f_3f_4,f_5}=\frac{1}{3}\left({\cal C}^{[2,1]}_{f_2f_3f_4,f_5}+ {\cal C}^{[2,1]}_{f_3f_2f_4,f_5}- {\cal C}^{[2,1]}_{f_4f_3f_2,f_5}- {\cal C}^{[2,1]}_{f_4f_2f_3,f_5}\right)\equiv{\cal C}^{[2,1]}_{f_2f_3f_4,f_5}.
\end{eqnarray}
On the other hand, for the operator form, we use the Young symmetrizer ${\cal Y}$ to indicate the permutation symmetry of the corresponding Wilson coefficient that contracts with the operator in the Lagrangian term. It is pointed out at the beginning of the section 4.4 in Ref.~\cite{Li:2020xlh} that, after contracting with the flavor indices to form a Lagrangian term, demanding the permutation symmetry on the Wilson coefficient is equivalent to demanding the permutation symmetry on the corresponding operator.   

Lastly, we comment that if the class has only one operator, then the operator and the amplitude are in one-to-one correspondence. For the operator class including fermions, if the operator of the particular permutation symmetry only has one component, then the operators on the right and the amplitudes on the left are one-to-one correspondence. Otherwise, the operator on the left may generate amplitude as a linear combination of the amplitude basis on the left. For example, for the operator class $C_{\rm L}\psi^2\psi^{\dagger 2}D^2$, we have only one operator that is totally symmetric under the exchange of flavor indices $pr$ and $st$, the amplitude it generates must be the one proportional to ${\cal C}^{[2],[2]}_{f_2f_3,f_3f_4}$ on the left after proper symmetrization. This is not the case for the operator that is anti-symmetric under the exchange of the same set of flavor indices in the last two lines in this operator class, the amplitude $\mc{C}_{f_2f_3,f_4f_5}^{[1,1],[1,1]}\vev{12}\vev{13}^3[34][35]+\text{anti-symm}(23)+\text{anti-symm}(45)$ should be generated by the linear combination of the operators in the last two lines in the right column.

\subsection{Pure Gravity}
First, we list the local amplitudes and effective operators for pure gravity. Note that $C_{\rm L}$ and $C_{\rm R}$ are viewed as different fields that annihilate the different particles in the amplitude. 

\begin{align*}
    \begin{array}{|c|c|l|l|}
    \multicolumn{4}{c}{\text{Pure Gravity}}\\
    \hline
    \hline
    \text{dimension} & \text{class} & \text{amplitude} & \text{operator} \\
    \hline
        \text{dim-}6 & C_{\rm L}^3 & \vev{12}^2\vev{13}^2\vev{23}^2 & C_{\rm L}{}_{\mu\nu\rho\sigma}C_{\rm L}^{\mu\nu\lambda\delta}C_{\rm L}{}_{\lambda\delta}{}^{\rho\sigma}\\
        \hline
        \multirow{2}*{dim-8} & C_{\rm L}^4 & \vev{12}^4\vev{34}^4 & C_{\rm L}{}_{\mu\nu\rho\sigma}C_{\rm L}^{\mu\nu\rho\sigma}C_{\rm L}{}_{\lambda\eta\xi\tau}C_{\rm L}^{\lambda\eta\xi\tau}\\
        \cline{2-4}
         & C_{\rm L}^2C_{\rm R}^2 & \vev{12}^4[34]^4 & C_{\rm L}{}_{\mu\nu\rho\sigma}C_{\rm L}^{\mu\nu\rho\sigma}C_{\rm R}{}_{\lambda\eta\xi\tau}C_{\rm R}^{\lambda\eta\xi\tau}\\
         \hline
        \multirow{4}*{dim-10} & C_{\rm L}^4D^2 & \vev{12}^4\vev{34}^4s_{34} & C_{\rm L}{}_{\mu\nu\rho\sigma}C_{\rm L}^{\mu\nu\rho\sigma}D_{\zeta}C_{\rm L}{}_{\lambda\eta\xi\tau}D^{\zeta}C_{\rm L}^{\lambda\eta\xi\tau}\\
        \cline{2-4}
         & C_{\rm L}^2C_{\rm R}^2D^2 & \vev{12}^2[34]^2s_{34} & C_{\rm L}{}_{\mu\nu\rho\sigma}C_{\rm L}^{\mu\nu\rho\sigma}D_{\zeta}C_{\rm R}{}_{\lambda\eta\xi\tau}D^{\zeta}C_{\rm R}^{\lambda\eta\xi\tau}\\
        \cline{2-4}
        & C_{\rm L}^5 & \vev{12}^4\vev{34}^2\vev{35}^2\vev{45}^2 & C_{\rm L}{}_{\mu\nu\lambda\rho}C_{\rm L}{}_{\tau\upsilon\phi\chi}C_{\rm L}^{\mu\nu\eta\xi}C_{\rm L}^{\tau\upsilon\phi\chi}C_{\rm L}{}_{\eta\xi}{}^{\lambda\rho} \\
        \cline{2-4}
         & C_{\rm L}^3C_{\rm R}^2 & \vev{12}^2\vev{13}^2\vev{23}^2[34]^4 & C_{\rm L}{}_{\mu\nu\lambda\rho}C_{\rm R}{}_{\tau\upsilon\phi\chi}C_{\rm L}^{\mu\nu\eta\xi}C_{\rm R}^{\tau\upsilon\phi\chi}C_{\rm L}{}_{\eta\xi}{}^{\lambda\rho}\\
         \hline
    \end{array}
\end{align*}

\subsection{Scalar Gravity (including Goldstone scalar)}
Second, we list the amplitudes and operators in Gravity EFT with an extra scalar. To concentrate on the Lorentz structure, the scalar is set to be hermitian and a singlet under any gauge group.
\begin{align*}
    \begin{array}{|c|c|l|l|}
        \multicolumn{4}{c}{\text{Scalar Gravity}}\\
    \hline
    \hline
    \text{dimension} & \text{class} & \text{amplitude} & \text{operator} \\
    \hline
    \text{dim-}5 & C_{\rm L}^2\phi & \vev{12}^4 & C_{\rm L}{}_{\mu\nu\rho\sigma}C_{\rm L}^{\mu\nu\rho\sigma}\phi\\
        \hline
        \text{dim-}6 & C_{\rm L}^2\phi^2 & \vev{12}^4 & C_{\rm L}{}_{\mu\nu\rho\sigma}C_{\rm L}^{\mu\nu\rho\sigma}\phi^2 \\
        \hline
        \multirow{2}*{dim-7} & C_{\rm L}^3\phi & \vev{12}^2\vev{13}^2\vev{23}^2 & C_{\rm L}{}_{\mu\nu\rho\sigma}C_{\rm L}^{\mu\nu\lambda\eta}C_{\rm L}{}_{\lambda\eta}{}^{\rho\sigma}\phi\\
        \cline{2-4}
         & C_{\rm L}^2\phi^3 & \vev{12}^4 & C_{\rm L}{}_{\mu\nu\rho\sigma}C_{\rm L}^{\mu\nu\rho\sigma}\phi^3\\
        \hline
        \multirow{3}*{dim-8} & C_{\rm L}^2\phi^2D^2 & \vev{12}^4s_{34} & C_{\rm L}{}_{\mu\nu\rho\sigma}C_{\rm L}^{\mu\nu\rho\sigma}D_{\lambda}\phi D^{\lambda}\phi\\
        \cline{2-4}
         & C_{\rm L}^3\phi^2 & \vev{12}^2\vev{13}^2\vev{23}^2 & C_{\rm L}{}_{\mu\nu\rho\sigma}C_{\rm L}^{\mu\nu\lambda\eta}C_{\rm L}{}_{\lambda\eta}{}^{\rho\sigma}\phi^2\\
        \cline{2-4}
         & C_{\rm L}^2\phi^4 & \vev{12}^4 & C_{\rm L}{}_{\mu\nu\rho\sigma}C_{\rm L}^{\mu\nu\rho\sigma}\phi^4\\
        \hline
        \multirow{6}*{dim-9} & C_{\rm L}\phi^3 D^4 & \vev{13}^2\vev{14}^2[34]^2 & C_{\rm L}^{\mu\nu\rho\sigma}\phi D_{\mu}D_{\nu}\phi D_{\rho}D_{\sigma}\phi \\
        \cline{2-4}
         & C_{\rm L}^4\phi & \vev{12}^4\vev{34}^4 & C_{\rm L}{}_{\mu\nu\rho\sigma}C_{\rm L}^{\mu\nu\rho\sigma}C_{\rm L}{}_{\eta\lambda\xi\tau}C_{\rm L}^{\eta\lambda\xi\tau}\phi \\
        \cline{2-4}
         & C_{\rm L}^2C_{\rm R}^2\phi & \vev{12}^4[34]^4 & C_{\rm L}{}_{\mu\nu\rho\sigma}C_{\rm L}^{\mu\nu\rho\sigma}C_{\rm R}{}_{\eta\lambda\xi\tau}C_{\rm R}^{\eta\lambda\xi\tau}\phi \\
        \cline{2-4}
         & C_{\rm L}^2\phi^3 D^2 & \vev{12}^4s_{45} & C_{\rm L}{}_{\mu\nu\rho\sigma}C_{\rm L}^{\mu\nu\rho\sigma}\phi D_{\lambda}\phi D^{\lambda}\phi \\
        \cline{2-4}
         & C_{\rm L}^3\phi^3 & \vev{12}^2\vev{13}^2\vev{23}^2 & C_{\rm L}{}_{\mu\nu\rho\sigma}C_{\rm L}^{\mu\nu\lambda\eta}C_{\rm L}{}_{\lambda\eta}{}^{\rho\sigma}\phi^3\\
        \cline{2-4}
        & C_{\rm L}^2\phi^5 & \vev{12}^4 & C_{\rm L}{}_{\mu\nu\rho\sigma}C_{\rm L}^{\mu\nu\rho\sigma}\phi^5 \\
        \hline
        \multirow{10}*{dim-10} & \multirow{2}*{$C_{\rm L}^2\phi^2D^4$} & \vev{12}^4s_{34}^2 & C_{\rm L}{}_{\mu\nu\rho\sigma}C_{\rm L}^{\mu\nu\rho\sigma} D_{\lambda}D_{\eta}\phi D^{\lambda}D^{\eta}\phi \\
         \cline{3-4}
         & & \vev{12}^2\vev{13}^2\vev{24}^2[34]^2 & C_{\rm L}{}_{\mu\nu\rho\sigma}C_{\rm L}^{\mu\nu\lambda\eta} D_{\lambda}D_{\eta}\phi D^{\rho}D^{\sigma}\phi \\
         \cline{2-4}
         & C_{\rm L}C_{\rm R}\phi^2D^4 & \vev{13}^4[23]^4 & C_{\rm L}{}_{\mu\nu\rho\sigma}C_{\rm R}^{\eta\nu\lambda\sigma} \phi D_{\lambda}D_{\eta}D^{\rho}D^{\mu}\phi \\
         \cline{2-4}
         & C_{\rm L}^3\phi^2D^2 & \vev{12}^2\vev{13}^2\vev{23}^2s_{45} & C_{\rm L}{}_{\mu\nu\rho\sigma}C_{\rm L}^{\mu\nu\lambda\eta}C_{\rm L}{}_{\lambda\eta}{}^{\rho\sigma}D_{\xi}\phi D^{\xi}\phi \\
         \cline{2-4}
         & C_{\rm L}\phi^4D^4 & \vev{14}^2\vev{15}^2[45]^2 & C_{\rm L}^{\mu\nu\rho\sigma}\phi^2 D_{\mu}D_{\nu}\phi D_{\rho}D_{\sigma}\phi \\
         \cline{2-4}
         & C_{\rm L}^4\phi^2 & \vev{12}^4\vev{34}^4 & C_{\rm L}{}_{\mu\nu\rho\sigma}C_{\rm L}^{\mu\nu\rho\sigma}C_{\rm L}{}_{\eta\lambda\xi\tau}C_{\rm L}^{\eta\lambda\xi\tau}\phi^2 \\
        \cline{2-4}
         & C_{\rm L}^2C_{\rm R}^2\phi^2 & \vev{12}^4[34]^4 & C_{\rm L}{}_{\mu\nu\rho\sigma}C_{\rm L}^{\mu\nu\rho\sigma}C_{\rm R}{}_{\eta\lambda\xi\tau}C_{\rm R}^{\eta\lambda\xi\tau}\phi^2 \\
         \cline{2-4}
         & C_{\rm L}^2\phi^4D^2 & \vev{12}^4s_{56} & C_{\rm L}{}_{\mu\nu\rho\sigma}C_{\rm L}^{\mu\nu\rho\sigma}\phi^2 D_{\lambda}\phi D^{\lambda}\phi \\
         \cline{2-4}
         & C_{\rm L}^3\phi^4 & \vev{12}^2\vev{13}^2\vev{23}^2 & C_{\rm L}{}_{\mu\nu\rho\sigma}C_{\rm L}^{\mu\nu\lambda\eta}C_{\rm L}{}_{\lambda\eta}{}^{\rho\sigma}\phi^4 \\
         \cline{2-4}
         & C_{\rm L}^2\phi^6 & \vev{12}^4 & C_{\rm L}{}_{\mu\nu\rho\sigma}C_{\rm L}^{\mu\nu\rho\sigma}\phi^6 \\
        \hline
    \end{array}
\end{align*}

In addition, we also consider the case when the scalar is a Goldstone boson for which amplitudes must satisfy the Adler's zero condition \cite{Dai:2020cpk,Sun:2022ssa,Low:2022iim,Sun:2022snw}. For simplicity, the Goldstone field $a$ comes from a broken $U(1)$ and thus has only one flavor, and we adopt the operator building block $U = e^{ia/f}$. In the leading order, the following operators should be expressed by the building block $a$ via the simple relation $D_\mu U \simeq D_\mu a$.
\begin{align*}
    \begin{array}{|c|c|l|l|}
    \multicolumn{4}{c}{\text{including Goldstone scalar}}\\
    \hline
    \hline
    \text{dimension} & \text{class} & \text{amplitude} & \text{operator} \\
    \hline
        \text{dim-}8 & C_{\rm L}^2U^2D^2 & \vev{12}^4s_{34} & C_{\rm L}{}_{\mu\nu\rho\sigma}C_{\rm L}^{\mu\nu\rho\sigma}D_{\lambda}U D^{\lambda}U\\
        \hline
        \text{dim-}9 & C_{\rm L}U^3D^4 & \vev{13}^2\vev{14}^2[34]^2 & C_{\rm L}{}_{\mu\nu\rho\sigma}D^{\mu}U D^{\nu}U D^{\rho}D^{\sigma}U \\
        \hline
        \multirow{5}*{dim-10} & \multirow{2}*{$C_{\rm L}^2U^2D^4$} & \vev{12}^4s_{34}^2 & C_{\rm L}{}_{\mu\nu\rho\sigma}C_{\rm L}^{\mu\nu\rho\sigma}D_{\lambda}D_{\eta}U D^{\lambda}D^{\eta}U \\
         \cline{3-4}
         & & \vev{12}^2\vev{13}^2\vev{24}^2[34]^2 & C_{\rm L}{}_{\mu\nu\rho\sigma}C_{\rm L}^{\mu\nu\lambda\eta} D_{\lambda}D_{\eta}U D^{\rho}D^{\sigma}U \\
         \cline{2-4}
        & C_{\rm L}C_{\rm R}U^2D^4 & \vev{13}^4[23]^4 & C_{\rm L}{}_{\mu\nu\rho\sigma}C_{\rm R}^{\eta\nu\lambda\sigma} D_{\lambda}D_{\eta}U D^{\rho}D^{\mu}U \\
         \cline{2-4}
         & C_{\rm L}^3U^2D^2 & \vev{12}^2\vev{13}^2\vev{23}^2s_{45} & C_{\rm L}{}_{\mu\nu\rho\sigma}C_{\rm L}^{\mu\nu\lambda\eta}C_{\rm L}{}_{\lambda\eta}{}^{\rho\sigma}D_{\xi}U D^{\xi}U\\
         \cline{2-4}
         & C_{\rm L}U^4D^4 & \vev{12}\vev{13}\vev{14}\vev{15}[23][45] & C_{\rm L}^{\mu\nu\rho\sigma}D_{\mu}U D_{\nu}U D_{\rho}U D_{\sigma}U \\
        \hline
    \end{array}
\end{align*}

\subsection{Einstein-Yang-Mills}
Here we list the amplitudes and operators in Gravity EFT with an extra $SU(3)$ group. The gauge boson in $SU(3)$ group is written as $G_{\rm L}$ and $G_{\rm R}$ responsible for the creation of gauge boson particles of two polarizations respectively. $A_i$ in amplitudes is the gauge index for particle $i$. They have the same formula for general $SU(N)$ group, since the structural constants $\delta^{AB}$, $f^{ABC}$ and $d^{ABC}$ also apply to $SU(N)$ via Fierz identity
$
T^A_{ab}T^A_{cd}=\delta_{ad}\delta_{cb}-\frac{1}{N}\delta_{ab}\delta_{cd}$.
Thus the above structural constants are enough In the case of not much gauge boson. 
\begin{align*}
    \begin{array}{|c|c|l|l|}
    \multicolumn{4}{c}{\text{including Yang Mills}}\\
    \hline
    \hline
    \text{dimension} & \text{class} & \text{amplitude} & \text{operator} \\
    \hline
    \text{dim-}6 & C_{\rm L}F_{\rm L}^2 & \delta^{A_2A_3}\langle 12\rangle^2\langle 13\rangle^2 & C_{\rm L}{}_{\mu\nu\rho\sigma}G_{\rm L}^{A\mu\nu}G_{\rm L}^{A\rho\sigma} \\
    \hline
    \multirow{3}*{dim-8} & \multirow{2}*{$C_{\rm L}^2F_{\rm L}^2$} & \delta^{A_3A_4}\langle 12\rangle^4\langle 34\rangle^2 & C_{\rm L}{}_{\mu\nu\rho\sigma}C_{\rm L}{}_{\mu\nu\rho\sigma}G^A_{\rm L}{}_{\lambda\zeta}G_{\rm L}^{A\lambda\zeta} \\
    \cline{3-4}
     & & \delta^{A_3A_4}\vev{12}^2\vev{13}^2\vev{24}^2 & C_{\rm L}{}_{\mu\nu\rho\sigma}C_{\rm L}{}_{\mu\nu\rho\sigma}G^A_{\rm L}{}_{\lambda\zeta}G_{\rm L}^{A\lambda\zeta} \\
     \cline{2-4}
     & C_{\rm R}^2F_{\rm L}^2 & \delta^{A_1A_2}\vev{12}^2[34]^4 & C_{\rm R}{}_{\mu\nu\rho\sigma}C_{\rm R}{}_{\mu\nu\rho\sigma}G^A_{\rm L}{}_{\lambda\zeta}G_{\rm L}^{A\lambda\zeta} \\
          \hline
     \multirow{20}*{dim-10} & C_{\rm L}F_{\rm L}^3D^2 & f^{A_2A_3A_4}\vev{12}^2\vev{13}\vev{14}\vev{34}s_{34} & f^{ABC}C_{\rm L}{}_{\mu\nu\rho\sigma}G_{\rm L}^{A\mu\nu}D^{\xi}G_{\rm L}^{B\rho\lambda}D_{\xi}G^C_{\rm L}{}_{\lambda}{}^{\sigma}\\
     \cline{2-4}
     & \multirow{2}*{$C_{\rm L}^2F_{\rm L}^2D^2$} & \delta^{A_3A_4}\vev{12}^4\vev{34}^2s_{34} & C_{\rm L}{}_{\mu\nu\rho\sigma}C_{\rm L}^{\mu\nu\rho\sigma} D_{\mu'}G^A_{\rm L}{}_{\nu'\rho'}D^{\mu'}G_{\rm L}^{A\nu'\rho'} \\
     & & \delta^{A_3A_4}\vev{12}^2\vev{13}^2\vev{24}^2s_{34} & C_{\rm L}{}_{\mu\nu\rho\sigma}C_{\rm L}^{\mu\nu\lambda\xi}D^{\eta}G_{\rm L}^{A\rho\sigma} D_{\eta}G^A_{\rm L}{}_{\lambda\xi}\\
    \cline{2-4}
     & C_{\rm L}F_{\rm L}^2F_{\rm R}D^2 & f^{A_2A_3A_4}\vev{12}\vev{13}^3\vev{23}[34]^2 & D_{\mu}C_{\rm L}{}_{\lambda\rho\eta\nu}D^{\nu}G_{\rm L}^{A\rho\mu}G^B_{\rm L}{}_{\xi}{}^{\eta}G_{\rm R}^{C\xi\lambda} \\
    \cline{2-4}
     & C_{\rm L}^2F_{\rm L}F_{\rm R}D^2 & \delta^{A_3A_4}\vev{12}^2\vev{13}^2\vev{23}^2[34]^2 & D_{\mu}D_{\nu}C_{\rm L}{}_{\lambda\rho\eta\xi}C_{\rm L}^{\rho\nu\xi\mu}G^A_{\rm L}{}_{\tau}{}^{\eta}G_{\rm R}^{A\tau\lambda} \\
     \cline{2-4}
     & C_{\rm L}^2F_{\rm R}^2D^2 & \delta^{A_3A_4}\vev{12}^4[34]^2s_{34} & C_{\rm L}{}_{\mu\nu\rho\sigma}C_{\rm L}^{\mu\nu\rho\sigma}D_{\lambda}G^A_{\rm L}{}_{\eta\xi}D^{\lambda}G_{\rm L}^{A\eta\xi}\\
     \cline{2-4}
     & F_{\rm L}^2F_{\rm R}C_{\rm R}D^2 & f^{A_1A_2A_3}\vev{12}\vev{13}\vev{23}[34]^4 & f^{ABC} D_{\mu}G^A_{\rm L}{}_{\rho\sigma}D_{\nu}G^B_{\rm L}{}_{\lambda\eta}G_{\rm R}^{C\mu\nu}C_{\rm R}^{\rho\sigma\lambda\eta}\\
     \cline{2-4}
     & C_{\rm L}C_{\rm R}F_{\rm L}F_{\rm R}D^2 & \delta^{A_2A_3}\vev{12}^2\vev{13}^2[34]^4 & C_{\rm L}{}_{\mu\nu\rho\sigma} D^{\mu}D^{\rho}G^A_{\rm L}{}_{\eta}{}^{\nu}G^A_{\rm R}{}_{\lambda\xi}C_{\rm R}^{\sigma\eta\lambda\xi}\\
     \cline{2-4}
     & \multirow{3}*{$C_{\rm L}F_{\rm L}^4$} & d^{A_2A_3E}d^{A_4A_5E}\vev{12}^2\vev{13}^2\vev{45}^2 & d^{ABE}d^{CDE}C_{\rm L}{}_{\mu\nu\rho\sigma}G_{\rm L}^{A\mu\nu}G_{\rm L}^{B\rho\sigma}G_{\rm L}^{C\lambda\eta}G^D_{\rm L}{}_{\lambda\eta}\\
     \cline{3-4}
     & & d^{A_2A_3E}d^{A_4A_5E}\vev{12}^2\vev{14}^2\vev{35}^2 & d^{ABE}d^{CDE}C_{\rm L}{}_{\mu\nu\rho\sigma}G_{\rm L}^{A\mu\nu}G_{\rm L}^{C\rho\sigma}G_{\rm L}^{B\lambda\eta}G^D_{\rm L}{}_{\lambda\eta}\\
     \cline{3-4}
     & & f^{A_2A_3E}f^{A_4A_5E}\vev{12}^2\vev{14}^2\vev{35}^2 & f^{ABE}f^{CDE}C_{\rm L}{}_{\mu\nu\rho\sigma}G_{\rm L}^{A\mu\nu}G_{\rm L}^{C\rho\sigma}G_{\rm L}^{B\lambda\eta}G^D_{\rm L}{}_{\lambda\eta}\\
     \cline{2-4}
     & C_{\rm L}^2F_{\rm L}^3 & f^{A_3A_4A_5}\vev{12}^4\vev{34}\vev{35}\vev{45} & f^{ABC}C_{\rm L}{}_{\mu\nu\rho\sigma}C_{\rm L}^{\mu\nu\rho\sigma}G^A_{\rm L}{}_{\lambda\eta}G_{\rm L}^{B\eta\xi}G^C_{\rm L}{}_{\xi}{}^{\lambda}\\
     \cline{2-4}
     & \multirow{2}*{$C_{\rm L}^3F_{\rm L}^2$} & \delta^{A_4A_5}\vev{12}^4\vev{34}^2\vev{35}^2 & C_{\rm L}{}_{\mu\nu\rho\sigma}C_{\rm L}^{\mu\nu\rho\sigma}C_{\rm L}{}_{\lambda\eta\tau\xi}G_{\rm L}^{A\lambda\eta}G_{\rm L}^{A\tau\xi}\\
     \cline{3-4}
     & & \delta^{A_4A_5}\vev{12}^2\vev{13}^2\vev{23}^2\vev{45}^2 & C_{\rm L}{}_{\mu\nu\rho\sigma}C_{\rm L}^{\mu\nu\lambda\eta}C_{\rm L}{}_{\lambda\eta}{}^{\rho\sigma}G_{\rm L}^{A\tau\xi}G^A_{\rm L}{}_{\tau\xi}\\
     \cline{2-4}
     & C_{\rm L}^2F_{\rm R}^3 & f^{A_3A_4A_5}\vev{12}^4[34][35][45] & f^{ABC}C_{\rm L}{}_{\mu\nu\rho\sigma}C_{\rm L}^{\mu\nu\rho\sigma}G^A_{\rm R}{}_{\lambda\eta}G_{\rm R}^{B\eta\xi}G^C_{\rm R}{}_{\xi}{}^{\lambda}\\
     \cline{2-4}
     & C_{\rm L}F_{\rm L}^2C_{\rm R}^2 & \delta^{A_2A_3}\vev{12}^2\vev{13}^2[45]^4 & C_{\rm L}{}_{\mu\nu\rho\sigma}G_{\rm L}^{A\mu\nu}G_{\rm L}^{A\rho\sigma}C_{\rm R}{}_{\lambda\eta\tau\xi}C_{\rm R}^{\lambda\eta\tau\xi}\\
     \cline{2-4}
     & \multirow{3}*{$C_{\rm L}F_{\rm L}^2F_{\rm R}^2$} & d^{A_2A_3E}d^{A_4A_5E}\vev{12}^2\vev{13}^2[45]^2 & d^{ABE}d^{CDE}C_{\rm L}{}_{\mu\nu\rho\sigma}G_{\rm L}^{A\mu\nu}G_{\rm L}^{B\rho\sigma}G^C_{\rm R}{}_{\lambda\eta}G_{\rm R}^{D\lambda\eta}\\
     \cline{3-4}
     & & \delta^{A_2A_3}\delta^{A_4A_5}\vev{12}^2\vev{13}^3[45]^2 & C_{\rm L}{}_{\mu\nu\rho\sigma}G_{\rm L}^{A\mu\nu}G_{\rm L}^{A\rho\sigma}G^B_{\rm R}{}_{\lambda\eta}G_{\rm R}^{B\lambda\eta}\\
     \cline{3-4}
     & & \delta^{A_2A_4}\delta^{A_3A_5}\vev{12}^2\vev{13}^3[45]^2 & C_{\rm L}{}_{\mu\nu\rho\sigma}G_{\rm L}^{A\mu\nu}G_{\rm L}^{B\rho\sigma}G^A_{\rm R}{}_{\lambda\eta}G_{\rm R}^{B\lambda\eta}\\
     \cline{2-4}
     & C_{\rm L}^3F_{\rm R}^2 & \delta^{A_4A_5}\vev{12}^2\vev{13}^2\vev{23}^2[45]^2 & C_{\rm L}{}_{\mu\nu\rho\sigma}C_{\rm L}^{\mu\nu\lambda\eta}C_{\rm L}{}_{\lambda\eta}{}^{\rho\sigma}G^A_{\rm R}{}_{\tau\xi}G_{\rm R}^{A\tau\xi}\\
     \hline
    \end{array}
\end{align*}

\subsection{Gravity + Fermion}
Here we list the amplitudes and operators in Gravity EFT with two extra neutral fermions with opposite helicities. Similarly, the fermions are singlet under any gauge group, but carry flavor indices $f_i$ for each fermion $i$. The Wilson coefficients $\cal C$s are also irreducible flavor tensors of the corresponding repeated fermion fields, and the superscripts denote the shapes of the Young diagrams representing the permutation symmetries of the corresponding flavor indices. The amplitudes and operators are similar for massive fermion, except the massive spinors have extra little group indices \cite{Li:2020tsi,Arkani-Hamed:2017jhn}. 
\begin{align*}
    \begin{array}{|c|c|l|l|}
    \multicolumn{4}{c}{\text{including Fermion}}\\
    \hline
    \hline
    \text{dimension} & \text{class} & \text{amplitude} & \text{operator} \\
    \hline
        \multirow{2}*{dim-7} & C_{\rm L}^2\psi^2 & \mc{C}_{f_3f_4}^{[2]}\vev{12}^4\vev{34} & \mc{Y}\left[{\tiny \young(pr)}\right]C_{\rm L}{}_{\mu\nu\rho\sigma}C_{\rm L}^{\mu\nu\rho\sigma} (\psi_p\psi_r) \\
         \cline{2-4}
         & C_{\rm L}^2\psi^{\dagger 2} & \mc{C}_{f_3f_4}^{[2]}\vev{12}^4[34] & \mc{Y}\left[{\tiny \young(pr)}\right]C_{\rm L}{}_{\mu\nu\rho\sigma}C_{\rm L}^{\mu\nu\rho\sigma} (\psi^{\dagger}_p\psi^{\dagger}_r)\\
        \hline
        \text{dim-8}
         & C_{\rm L}\psi^{4} & \mc{C}_{f_2f_3f_4f_5}^{[4]}\vev{12}\vev{13}\vev{14}\vev{15} & \mc{Y}\left[{\tiny \young(p,r,s,t)}\right]C_{\rm L}^{\mu\nu\rho\sigma}\left( \psi_p\sigma_{\mu\nu} \psi_r\right)\left( \psi_s\sigma_{\rho\sigma} \psi_t\right) \\
        \hline
        \multirow{6}*{dim-9} & \multirow{2}*{$C_{\rm L}^2\psi^2D^2$} & \mc{C}_{f_3f_4}^{[2]}\vev{12}^4\vev{34}s_{34} & \mc{Y}\left[{\tiny \young(pr)}\right]C_{\rm L}{}_{\mu\nu\rho\sigma}C_{\rm L}^{\mu\nu\rho\sigma}\left(D_{\lambda}\psi_p D^{\lambda}\psi_r\right) \\
        \cline{3-4}
        & & \mc{C}_{f_3f_4}^{[2]}\vev{12}^2\vev{13}^2\vev{24}^2[34] & \mc{Y}\left[{\tiny \young(pr)}\right]C_{\rm L}{}_{\mu\nu\rho}{}^{\eta}D^{\lambda}C_{\rm L}^{\mu\nu\rho\sigma}\left(D_{\lambda}\psi_p \sigma_{\eta\sigma}\psi_r\right) \\
         \cline{2-4}
         & C_{\rm L}^2\psi^{\dagger 2}D^2 & \mc{C}_{f_3f_4}^{[2]}\vev{12}^4[34]s_{34} & \mc{Y}\left[{\tiny \young(pr)}\right]C_{\rm L}{}_{\mu\nu\rho\sigma}C_{\rm L}^{\mu\nu\rho\sigma}\left(D_{\lambda}\psi^{\dagger}_p D^{\lambda}\psi^{\dagger}_r\right)\\
         \cline{2-4}
         & C_{\rm L}^3\psi^2 & \mc{C}_{f_4f_5}^{[2]}\vev{12}^2\vev{13}^2\vev{23}^2\vev{45} & \mc{Y}\left[{\tiny \young(pr)}\right]C_{\rm L}{}_{\mu\nu\rho\sigma}C_{\rm L}^{\mu\nu\lambda\eta}C_{\rm L}{}_{\lambda\eta}{}^{\rho\sigma}\left(\psi_p \psi_r\right) \\
         \cline{2-4}
         & C_{\rm L}^3\psi^{\dagger 2} & \mc{C}_{f_4f_5}^{[2]}\vev{12}^2\vev{13}^2\vev{23}^2[45] & \mc{Y}\left[{\tiny \young(pr)}\right]C_{\rm L}{}_{\mu\nu\rho\sigma}C_{\rm L}^{\mu\nu\lambda\eta}C_{\rm L}{}_{\lambda\eta}{}^{\rho\sigma}\left(\psi^{\dagger}_p \psi^{\dagger}_r\right) \\
         \cline{2-4}
         & C_{\rm L}\psi^3\psi^{\dagger}D & \mc{C}_{f_2f_3f_4,f_5}^{[2,1]}\vev{12}\vev{13}^2\vev{14}[35] & \mc{Y}\left[{\tiny \young(pr,s)}\right] C_{\rm L}{}_{\mu\nu\rho\eta}\left(\psi_p\sigma^{\mu\nu}\psi_r\right)\left(D^{\rho}\psi_s\sigma^{\eta}\psi^{\dagger}_t\right) \\
        \hline
        \multirow{13}*{dim-10} & C_{\rm L}^2\psi\psi^{\dagger}D^3 & {\cal C}_{f_3,f_4}\vev{12}^2\vev{13}^2\vev{23}\vev{24}[34]^2 & C_{\rm L}{}_{\mu\nu\rho\eta}C_{\rm L}^{\mu\nu\lambda\xi}\left(D^{\rho}D^{\eta}\psi_p\sigma_{\lambda}D_{\xi}\psi^{\dagger}_r \right) \\
        \cline{2-4}
        & C_{\rm L}C_{\rm R}\psi\psi^{\dagger}D^3 & {\cal C}_{f_3,f_4}\vev{13}\vev{14}^3[24]^4 & C_{\rm L}{}_{\mu\nu\rho\eta}C_{\rm R}^{\lambda\nu\xi\eta}\left(D^{\mu}D^{\rho}\psi_p\sigma_{\lambda}D_{\xi}\psi^{\dagger}_r \right) \\
        \cline{2-4}
        &\multirow{4}*{$C_{\rm L}\psi^4D^2$} & \mc{C}_{f_2f_3f_4f_5}^{[4]}\vev{12}\vev{13}^2\vev{15}\vev{45}[35] & \mc{Y}\left[{\tiny \young(prst)}\right]C_{\rm L}{}_{\mu\nu\rho\eta}\left(D^{\mu}\psi_p  \sigma^{\rho\eta}D^{\nu}\psi_r\right)\left(\psi_s\psi_t\right) \\
        \cline{3-4}
        & & \mc{C}_{f_2f_3f_4f_5}^{[3,1]}\vev{12}\vev{13}^2\vev{15}\vev{45}[35] & \mc{Y}\left[{\tiny \young(prs,t)}\right]C_{\rm L}{}_{\mu\nu\rho\eta}\left(D^{\mu}\psi_p  \sigma^{\rho\eta}D^{\nu}\psi_r\right)\left(\psi_s\psi_t\right)\\
        \cline{3-4}
        & & \mc{C}_{f_2f_3f_4f_5}^{[2,2]}\vev{12}\vev{13}^2\vev{15}\vev{45}[35] & \mc{Y}\left[{\tiny \young(pr,st)}\right]C_{\rm L}{}_{\mu\nu\rho\eta}\left(D^{\mu}\psi_p  \sigma^{\rho\eta}D^{\nu}\psi_r\right)\left(\psi_s\psi_t\right) \\
        \cline{3-4}
        & & \mc{C}_{f_2f_3f_4f_5}^{[2,1,1]}\vev{12}\vev{13}^2\vev{15}\vev{45}[35] & \mc{Y}\left[{\tiny \young(pr,s,t)}\right]C_{\rm L}{}_{\mu\nu\rho\eta}\left(D^{\mu}\psi_p  \sigma^{\rho\eta}\psi_r\right)\left(D^{\nu}\psi_s\psi_t\right) \\
        \cline{2-4}
        &\multirow{4}*{$C_{\rm L}\psi^2\psi^{\dagger 2}D^2$} & \mc{C}_{f_2f_3,f_4f_5}^{[2],[2]}\vev{12}\vev{13}^2\vev{15}[35][45] & \mc{Y}\left[{\tiny \young(pr)},{\tiny \young(st)}\right]C_{\rm L}{}_{\mu\nu\rho\eta}\left(D^{\mu}\psi_p\sigma^{\rho}\psi^{\dagger}_s\right)\left(D^{\nu}\psi_r\sigma^{\eta}\psi^{\dagger}_t\right) \\
        \cline{3-4}
        & & \mc{C}_{f_2f_3,f_4f_5}^{[2],[1,1]}\vev{12}\vev{13}^3[34][35] & \mc{Y}\left[{\tiny \young(pr)},{\tiny \young(s,t)}\right]C_{\rm L}{}_{\mu\nu\rho\eta}\left(D^{\mu}D^{\rho}\psi_p\sigma^{\nu}\psi^{\dagger}_s\right)\left(\psi_r\sigma^{\eta}\psi^{\dagger}_t\right) \\
        \cline{3-4}
        & & \mc{C}_{f_2f_3,f_4f_5}^{[1,1],[1,1]}\vev{12}\vev{13}^3[34][35] & \mc{Y}\left[{\tiny \young(p,r)},{\tiny \young(s,t)}\right]C_{\rm L}{}_{\mu\nu\rho\eta}\left(D^{\mu}D^{\rho}\psi_p\sigma^{\nu}\psi^{\dagger}_s\right)\left(\psi_r\sigma^{\eta}\psi^{\dagger}_t\right) \\
        \cline{3-4}
        & & \mc{C}_{f_2f_3,f_4f_5}^{[1,1],[1,1]}\vev{12}\vev{13}\vev{14}\vev{15}[45]^2 & \mc{Y}\left[{\tiny \young(p,r)},{\tiny \young(s,t)}\right]C_{\rm L}{}_{\mu\nu\rho\eta}\left(D^{\mu}\psi_p\sigma^{\nu}\psi^{\dagger}_s\right)\left(D^{\rho}\psi_r\sigma^{\eta}\psi^{\dagger}_t\right)\\
        \cline{2-4}
        & \multirow{2}*{$C_{\rm L}^2\psi^4$} & \mc{C}_{f_3f_4f_5f_6}^{[2,2]}\vev{12}^4\vev{34}\vev{56} & \mc{Y}\left[{\tiny \young(pr,st)}\right]C_{\rm L}{}_{\mu\nu\rho\eta}C_{\rm L}^{\mu\nu\rho\eta}\left(\psi_p\psi_r\right)\left(\psi_s\psi_t\right) \\
        \cline{3-4}
        & & \mc{C}_{f_3f_4f_5f_6}^{[1,1,1,1]}\vev{12}^2\vev{13}\vev{14}\vev{25}\vev{26} & \mc{Y}\left[{\tiny \young(p,r,s,t)}\right]C_{\rm L}{}_{\mu\nu\lambda\rho}C_{\rm L}{}_{\eta\xi}{}^{\lambda\rho}\left(\psi_p\sigma^{\mu\nu}\psi_r\right)\left(\psi_s\sigma^{\eta\xi}\psi_t\right) \\
        \cline{2-4}
        & C_{\rm L}^2\psi^2\psi^{\dagger 2} & \mc{C}_{f_3f_4,f_5f_6}^{[2],[2]}\vev{12}^4\vev{34}[56] & \mc{Y}\left[{\tiny \young(pr),\young(st)}\right]C_{\rm L}{}_{\mu\nu\rho\eta}C_{\rm L}^{\mu\nu\rho\eta}\left(\psi_p\psi_r\right)\left(\psi^{\dagger}_s\psi^{\dagger}_t\right)\\
        \cline{2-4}
        & C_{\rm R}^2\psi^4 & \mc{C}_{f_3f_4f_5f_6}^{[2,2]}[12]^4\vev{34}\vev{56} & \mc{Y}\left[{\tiny \young(pr,st)}\right]C_{\rm R}{}_{\mu\nu\rho\eta}C_{\rm R}^{\mu\nu\rho\eta}\left(\psi_p\psi_r\right)\left(\psi_s\psi_t\right) \\
        \hline
    \end{array}
\end{align*}

\section{GRSMEFT Operator/Amplitude Bases up to Dim-10}\label{sec:GRSMEFTopes}
In this section, we list the on-shell amplitude bases and the operator bases in the GRSMEFT up to the mass dimension 10. Note that the amplitudes are not in one-to-one correspondence with the operators even though they are put side-by-side, but these are both complete and independent bases related by linear transformations from the operator-amplitude correspondence perspective. Similar to the amplitude in the previous section, the amplitude in the GRSMEFT should totally symmetrize over the particle number indices associated with the identical particles of bosons (including graviton, gauge bosons, and Higgs boson), and totally anti-symmetrize over the particle number indices associated with the identical fermions as explained by examples at the beginning of section.~\ref{sec:grEFTs}

\subsection{Field Contents}

The field content of GRSMEFT are listed in table~\ref{tab:GRSMEFT-field-content}. The operators are constructed in two-component Weyl spinor notation and then converted to four-component Dirac spinor notation using the following relations:
\begin{align}
q_{\rm{L}}=\begin{pmatrix}Q\\0\end{pmatrix},\quad u_{\rm{R}}=\left(\begin{array}{c}0\\u_{_\mathbb{C}}^{\dagger}\end{array}\right),\quad d_{\rm R}=\left(\begin{array}{c}0\\d_{_\mathbb{C}}^{\dagger}\end{array}\right),\quad l_{\rm L}=\left(\begin{array}{c}L\\0\end{array}\right),\quad e_{\rm R}=\left(\begin{array}{c}0\\e_{_\mathbb{C}}^{\dagger}\end{array}\right).\\
\bar{q}_{\rm{L}}=\left(0\,,\,Q^{\dagger} \right),\quad \bar{u}_{\rm{R}}=\left(u_{_\mathbb{C}}\,,\,0 \right),\quad \bar{d}_{\rm R}=\left(d_{_\mathbb{C}}\,,\,0\right),\quad \bar{l}_{\rm L}=\left(0\,,\,L^{\dagger}\right),\quad \bar{e}_{\rm R}=\left(e_{_\mathbb{C}}\,,\,0\right).
\end{align}

\begin{table}[h]
    	\begin{center}
    		\begin{tabular}{|c|cc|ccc|c|}
    			\hline
    			\text{Fields} & $SU(2)_{l}\times SU(2)_{r}$	& $h$ & $SU(3)_{C}$ & $SU(2)_{W}$ & $U(1)_{Y}$ &  Flavor\tabularnewline
    			\hline
    			$C_{\rm L \alpha\beta\gamma\delta}$ & $\left(2,0\right)$ & $-2$ & $\boldsymbol{1}$ & $\boldsymbol{1}$ & 0  & $1$\tabularnewline
    			\hline
    			$G_{\rm L\alpha\beta}^A$   & $\left(1,0\right)$  & $-1$    & $\boldsymbol{8}$ & $\boldsymbol{1}$ & 0  & $1$\tabularnewline
    			$W_{\rm L\alpha\beta}^I$   & $\left(1,0\right)$  & $-1$           & $\boldsymbol{1}$ & $\boldsymbol{3}$ & 0  & $1$\tabularnewline
    			$B_{\rm L\alpha\beta}$   & $\left(1,0\right)$    & $-1$        & $\boldsymbol{1}$ & $\boldsymbol{1}$ & 0  & $1$\tabularnewline
    			\hline
    			$L_{\alpha i}$     & $\left(\frac{1}{2},0\right)$  & $-1/2$  & $\boldsymbol{1}$ & $\boldsymbol{2}$ & $-1/2$  & $n_f$\tabularnewline
    			$e_{_\mathbb{C}\alpha}$ & $\left(\frac{1}{2},0\right)$ & $-1/2$   & $\boldsymbol{1}$ & $\boldsymbol{1}$ & $1$  & $n_f$\tabularnewline
    			$Q_{\alpha ai}$     & $\left(\frac{1}{2},0\right)$ & $-1/2$   & $\boldsymbol{3}$ & $\boldsymbol{2}$ & $1/6$  & $n_f$\tabularnewline
    			$u_{_\mathbb{C}\alpha}^a$ & $\left(\frac{1}{2},0\right)$ & $-1/2$   & $\overline{\boldsymbol{3}}$ & $\boldsymbol{1}$ & $-2/3$  & $n_f$\tabularnewline
    			$d_{_\mathbb{C}\alpha}^a$ & $\left(\frac{1}{2},0\right)$ & $-1/2$   & $\overline{\boldsymbol{3}}$ & $\boldsymbol{1}$ & $1/3$  & $n_f$\tabularnewline
    			\hline
    			$H_i$     & $\left(0,0\right)$&  0     & $\boldsymbol{1}$ & $\boldsymbol{2}$ & $1/2$  & $1$\tabularnewline
    			\hline
    		\end{tabular}
    		\caption{\label{tab:GRSMEFT-field-content}
    			The field content of the GRSMEFT, along with their representations under the Lorentz and gauge groups. The representation under Lorentz group is denoted by $(j_l,j_r)$, while the helicity of the field is given by $h = j_r-j_l$ .
    		The number of fermion flavors is denoted as $n_f$, and $n_f=3$ in the GRSMEFT. All of the fields are accompanied with their Hermitian conjugates that are omitted, $(C_{\rm L \alpha\beta\gamma\delta})^\dagger = C_{\rm R \dot{\alpha}\dot{\beta}\dot{\gamma}\dot{\delta}}$ for gravitons, $(F_{\rm L \alpha\beta})^\dagger = F_{\rm R \dot\alpha\dot\beta}$ for gauge bosons, $(\psi_\alpha)^\dagger = (\psi^\dagger)_{\dot\alpha}$ for fermions, and $H^\dagger$ for the Higgs, which are under the conjugate representations of all the Lorentz and gauge groups. }
    	\end{center}
    \end{table}

\subsection{Dim-6 Operator and Amplitude Bases}
\underline{Class $C_{\rm{L}} F_{\rm{L}}{}^2      $}: 3 types

\begin{align}

&
\begin{array}{c|l|l|}

\mathcal{O}_{C_{\rm{L}} B_{\rm{L}}{}^2      }

& B_{\rm L}{}_{\lambda\rho}B_{\rm L}{}_{\mu\nu}C_{\rm L}^{\lambda\rho\mu\nu}

&\langle 12\rangle ^2 \langle 13\rangle ^2

\end{array}\\

&\begin{array}{c|l|l|}

\mathcal{O}_{C_{\rm{L}} W_{\rm{L}}{}^2      }

& C_{\rm L}^{\lambda\rho\mu\nu}W_{\rm L}^I{}_{\lambda\rho}W_{\rm L}^I{}_{\mu\nu}

&\langle 12\rangle ^2 \langle 13\rangle ^2 \delta^{I_2I_3}

\end{array}\\

&\begin{array}{c|l|l|}

\mathcal{O}_{C_{\rm{L}} G_{\rm{L}}{}^2      }

& C_{\rm L}^{\lambda\rho\mu\nu}G_{\rm L}^A{}_{\lambda\rho}G_{\rm L}^A{}_{\mu\nu}

&\langle 12\rangle ^2 \langle 13\rangle ^2 \delta^{A_2A_3}

\end{array}

\end{align}

\underline{Class $C_{\rm{L}}{}^3        $}: 1 types

\begin{align}

&\begin{array}{c|l|l|}

\mathcal{O}_{C_{\rm{L}}{}^3        }

& C_{\rm L}{}_{\mu\nu\lambda\rho}C_{\rm L}^{\mu\nu\eta\xi}C_{\rm L}{}_{\eta\xi}{}^{\lambda\rho}

&\langle 12\rangle ^2 \langle 13\rangle ^2 \langle 23\rangle ^2

\end{array}

\end{align}

\underline{Class $C_{\rm{L}}{}^2     \phi {}^2  $}: 1 types

\begin{align}

&\begin{array}{c|l|l|}

\mathcal{O}_{C_{\rm{L}}{}^2     H H^{\dagger}   }

& H_iH^{\dagger i}C_{\rm L}{}_{\mu\nu\lambda\rho}C_{\rm L}^{\mu\nu\lambda\rho}

&\langle 12\rangle ^4 \delta^{i_3}_{i_4}

\end{array}

\end{align}

\subsection{Dim-8 Operator and Amplitude Bases}
\underline{Class $C_{\rm{L}} F_{\rm{L}}{}^3      $}: 2 types

\begin{align}

&\begin{array}{c|l|l|}

\mathcal{O}_{C_{\rm{L}} B_{\rm{L}} W_{\rm{L}}{}^2      }

& C_{\rm L}{}_{\mu\nu\lambda\rho}W_{\rm L}^I{}^{\lambda\eta}W_{\rm L}^I{}^{\mu\nu}B_{\rm L}{}_{\eta}{}^{\rho}

&\langle 12\rangle  \langle 13\rangle ^2 \langle 14\rangle  \langle 24\rangle  \delta^{I_3I_4}

\end{array}\\

&\begin{array}{c|l|l|}

\mathcal{O}_{C_{\rm{L}} B_{\rm{L}} G_{\rm{L}}{}^2      }

& C_{\rm L}{}_{\mu\nu\lambda\rho}G_{\rm L}^A{}^{\lambda\eta}G_{\rm L}^A{}^{\mu\nu}B_{\rm L}{}_{\eta}{}^{\rho}

&\langle 12\rangle  \langle 13\rangle ^2 \langle 14\rangle  \langle 24\rangle  \delta^{A_3A_4}

\end{array}

\end{align}

\underline{Class $C_{\rm{L}}{}^2 F_{\rm{L}}{}^2      $}: 3 types

\begin{align}

&\begin{array}{c|l|l|}

\multirow{2}*{$\mathcal{O}_{C_{\rm{L}}{}^2 B_{\rm{L}}{}^2      }^{(1,2)}$}

& B_{\rm L}{}_{\eta\xi}B_{\rm L}^{\eta\xi}C_{\rm L}{}_{\mu\nu\lambda\rho}C_{\rm L}^{\mu\nu\lambda\rho}

&\langle 12\rangle ^4 \langle 34\rangle ^2

\\& B_{\rm L}^{\eta\xi}B_{\rm L}^{\mu\nu}C_{\rm L}{}_{\mu\nu\lambda\rho}C_{\rm L}{}_{\eta\xi}{}^{\lambda\rho}

&\langle 12\rangle ^2 \langle 13\rangle ^2 \langle 24\rangle ^2

\end{array}\\

&\begin{array}{c|l|l|}

\multirow{2}*{$\mathcal{O}_{C_{\rm{L}}{}^2 W_{\rm{L}}{}^2      }^{(1,2)}$}

& C_{\rm L}{}_{\mu\nu\lambda\rho}C_{\rm L}^{\mu\nu\lambda\rho}W_{\rm L}^I{}_{\eta\xi}W_{\rm L}^I{}^{\eta\xi}

&\langle 12\rangle ^4 \langle 34\rangle ^2 \delta^{I_3I_4}

\\& C_{\rm L}{}_{\mu\nu\lambda\rho}W_{\rm L}^I{}^{\eta\xi}W_{\rm L}^I{}^{\mu\nu}C_{\rm L}{}_{\eta\xi}{}^{\lambda\rho}

&\langle 12\rangle ^2 \langle 13\rangle ^2 \langle 24\rangle ^2 \delta^{I_3I_4}

\end{array}\\

&\begin{array}{c|l|l|}

\multirow{2}*{$\mathcal{O}_{C_{\rm{L}}{}^2 G_{\rm{L}}{}^2      }^{(1,2)}$}

& C_{\rm L}{}_{\mu\nu\lambda\rho}C_{\rm L}^{\mu\nu\lambda\rho}G_{\rm L}^A{}_{\eta\xi}G_{\rm L}^A{}^{\eta\xi}

&\langle 12\rangle ^4 \langle 34\rangle ^2 \delta^{A_3A_4}

\\& C_{\rm L}{}_{\mu\nu\lambda\rho}G_{\rm L}^A{}^{\eta\xi}G_{\rm L}^A{}^{\mu\nu}C_{\rm L}{}_{\eta\xi}{}^{\lambda\rho}

&\langle 12\rangle ^2 \langle 13\rangle ^2 \langle 24\rangle ^2 \delta^{A_3A_4}

\end{array}

\end{align}

\underline{Class $C_{\rm{L}}{}^4        $}: 1 types

\begin{align}

&\begin{array}{c|l|l|}

\mathcal{O}_{C_{\rm{L}}{}^4        }

& C_{\rm L}{}_{\eta\xi\tau\upsilon}C_{\rm L}{}_{\mu\nu\lambda\rho}C_{\rm L}^{\eta\xi\tau\upsilon}C_{\rm L}^{\mu\nu\lambda\rho}

&\langle 12\rangle ^4 \langle 34\rangle ^4

\end{array}

\end{align}

\underline{Class $C_{\rm{L}} F_{\rm{L}} \psi  \bar{\psi }   D$}: 10 types

\begin{align}

&\begin{array}{c|l|l|}

\mathcal{O}_{C_{\rm{L}} B_{\rm{L}} \bar{e}_{\rm R} e_{\rm R}    D}

& C_{\rm L}{}_{\nu\lambda\rho}{}^{\mu}\left(D_{\mu}B_{\rm L}^{\nu\lambda}\right)\left(\bar{e}_{\rm R}{}_{p}\gamma^{\rho}e_{\rm R}{}_r{}\right)

&\mc{C}_{f_3,f_4} [34]  \langle 12\rangle ^2 \langle 13\rangle ^2

\end{array}\\

&\begin{array}{c|l|l|}

\mathcal{O}_{C_{\rm{L}} B_{\rm{L}} l \bar{l}    D}

& C_{\rm L}{}_{\nu\lambda\rho}{}^{\mu}\left(D_{\mu}B_{\rm L}^{\nu\lambda}\right)\left(\bar{l}_{\rm L}{}_{r}^{i}\gamma^{\rho}l_{\rm L}{}_p{}_{i}\right)

&\mc{C}_{f_3,f_4} [34]  \langle 12\rangle ^2 \langle 13\rangle ^2 \delta^{i_3}_{i_4}

\end{array}\\

&\begin{array}{c|l|l|}

\mathcal{O}_{C_{\rm{L}} B_{\rm{L}} \bar{d}_{\rm R} d_{\rm R}    D}

& C_{\rm L}{}_{\nu\lambda\rho}{}^{\mu}\left(D_{\mu}B_{\rm L}^{\nu\lambda}\right)\left(\bar{d}_{\rm R}{}_{p}^{a}\gamma^{\rho}d_{\rm R}{}_r{}_{a}\right)

&\mc{C}_{f_3,f_4} [34]  \langle 12\rangle ^2 \langle 13\rangle ^2 \delta^{a_4}_{a_3}

\end{array}\\

&\begin{array}{c|l|l|}

\mathcal{O}_{C_{\rm{L}} B_{\rm{L}} \bar{u}_{\rm R} u_{\rm R}    D}

& C_{\rm L}{}_{\nu\lambda\rho}{}^{\mu}\left(D_{\mu}B_{\rm L}^{\nu\lambda}\right)\left(\bar{u}_{\rm R}{}_{p}^{a}\gamma^{\rho}u_{\rm R}{}_r{}_{a}\right)

&\mc{C}_{f_3,f_4} [34]  \langle 12\rangle ^2 \langle 13\rangle ^2 \delta^{a_4}_{a_3}

\end{array}\\

&\begin{array}{c|l|l|}

\mathcal{O}_{C_{\rm{L}} B_{\rm{L}} q \bar{q}    D}

& C_{\rm L}{}_{\nu\lambda\rho}{}^{\mu}\left(D_{\mu}B_{\rm L}^{\nu\lambda}\right)\left(\bar{q}_{\rm L}{}_{r}^{ai}\gamma^{\rho}q_{\rm L}{}_p{}_{ai}\right)

&\mc{C}_{f_3,f_4} [34]  \langle 12\rangle ^2 \langle 13\rangle ^2 \delta^{i_3}_{i_4} \delta^{a_3}_{a_4}

\end{array}\\

&\begin{array}{c|l|l|}

\mathcal{O}_{C_{\rm{L}} W_{\rm{L}} l \bar{l}    D}

& \tau^I{}_{j}^{i}C_{\rm L}{}_{\nu\lambda\rho}{}^{\mu}\left(D_{\mu}W_{\rm L}^I{}^{\nu\lambda}\right)\left(\bar{l}_{\rm L}{}_{r}^{j}\gamma^{\rho}l_{\rm L}{}_p{}_{i}\right)

&\mc{C}_{f_3,f_4} [34]  \langle 12\rangle ^2 \langle 13\rangle ^2 \left(\tau^{I_2}\right)^{i_3}_{i_4}

\end{array}\\

&\begin{array}{c|l|l|}

\mathcal{O}_{C_{\rm{L}} W_{\rm{L}} q \bar{q}    D}

& \tau^I{}_{j}^{i}C_{\rm L}{}_{\nu\lambda\rho}{}^{\mu}\left(D_{\mu}W_{\rm L}^I{}^{\nu\lambda}\right)\left(\bar{q}_{\rm L}{}_{r}^{aj}\gamma^{\rho}q_{\rm L}{}_p{}_{ai}\right)

&\mc{C}_{f_3,f_4} [34]  \langle 12\rangle ^2 \langle 13\rangle ^2 \delta^{a_3}_{a_4} \left(\tau^{I_2}\right)^{i_3}_{i_4}

\end{array}\\

&\begin{array}{c|l|l|}

\mathcal{O}_{C_{\rm{L}} G_{\rm{L}} \bar{d}_{\rm R} d_{\rm R}    D}

& \lambda^A{}_{a}^{b}C_{\rm L}{}_{\nu\lambda\rho}{}^{\mu}\left(D_{\mu}G_{\rm L}^A{}^{\nu\lambda}\right)\left(\bar{d}_{\rm R}{}_{p}^{a}\gamma^{\rho}d_{\rm R}{}_r{}_{b}\right)

&\mc{C}_{f_3,f_4} [34]  \langle 12\rangle ^2 \langle 13\rangle ^2 \left(\lambda^{A_2}\right)^{a_4}_{a_3}

\end{array}\\

&\begin{array}{c|l|l|}

\mathcal{O}_{C_{\rm{L}} G_{\rm{L}} \bar{u}_{\rm R} u_{\rm R}    D}

& \lambda^A{}_{a}^{b}C_{\rm L}{}_{\nu\lambda\rho}{}^{\mu}\left(D_{\mu}G_{\rm L}^A{}^{\nu\lambda}\right)\left(\bar{u}_{\rm R}{}_{p}^{a}\gamma^{\rho}u_{\rm R}{}_r{}_{b}\right)

&\mc{C}_{f_3,f_4} [34]  \langle 12\rangle ^2 \langle 13\rangle ^2 \left(\lambda^{A_2}\right)^{a_4}_{a_3}

\end{array}\\

&\begin{array}{c|l|l|}

\mathcal{O}_{C_{\rm{L}} G_{\rm{L}} q \bar{q}    D}

& \lambda^A{}_{b}^{a}C_{\rm L}{}_{\nu\lambda\rho}{}^{\mu}\left(D_{\mu}G_{\rm L}^A{}^{\nu\lambda}\right)\left(\bar{q}_{\rm L}{}_{r}^{bi}\gamma^{\rho}q_{\rm L}{}_p{}_{ai}\right)

&\mc{C}_{f_3,f_4} [34]  \langle 12\rangle ^2 \langle 13\rangle ^2 \delta^{i_3}_{i_4} \left(\lambda^{A_2}\right)^{a_3}_{a_4}

\end{array}

\end{align}

\underline{Class $C_{\rm{L}}   \psi {}^2 \phi D^2$}: 3 types

\begin{align}

&\begin{array}{c|l|l|}

\mathcal{O}_{C_{\rm{L}}   \bar{e}_{\rm R} l H^{\dagger}  D^2}

& H^{\dagger i}C_{\rm L}{}_{\lambda\rho}{}^{\nu\mu}\left(\left(D_{\mu}\bar{e}_{\rm R}{}_{p}\right)\sigma^{\lambda \rho}\left(D_{\nu}l_{\rm L}{}_r{}_{i}\right)\right)

&\mc{C}_{f_2,f_3} [34]  \langle 12\rangle  \langle 13\rangle ^2 \langle 14\rangle  \delta^{i_3}_{i_4}

\end{array}\\

&\begin{array}{c|l|l|}

\mathcal{O}_{C_{\rm{L}}   \bar{d}_{\rm R} q H^{\dagger}  D^2}

& H^{\dagger i}C_{\rm L}{}_{\lambda\rho}{}^{\nu\mu}\left(\left(D_{\mu}\bar{d}_{\rm R}{}_{p}^{a}\right)\sigma^{\lambda \rho}\left(D_{\nu}q_{\rm L}{}_r{}_{ai}\right)\right)

&\mc{C}_{f_2,f_3} [34]  \langle 12\rangle  \langle 13\rangle ^2 \langle 14\rangle  \delta^{i_3}_{i_4} \delta^{a_3}_{a_2}

\end{array}\\

&\begin{array}{c|l|l|}

\mathcal{O}_{C_{\rm{L}}   q \bar{u}_{\rm R} H D^2}

& \epsilon^{ij}H_jC_{\rm L}{}_{\lambda\rho}{}^{\nu\mu}\left(\left(D_{\nu}\bar{u}_{\rm R}{}_{r}^{a}\right)\sigma^{\lambda \rho}\left(D_{\mu}q_{\rm L}{}_p{}_{ai}\right)\right)

&\mc{C}_{f_2,f_3} [34]  \langle 12\rangle  \langle 13\rangle ^2 \langle 14\rangle  \delta^{a_2}_{a_3} \epsilon^{i_2i_4}

\end{array}

\end{align}

\underline{Class $C_{\rm{L}} F_{\rm{L}}   \phi {}^2 D^2$}: 2 types

\begin{align}

&\begin{array}{c|l|l|}

\mathcal{O}_{C_{\rm{L}} B_{\rm{L}}   H H^{\dagger}  D^2}

& H^{\dagger i}C_{\rm L}{}_{\lambda\rho}{}^{\nu\mu}\left(D_{\nu}H_i\right)\left(D_{\mu}B_{\rm L}^{\lambda\rho}\right)

&[34] \langle 12\rangle ^2 \langle 13\rangle  \langle 14\rangle  \delta^{i_3}_{i_4}

\end{array}\\

&\begin{array}{c|l|l|}

\mathcal{O}_{C_{\rm{L}} W_{\rm{L}}   H H^{\dagger}  D^2}

& \tau^I{}_{j}^{i}H^{\dagger j}C_{\rm L}{}_{\lambda\rho}{}^{\nu\mu}\left(D_{\nu}H_i\right)\left(D_{\mu}W_{\rm L}^I{}^{\lambda\rho}\right)

&[34] \langle 12\rangle ^2 \langle 13\rangle  \langle 14\rangle  \left(\tau^{I_2}\right)^{i_3}_{i_4}

\end{array}

\end{align}

\underline{Class $C_{\rm{L}}{}^2     \phi {}^2 D^2$}: 1 types

\begin{align}

&\begin{array}{c|l|l|}

\mathcal{O}_{C_{\rm{L}}{}^2     H H^{\dagger}  D^2}

& H_iH^{\dagger i}\left(D_{\mu}C_{\rm L}{}_{\nu\lambda\rho\eta}\right)\left(D^{\mu}C_{\rm L}^{\nu\lambda\rho\eta}\right)

&s_{34} \langle 12\rangle ^4 \delta^{i_3}_{i_4}

\end{array}

\end{align}

\underline{Class $C_{\rm{R}}{}^2 F_{\rm{L}}{}^2      $}: 3 types

\begin{align}

&\begin{array}{c|l|l|}

\mathcal{O}_{C_{\rm{R}}{}^2 B_{\rm{L}}{}^2      }

& B_{\rm L}{}_{\mu\nu}B_{\rm L}^{\mu\nu}C_{\rm R}{}_{\lambda\rho\eta\xi}C_{\rm R}^{\lambda\rho\eta\xi}

&[34]^4 \langle 12\rangle ^2

\end{array}\\

&\begin{array}{c|l|l|}

\mathcal{O}_{C_{\rm{R}}{}^2 W_{\rm{L}}{}^2      }

& C_{\rm R}{}_{\lambda\rho\eta\xi}C_{\rm R}^{\lambda\rho\eta\xi}W_{\rm L}^I{}_{\mu\nu}W_{\rm L}^I{}^{\mu\nu}

&[34]^4 \langle 12\rangle ^2 \delta^{I_1I_2}

\end{array}\\

&\begin{array}{c|l|l|}

\mathcal{O}_{C_{\rm{R}}{}^2 G_{\rm{L}}{}^2      }

& C_{\rm R}{}_{\lambda\rho\eta\xi}C_{\rm R}^{\lambda\rho\eta\xi}G_{\rm L}^A{}_{\mu\nu}G_{\rm L}^A{}^{\mu\nu}

&[34]^4 \langle 12\rangle ^2 \delta^{A_1A_2}

\end{array}

\end{align}

\underline{Class $C_{\rm{L}}{}^2 C_{\rm{R}}{}^2        $}: 1 types

\begin{align}

&\begin{array}{c|l|l|}

\mathcal{O}_{C_{\rm{L}}{}^2 C_{\rm{R}}{}^2        }

& C_{\rm L}{}_{\mu\nu\lambda\rho}C_{\rm L}^{\mu\nu\lambda\rho}C_{\rm R}{}_{\eta\xi\tau\upsilon}C_{\rm R}^{\eta\xi\tau\upsilon}

&[34]^4 \langle 12\rangle ^4

\end{array}

\end{align}

\underline{Class $C_{\rm{L}}   \psi {}^4    $}: 4 types

\begin{align}

&\begin{array}{c|l|l|}

\mathcal{O}_{C_{\rm{L}}   \bar{d}_{\rm R} \bar{e}_{\rm R} \bar{u}_{\rm R}^2    }

&\mathcal{Y}\left[\tiny{\young(st)}\right]\epsilon_{abc}C_{\rm L}{}_{\mu\nu\lambda\rho}\left(\bar{d}_{\rm R}{}_{p}^{a}\sigma^{\lambda \rho}C\bar{e}_{\rm R}{}_{r}\right)\left(\bar{u}_{\rm R}{}_{s}^{b}\sigma^{\mu \nu}C\bar{u}_{\rm R}{}_{t}^{c}\right)

&\mc{C}_{f_2,f_3,f_4f_5}^{[2]} \langle 12\rangle  \langle 13\rangle  \langle 14\rangle  \langle 15\rangle  \epsilon_{a_2a_4a_5}

\end{array}\\

&\begin{array}{c|l|l|}

\mathcal{O}_{C_{\rm{L}}   \bar{e}_{\rm R} l q \bar{u}_{\rm R}    }

& \epsilon^{ij}C_{\rm L}{}_{\mu\nu\lambda\rho}\left(\bar{e}_{\rm R}{}_{p}\sigma^{\lambda \rho}l_{\rm L}{}_r{}_{i}\right)\left(\bar{u}_{\rm R}{}_{t}^{a}\sigma^{\mu \nu}q_{\rm L}{}_s{}_{aj}\right)

&\mc{C}_{f_2,f_3,f_4,f_5} \langle 12\rangle  \langle 13\rangle  \langle 14\rangle  \langle 15\rangle  \delta^{a_4}_{a_5} \epsilon^{i_3i_4}

\end{array}\\

&\begin{array}{c|l|l|}

\multirow{2}*{$\mathcal{O}_{C_{\rm{L}}   \bar{d}_{\rm R} q^2 \bar{u}_{\rm R}    }^{(1,2)}$}

&\mathcal{Y}\left[\tiny{\young(rs)}\right]\epsilon^{ij}C_{\rm L}{}_{\mu\nu\lambda\rho}\left(\bar{d}_{\rm R}{}_{p}^{a}\sigma^{\lambda \rho}q_{\rm L}{}_r{}_{bi}\right)\left(\bar{u}_{\rm R}{}_{t}^{b}\sigma^{\mu \nu}q_{\rm L}{}_s{}_{aj}\right)

&\mc{C}_{f_2,f_3f_4,f_5}^{[2]} \langle 12\rangle  \langle 13\rangle  \langle 14\rangle  \langle 15\rangle  \delta^{a_4}_{a_2} \delta^{a_3}_{a_5} \epsilon^{i_3i_4}

\\&\mathcal{Y}\left[\tiny{\young(r,s)}\right]\epsilon^{ij}C_{\rm L}{}_{\mu\nu\lambda\rho}\left(\bar{d}_{\rm R}{}_{p}^{a}\sigma^{\lambda \rho}q_{\rm L}{}_r{}_{bi}\right)\left(\bar{u}_{\rm R}{}_{t}^{b}\sigma^{\mu \nu}q_{\rm L}{}_s{}_{aj}\right)

&\mc{C}_{f_2,f_3f_4,f_5}^{[1,1]} \langle 12\rangle  \langle 13\rangle  \langle 14\rangle  \langle 15\rangle  \delta^{a_4}_{a_2} \delta^{a_3}_{a_5} \epsilon^{i_3i_4}

\end{array}\\

&\begin{array}{c|l|l|}

\mathcal{O}_{C_{\rm{L}}   l q^3    }

&\mathcal{Y}\left[\tiny{\young(rs,t)}\right]\epsilon^{abc}\epsilon^{ik}\epsilon^{jl}C_{\rm L}{}_{\mu\nu\lambda\rho}\left(l_{\rm L}{}_p{}_{i}C\sigma^{\lambda \rho}q_{\rm L}{}_r{}_{aj}\right)\left(q_{\rm L}{}_s{}_{bk}C\sigma^{\mu \nu}q_{\rm L}{}_t{}_{cl}\right)

&\mc{C}_{f_2,f_3f_4f_5}^{[2,1]} \langle 12\rangle  \langle 13\rangle  \langle 14\rangle  \langle 15\rangle  \epsilon^{i_2i_4} \epsilon^{i_3i_5} \epsilon^{a_3a_4a_5}

\end{array}

\end{align}

\underline{Class $C_{\rm{L}} F_{\rm{L}} \psi {}^2 \phi  $}: 8 types

\begin{align}

&\begin{array}{c|l|l|}

\mathcal{O}_{C_{\rm{L}} B_{\rm{L}} \bar{e}_{\rm R} l H^{\dagger}   }

& H^{\dagger i}B_{\rm L}^{\lambda\rho}C_{\rm L}{}_{\mu\nu\lambda\rho}\left(\bar{e}_{\rm R}{}_{p}\sigma^{\mu \nu}l_{\rm L}{}_r{}_{i}\right)

&\mc{C}_{f_3,f_4} \langle 12\rangle ^2 \langle 13\rangle  \langle 14\rangle  \delta^{i_4}_{i_5}

\end{array}\\

&\begin{array}{c|l|l|}

\mathcal{O}_{C_{\rm{L}} B_{\rm{L}} \bar{d}_{\rm R} q H^{\dagger}   }

& H^{\dagger i}B_{\rm L}^{\lambda\rho}C_{\rm L}{}_{\mu\nu\lambda\rho}\left(\bar{d}_{\rm R}{}_{p}^{a}\sigma^{\mu \nu}q_{\rm L}{}_r{}_{ai}\right)

&\mc{C}_{f_3,f_4} \langle 12\rangle ^2 \langle 13\rangle  \langle 14\rangle  \delta^{i_4}_{i_5} \delta^{a_4}_{a_3}

\end{array}\\

&\begin{array}{c|l|l|}

\mathcal{O}_{C_{\rm{L}} B_{\rm{L}} q \bar{u}_{\rm R} H  }

& \epsilon^{ij}H_jB_{\rm L}^{\lambda\rho}C_{\rm L}{}_{\mu\nu\lambda\rho}\left(\bar{u}_{\rm R}{}_{r}^{a}\sigma^{\mu \nu}q_{\rm L}{}_p{}_{ai}\right)

&\mc{C}_{f_3,f_4} \langle 12\rangle ^2 \langle 13\rangle  \langle 14\rangle  \delta^{a_3}_{a_4} \epsilon^{i_3i_5}

\end{array}\\

&\begin{array}{c|l|l|}

\mathcal{O}_{C_{\rm{L}} W_{\rm{L}} \bar{e}_{\rm R} l H^{\dagger}   }

& \tau^I{}_{j}^{i}H^{\dagger j}C_{\rm L}{}_{\mu\nu\lambda\rho}W_{\rm L}^I{}^{\lambda\rho}\left(\bar{e}_{\rm R}{}_{p}\sigma^{\mu \nu}l_{\rm L}{}_r{}_{i}\right)

&\mc{C}_{f_3,f_4} \langle 12\rangle ^2 \langle 13\rangle  \langle 14\rangle  \left(\tau^{I_2}\right)^{i_4}_{i_5}

\end{array}\\

&\begin{array}{c|l|l|}

\mathcal{O}_{C_{\rm{L}} W_{\rm{L}} \bar{d}_{\rm R} q H^{\dagger}   }

& \tau^I{}_{j}^{i}H^{\dagger j}C_{\rm L}{}_{\mu\nu\lambda\rho}W_{\rm L}^I{}^{\lambda\rho}\left(\bar{d}_{\rm R}{}_{p}^{a}\sigma^{\mu \nu}q_{\rm L}{}_r{}_{ai}\right)

&\mc{C}_{f_3,f_4} \langle 12\rangle ^2 \langle 13\rangle  \langle 14\rangle  \delta^{a_4}_{a_3} \left(\tau^{I_2}\right)^{i_4}_{i_5}

\end{array}\\

&\begin{array}{c|l|l|}

\mathcal{O}_{C_{\rm{L}} W_{\rm{L}} q \bar{u}_{\rm R} H  }

& \tau^I{}_{k}^{i}\epsilon^{jk}H_jC_{\rm L}{}_{\mu\nu\lambda\rho}W_{\rm L}^I{}^{\lambda\rho}\left(\bar{u}_{\rm R}{}_{r}^{a}\sigma^{\mu \nu}q_{\rm L}{}_p{}_{ai}\right)

&\mc{C}_{f_3,f_4} \langle 12\rangle ^2 \langle 13\rangle  \langle 14\rangle  \delta^{a_3}_{a_4} \epsilon^{i_5\text{k}} \left(\tau^{I_2}\right)^{i_3}_{\text{k}}

\end{array}\\

&\begin{array}{c|l|l|}

\mathcal{O}_{C_{\rm{L}} G_{\rm{L}} \bar{d}_{\rm R} q H^{\dagger}   }

& \lambda^A{}_{a}^{b}H^{\dagger i}C_{\rm L}{}_{\mu\nu\lambda\rho}G_{\rm L}^A{}^{\lambda\rho}\left(\bar{d}_{\rm R}{}_{p}^{a}\sigma^{\mu \nu}q_{\rm L}{}_r{}_{bi}\right)

&\mc{C}_{f_3,f_4} \langle 12\rangle ^2 \langle 13\rangle  \langle 14\rangle  \delta^{i_4}_{i_5} \left(\lambda^{A_2}\right)^{a_4}_{a_3}

\end{array}\\

&\begin{array}{c|l|l|}

\mathcal{O}_{C_{\rm{L}} G_{\rm{L}} q \bar{u}_{\rm R} H  }

& \lambda^A{}_{b}^{a}\epsilon^{ij}H_jC_{\rm L}{}_{\mu\nu\lambda\rho}G_{\rm L}^A{}^{\lambda\rho}\left(\bar{u}_{\rm R}{}_{r}^{b}\sigma^{\mu \nu}q_{\rm L}{}_p{}_{ai}\right)

&\mc{C}_{f_3,f_4} \langle 12\rangle ^2 \langle 13\rangle  \langle 14\rangle  \epsilon^{i_3i_5} \left(\lambda^{A_2}\right)^{a_3}_{a_4}

\end{array}

\end{align}

\underline{Class $C_{\rm{L}} F_{\rm{L}}{}^2   \phi {}^2  $}: 4 types

\begin{align}

&\begin{array}{c|l|l|}

\mathcal{O}_{C_{\rm{L}} B_{\rm{L}}{}^2   H H^{\dagger}   }

& H_iH^{\dagger i}B_{\rm L}^{\lambda\rho}B_{\rm L}^{\mu\nu}C_{\rm L}{}_{\mu\nu\lambda\rho}

&\langle 12\rangle ^2 \langle 13\rangle ^2 \delta^{i_4}_{i_5}

\end{array}\\

&\begin{array}{c|l|l|}

\mathcal{O}_{C_{\rm{L}} B_{\rm{L}} W_{\rm{L}}   H H^{\dagger}   }

& \tau^I{}_{j}^{i}H_iH^{\dagger j}B_{\rm L}^{\lambda\rho}C_{\rm L}{}_{\mu\nu\lambda\rho}W_{\rm L}^I{}^{\mu\nu}

&\langle 12\rangle ^2 \langle 13\rangle ^2 \left(\tau^{I_3}\right)^{i_4}_{i_5}

\end{array}\\

&\begin{array}{c|l|l|}

\mathcal{O}_{C_{\rm{L}} W_{\rm{L}}{}^2   H H^{\dagger}   }

& H_iH^{\dagger i}C_{\rm L}{}_{\mu\nu\lambda\rho}W_{\rm L}^I{}^{\lambda\rho}W_{\rm L}^I{}^{\mu\nu}

&\langle 12\rangle ^2 \langle 13\rangle ^2 \delta^{i_4}_{i_5} \delta^{I_2I_3}

\end{array}\\

&\begin{array}{c|l|l|}

\mathcal{O}_{C_{\rm{L}} G_{\rm{L}}{}^2   H H^{\dagger}   }

& H_iH^{\dagger i}C_{\rm L}{}_{\mu\nu\lambda\rho}G_{\rm L}^A{}^{\lambda\rho}G_{\rm L}^A{}^{\mu\nu}

&\langle 12\rangle ^2 \langle 13\rangle ^2 \delta^{i_4}_{i_5} \delta^{A_2A_3}

\end{array}

\end{align}

\underline{Class $C_{\rm{L}}{}^2   \psi {}^2 \phi  $}: 3 types

\begin{align}

&\begin{array}{c|l|l|}

\mathcal{O}_{C_{\rm{L}}{}^2   \bar{e}_{\rm R} l H^{\dagger}   }

& \left(\bar{e}_{\rm R}{}_{p}l_{\rm L}{}_r{}_{i}\right)H^{\dagger i}C_{\rm L}{}_{\mu\nu\lambda\rho}C_{\rm L}^{\mu\nu\lambda\rho}

&\mc{C}_{f_3,f_4} \langle 12\rangle ^4 \langle 34\rangle  \delta^{i_4}_{i_5}

\end{array}\\

&\begin{array}{c|l|l|}

\mathcal{O}_{C_{\rm{L}}{}^2   \bar{d}_{\rm R} q H^{\dagger}   }

& \left(\bar{d}_{\rm R}{}_{p}^{a}q_{\rm L}{}_r{}_{ai}\right)H^{\dagger i}C_{\rm L}{}_{\mu\nu\lambda\rho}C_{\rm L}^{\mu\nu\lambda\rho}

&\mc{C}_{f_3,f_4} \langle 12\rangle ^4 \langle 34\rangle  \delta^{i_4}_{i_5} \delta^{a_4}_{a_3}

\end{array}\\

&\begin{array}{c|l|l|}

\mathcal{O}_{C_{\rm{L}}{}^2   q \bar{u}_{\rm R} H  }

& \left(\bar{u}_{\rm R}{}_{r}^{a}q_{\rm L}{}_p{}_{ai}\right)\epsilon^{ij}H_jC_{\rm L}{}_{\mu\nu\lambda\rho}C_{\rm L}^{\mu\nu\lambda\rho}

&\mc{C}_{f_3,f_4} \langle 12\rangle ^4 \langle 34\rangle  \delta^{a_3}_{a_4} \epsilon^{i_3i_5}

\end{array}

\end{align}

\underline{Class $C_{\rm{L}}{}^3     \phi {}^2  $}: 1 types

\begin{align}

&\begin{array}{c|l|l|}

\mathcal{O}_{C_{\rm{L}}{}^3     H H^{\dagger}   }

& H_iH^{\dagger i}C_{\rm L}{}_{\mu\nu\lambda\rho}C_{\rm L}^{\mu\nu\eta\xi}C_{\rm L}{}_{\eta\xi}{}^{\lambda\rho}

&\langle 12\rangle ^2 \langle 13\rangle ^2 \langle 23\rangle ^2 \delta^{i_4}_{i_5}

\end{array}

\end{align}

\underline{Class $C_{\rm{L}}{}^2   \bar{\psi }{}^2 \phi  $}: 3 types

\begin{align}

&\begin{array}{c|l|l|}

\mathcal{O}_{C_{\rm{L}}{}^2   \bar{q}  u_{\rm R}  H^{\dagger}   }

& \left(\bar{q}_{\rm L}{}_{p}^{aj}u_{\rm R}{}_r{}_{a}\right)\epsilon_{ij}H^{\dagger i}C_{\rm L}{}_{\mu\nu\lambda\rho}C_{\rm L}^{\mu\nu\lambda\rho}

&\mc{C}_{f_4,f_5} [45]  \langle 12\rangle ^4 \delta^{a_5}_{a_4} \epsilon_{i_3i_4}

\end{array}\\

&\begin{array}{c|l|l|}

\mathcal{O}_{C_{\rm{L}}{}^2   e_{\rm R}  \bar{l}  H  }

& \left(\bar{l}_{\rm L}{}_{r}^{i}e_{\rm R}{}_p{}\right)H_iC_{\rm L}{}_{\mu\nu\lambda\rho}C_{\rm L}^{\mu\nu\lambda\rho}

&\mc{C}_{f_4,f_5} [45]  \langle 12\rangle ^4 \delta^{i_3}_{i_5}

\end{array}\\

&\begin{array}{c|l|l|}

\mathcal{O}_{C_{\rm{L}}{}^2   d_{\rm R}  \bar{q}  H  }

& \left(\bar{q}_{\rm L}{}_{r}^{ai}d_{\rm R}{}_p{}_{a}\right)H_iC_{\rm L}{}_{\mu\nu\lambda\rho}C_{\rm L}^{\mu\nu\lambda\rho}

&\mc{C}_{f_4,f_5} [45]  \langle 12\rangle ^4 \delta^{i_3}_{i_5} \delta^{a_4}_{a_5}

\end{array}

\end{align}

\underline{Class $C_{\rm{L}}{}^2     \phi {}^4  $}: 1 types

\begin{align}

&\begin{array}{c|l|l|}

\mathcal{O}_{C_{\rm{L}}{}^2     H{}^2 H^{\dagger} {}^2  }

& H_iH_jH^{\dagger i}H^{\dagger j}C_{\rm L}{}_{\mu\nu\lambda\rho}C_{\rm L}^{\mu\nu\lambda\rho}

&\langle 12\rangle ^4 \delta^{i_3}_{i_5} \delta^{i_4}_{i_6}

\end{array}

\end{align}

\subsection{Dim-9 Operator and Amplitude Bases}
\underline{Class $C_{\rm{L}}   \psi {}^3 \bar{\psi }   D$}: 3 types

\begin{align}

&\begin{array}{c|l|l|}

\multirow{2}*{$\mathcal{O}_{C_{\rm{L}}   \bar{d}_{\rm R} l^2 u_{\rm R}    D}^{(1,2)}$}

&\mathcal{Y}\left[\tiny{\young(rs)}\right]\epsilon^{ij}C_{\rm L}{}_{\nu\lambda\rho}{}^{\mu}\left(\left(D_{\mu}\bar{d}_{\rm R}{}_{p}^{a}\right)\gamma^{\rho}u_{\rm R}{}_t{}_{a}\right)\left(l_{\rm L}{}_r{}_{i}C\sigma^{\nu \lambda}l_{\rm L}{}_s{}_{j}\right)

&\mc{C}_{f_2,f_3f_4,f_5}^{[2]} [35]  \langle 12\rangle  \langle 13\rangle ^2 \langle 14\rangle  \delta^{a_5}_{a_2} \epsilon^{i_3i_4}

\\&\mathcal{Y}\left[\tiny{\young(r,s)}\right]\epsilon^{ij}C_{\rm L}{}_{\nu\lambda\rho}{}^{\mu}\left(\bar{d}_{\rm R}{}_{p}^{a}\gamma^{\rho}u_{\rm R}{}_t{}_{a}\right)\left(\left(D_{\mu}l_{\rm L}{}_r{}_{i}\right)C\sigma^{\nu \lambda}l_{\rm L}{}_s{}_{j}\right)

&\mc{C}_{f_2,f_3f_4,f_5}^{[1,1]} [35]  \langle 12\rangle  \langle 13\rangle ^2 \langle 14\rangle  \delta^{a_5}_{a_2} \epsilon^{i_3i_4}

\end{array}\\

&\begin{array}{c|l|l|}

\multirow{2}*{$\mathcal{O}_{C_{\rm{L}}   \bar{d}_{\rm R}^2 l \bar{q}    D}^{(1,2)}$}

&\mathcal{Y}\left[\tiny{\young(pr)}\right]\epsilon_{abc}C_{\rm L}{}_{\nu\lambda\rho}{}^{\mu}\left(\bar{d}_{\rm R}{}_{r}^{b}\sigma^{\nu \lambda}l_{\rm L}{}_s{}_{i}\right)\left(\bar{q}_{\rm L}{}_{t}^{ci}\gamma^{\rho}C\left(D_{\mu}\bar{d}_{\rm R}{}_{p}^{a}\right)\right)

&\mc{C}_{f_2f_3,f_4,f_5}^{[2]} [35]  \langle 12\rangle  \langle 13\rangle ^2 \langle 14\rangle  \delta^{i_4}_{i_5} \epsilon_{a_2a_3a_5}

\\&\mathcal{Y}\left[\tiny{\young(p,r)}\right]\epsilon_{abc}C_{\rm L}{}_{\nu\lambda\rho}{}^{\mu}\left(\bar{d}_{\rm R}{}_{r}^{b}\sigma^{\nu \lambda}l_{\rm L}{}_s{}_{i}\right)\left(\bar{q}_{\rm L}{}_{t}^{ci}\gamma^{\rho}C\left(D_{\mu}\bar{d}_{\rm R}{}_{p}^{a}\right)\right)

&\mc{C}_{f_2f_3,f_4,f_5}^{[1,1]} [35]  \langle 12\rangle  \langle 13\rangle ^2 \langle 14\rangle  \delta^{i_4}_{i_5} \epsilon_{a_2a_3a_5}

\end{array}\\

&\begin{array}{c|l|l|}

\mathcal{O}_{C_{\rm{L}}   \bar{d}_{\rm R}^3 e_{\rm R}    D}

&\mathcal{Y}\left[\tiny{\young(pr,s)}\right]\epsilon_{abc}C_{\rm L}{}_{\nu\lambda\rho}{}^{\mu}\left(\left(D_{\mu}\bar{d}_{\rm R}{}_{p}^{a}\right)\gamma^{\rho}e_{\rm R}{}_t{}\right)\left(\bar{d}_{\rm R}{}_{r}^{b}\sigma^{\nu \lambda}C\bar{d}_{\rm R}{}_{s}^{c}\right)

&\mc{C}_{f_2f_3f_4,f_5}^{[2,1]} [35]  \langle 12\rangle  \langle 13\rangle ^2 \langle 14\rangle  \epsilon_{a_2a_3a_4}

\end{array}

\end{align}

\underline{Class $C_{\rm{L}}   \psi {}^2 \phi {}^2 D^2$}: 1 types

\begin{align}

&\begin{array}{c|l|l|}

\multirow{3}*{$\mathcal{O}_{C_{\rm{L}}   l^2 H{}^2 D^2}^{(1\sim 3)}$}

&\mathcal{Y}\left[\tiny{\young(pr)}\right]\epsilon^{ik}\epsilon^{jl}H_kH_lC_{\rm L}{}_{\lambda\rho}{}^{\nu\mu}\left(\left(D_{\mu}l_{\rm L}{}_p{}_{i}\right)C\sigma^{\lambda \rho}\left(D_{\nu}l_{\rm L}{}_r{}_{j}\right)\right)

&\mc{C}_{f_2f_3}^{[2]} [35]  \langle 12\rangle  \langle 13\rangle ^2 \langle 15\rangle  \epsilon^{i_2i_4} \epsilon^{i_3i_5}

\\&\mathcal{Y}\left[\tiny{\young(pr)}\right]\epsilon^{ik}\epsilon^{jl}H_lC_{\rm L}{}_{\lambda\rho}{}^{\nu\mu}\left(D_{\nu}H_k\right)\left(\left(D_{\mu}l_{\rm L}{}_p{}_{i}\right)C\sigma^{\lambda \rho}l_{\rm L}{}_r{}_{j}\right)

&\mc{C}_{f_2f_3}^{[2]} [34]  \langle 12\rangle  \langle 13\rangle ^2 \langle 14\rangle  \epsilon^{i_2i_4} \epsilon^{i_3i_5}

\\&\mathcal{Y}\left[\tiny{\young(p,r)}\right]\epsilon^{ik}\epsilon^{jl}H_lC_{\rm L}{}_{\lambda\rho}{}^{\nu\mu}\left(D_{\nu}H_k\right)\left(\left(D_{\mu}l_{\rm L}{}_p{}_{i}\right)C\sigma^{\lambda \rho}l_{\rm L}{}_r{}_{j}\right)

&\mc{C}_{f_2f_3}^{[1,1]} [35]  \langle 12\rangle  \langle 13\rangle ^2 \langle 15\rangle  \epsilon^{i_2i_4} \epsilon^{i_3i_5}

\end{array}

\end{align}

\underline{Class $C_{\rm{L}}   \psi {}^4 \phi  $}: 4 types

\begin{align}

&\begin{array}{c|l|l|}

\mathcal{O}_{C_{\rm{L}}   \bar{e}_{\rm R} l^3 H  }

&\mathcal{Y}\left[\tiny{\young(rs,t)}\right]\epsilon^{ij}\epsilon^{kl}H_lC_{\rm L}{}_{\mu\nu\lambda\rho}\left(\bar{e}_{\rm R}{}_{p}\sigma^{\lambda \rho}l_{\rm L}{}_r{}_{i}\right)\left(l_{\rm L}{}_s{}_{j}C\sigma^{\mu \nu}l_{\rm L}{}_t{}_{k}\right)

&\mc{C}_{f_2,f_3f_4f_5}^{[2,1]} \langle 12\rangle  \langle 13\rangle  \langle 14\rangle  \langle 15\rangle  \epsilon^{i_3i_5} \epsilon^{i_4i_6}

\end{array}\\

&\begin{array}{c|l|l|}

\mathcal{O}_{C_{\rm{L}}   \bar{d}_{\rm R}^3 l H^{\dagger}   }

&\mathcal{Y}\left[\tiny{\young(prs)}\right]\epsilon_{abc}H^{\dagger i}C_{\rm L}{}_{\mu\nu\lambda\rho}\left(\bar{d}_{\rm R}{}_{s}^{c}\sigma^{\mu \nu}l_{\rm L}{}_t{}_{i}\right)\left(\bar{d}_{\rm R}{}_{p}^{a}\sigma^{\lambda \rho}C\bar{d}_{\rm R}{}_{r}^{b}\right)

&\mc{C}_{f_2f_3f_4,f_5}^{[3]} \langle 12\rangle  \langle 13\rangle  \langle 14\rangle  \langle 15\rangle  \delta^{i_5}_{i_6} \epsilon_{a_2a_3a_4}

\end{array}\\

&\begin{array}{c|l|l|}

\mathcal{O}_{C_{\rm{L}}   \bar{d}_{\rm R}^2 l \bar{u}_{\rm R} H  }

&\mathcal{Y}\left[\tiny{\young(pr)}\right]\epsilon_{abc}\epsilon^{ij}H_jC_{\rm L}{}_{\mu\nu\lambda\rho}\left(\bar{u}_{\rm R}{}_{t}^{c}\sigma^{\mu \nu}l_{\rm L}{}_s{}_{i}\right)\left(\bar{d}_{\rm R}{}_{p}^{a}\sigma^{\lambda \rho}C\bar{d}_{\rm R}{}_{r}^{b}\right)

&\mc{C}_{f_2f_3,f_4,f_5}^{[2]} \langle 12\rangle  \langle 13\rangle  \langle 14\rangle  \langle 15\rangle  \epsilon^{i_4i_6} \epsilon_{a_2a_3a_5}

\end{array}\\

&\begin{array}{c|l|l|}

\multirow{2}*{$\mathcal{O}_{C_{\rm{L}}   \bar{d}_{\rm R} l^2 q H  }^{(1,2)}$}

&\mathcal{Y}\left[\tiny{\young(rs)}\right]\epsilon^{ik}\epsilon^{jl}H_lC_{\rm L}{}_{\mu\nu\lambda\rho}\left(\bar{d}_{\rm R}{}_{p}^{a}\sigma^{\lambda \rho}l_{\rm L}{}_r{}_{i}\right)\left(l_{\rm L}{}_s{}_{j}C\sigma^{\mu \nu}q_{\rm L}{}_t{}_{ak}\right)

&\mc{C}_{f_2,f_3f_4,f_5}^{[2]} \langle 12\rangle  \langle 13\rangle  \langle 14\rangle  \langle 15\rangle  \delta^{a_5}_{a_2} \epsilon^{i_3i_5} \epsilon^{i_4i_6}

\\&\mathcal{Y}\left[\tiny{\young(r,s)}\right]\epsilon^{ik}\epsilon^{jl}H_lC_{\rm L}{}_{\mu\nu\lambda\rho}\left(\bar{d}_{\rm R}{}_{p}^{a}\sigma^{\lambda \rho}l_{\rm L}{}_r{}_{i}\right)\left(l_{\rm L}{}_s{}_{j}C\sigma^{\mu \nu}q_{\rm L}{}_t{}_{ak}\right)

&\mc{C}_{f_2,f_3f_4,f_5}^{[1,1]} \langle 12\rangle  \langle 13\rangle  \langle 14\rangle  \langle 15\rangle  \delta^{a_5}_{a_2} \epsilon^{i_3i_5} \epsilon^{i_4i_6}

\end{array}

\end{align}

\underline{Class $C_{\rm{L}} F_{\rm{L}} \psi {}^2 \phi {}^2  $}: 2 types

\begin{align}

&\begin{array}{c|l|l|}

\mathcal{O}_{C_{\rm{L}} B_{\rm{L}} l^2 H{}^2  }

&\mathcal{Y}\left[\tiny{\young(p,r)}\right]\epsilon^{ik}\epsilon^{jl}H_kH_lB_{\rm L}^{\lambda\rho}C_{\rm L}{}_{\mu\nu\lambda\rho}\left(l_{\rm L}{}_p{}_{i}C\sigma^{\mu \nu}l_{\rm L}{}_r{}_{j}\right)

&\mc{C}_{f_3f_4}^{[1,1]} \langle 12\rangle ^2 \langle 13\rangle  \langle 14\rangle  \epsilon^{i_3i_5} \epsilon^{i_4i_6}

\end{array}\\

&\begin{array}{c|l|l|}

\multirow{2}*{$\mathcal{O}_{C_{\rm{L}} W_{\rm{L}} l^2 H{}^2  }^{(1,2)}$}

&\mathcal{Y}\left[\tiny{\young(pr)}\right]\tau^I{}_{m}^{j}\epsilon^{il}\epsilon^{km}H_kH_lC_{\rm L}{}_{\mu\nu\lambda\rho}W_{\rm L}^I{}^{\lambda\rho}\left(l_{\rm L}{}_p{}_{i}C\sigma^{\mu \nu}l_{\rm L}{}_r{}_{j}\right)

&\mc{C}_{f_3f_4}^{[2]} \langle 12\rangle ^2 \langle 13\rangle  \langle 14\rangle  \epsilon^{i_5\text{m}} \epsilon^{i_3i_6} \left(\tau^{I_2}\right)^{i_4}_{\text{m}}

\\&\mathcal{Y}\left[\tiny{\young(p,r)}\right]\tau^I{}_{m}^{j}\epsilon^{il}\epsilon^{km}H_kH_lC_{\rm L}{}_{\mu\nu\lambda\rho}W_{\rm L}^I{}^{\lambda\rho}\left(l_{\rm L}{}_p{}_{i}C\sigma^{\mu \nu}l_{\rm L}{}_r{}_{j}\right)

&\mc{C}_{f_3f_4}^{[1,1]} \langle 12\rangle ^2 \langle 13\rangle  \langle 14\rangle  \epsilon^{i_5\text{m}} \epsilon^{i_3i_6} \left(\tau^{I_2}\right)^{i_4}_{\text{m}}

\end{array}

\end{align}

\underline{Class $C_{\rm{L}}{}^2   \psi {}^2 \phi {}^2  $}: 1 types

\begin{align}

&\begin{array}{c|l|l|}

\mathcal{O}_{C_{\rm{L}}{}^2   l^2 H{}^2  }

&\mathcal{Y}\left[\tiny{\young(pr)}\right]\epsilon^{ik}\epsilon^{jl}H_kH_lC_{\rm L}{}_{\mu\nu\lambda\rho}C_{\rm L}^{\mu\nu\lambda\rho}l_{\rm L}{}_p{}_{i}Cl_{\rm L}{}_r{}_{j}

&\mc{C}_{f_3f_4} \langle 12\rangle ^4 \langle 34\rangle  \epsilon^{i_3i_5} \epsilon^{i_4i_6}

\end{array}

\end{align}

\underline{Class $C_{\rm{L}}{}^2   \bar{\psi }{}^2 \phi {}^2  $}: 1 types

\begin{align}

&\begin{array}{c|l|l|}

\mathcal{O}_{C_{\rm{L}}{}^2   \bar{l} ^2 H^{\dagger} {}^2  }

&\mathcal{Y}\left[\tiny{\young(pr)}\right]\epsilon_{ik}\epsilon_{jl}H^{\dagger i}H^{\dagger j}C_{\rm L}{}_{\mu\nu\lambda\rho}C_{\rm L}^{\mu\nu\lambda\rho}\bar{l}_{\rm L}{}_{p}^{k}C\bar{l}_{\rm L}{}_{r}^{l}

&\mc{C}_{f_5f_6}^{[2]} [56]  \langle 12\rangle ^4 \epsilon_{i_3i_5} \epsilon_{i_4i_6}

\end{array}

\end{align}

\subsection{Dim-10 Operator and Amplitude Bases}
\input{GRSMEFT10}

\section{GRLEFT Operator/Amplitude Bases up to Dim-10}\label{sec:GRLEFTopes}

\subsection{Field Content}
The field content of the GRLEFT are listed in table~\ref{tab:GRLEFT-field-content}. Similar as these in the GRSMEFT, the operators are constructed in two-component Weyl spinor notation and then converted to four-component Dirac spinor notation using the following relations:
\begin{align}
\nu_{\rm L}=\begin{pmatrix}\nu\\0\end{pmatrix},\quad
e_{\rm L}=\left(\begin{array}{c}e\\0\end{array}\right),\quad 
e_{\rm R}=\left(\begin{array}{c}0\\e_{_\mathbb{C}}^{\dagger}\end{array}\right),\quad
u_{\rm{L}}=\begin{pmatrix}u\\0\end{pmatrix},\quad u_{\rm{R}}=\left(\begin{array}{c}0\\u_{_\mathbb{C}}^{\dagger}\end{array}\right),\quad 
d_{\rm{L}}=\begin{pmatrix}d\\0\end{pmatrix},\quad
d_{\rm R}=\left(\begin{array}{c}0\\d_{_\mathbb{C}}^{\dagger}\end{array}\right).\\
\bar{\nu}_{\rm L}=\left(0\,,\,\nu^{\dagger}\right),\quad
\bar{e}_{\rm L}=\left(0\,,\,e^{\dagger}\right),\quad 
\bar{e}_{\rm R}=\left(e_{_\mathbb{C}}\,,\,0\right),\quad
\bar{u}_{\rm{L}}=\left(0\,,\,u^{\dagger} \right),\quad \bar{u}_{\rm{R}}=\left(u_{_\mathbb{C}}\,,\,0 \right),\quad
\bar{d}_{\rm{L}}=\left(0\,,\,d^{\dagger} \right),\quad 
\bar{d}_{\rm R}=\left(d_{_\mathbb{C}}\,,\,0\right).
\end{align}

\begin{table}[t]
	\begin{center}
		\begin{tabular}{|c|cc|cc|ccc|}
			\hline
			\text{Fields} & $SU(2)_{l}\times SU(2)_{r}$	& $h$ & $SU(3)_{C}$ & $U(1)_{\rm EM}$ &  Flavor & $B$ & $L$ \tabularnewline
                \hline
    			$C_{\rm L \alpha\beta\gamma\delta}$ & $\left(2,0\right)$ & $-2$ & $\boldsymbol{1}$ & 0 & 1  & 0 & 0\tabularnewline
			\hline
			$G_{\rm L\alpha\beta}^A$   & $\left(1,0\right)$  & $-1$    & $\boldsymbol{8}$ & 0  & $1$ & 0 & 0 \tabularnewline
			$F_{\rm L\alpha\beta}$   & $\left(1,0\right)$    & $-1$        & $\boldsymbol{1}$ & 0  & $1$ & 0 & 0 \tabularnewline
			\hline
			$\nu_{\alpha}$     & $\left(\frac{1}{2},0\right)$  & $-\frac12$  & $\boldsymbol{1}$ & $0$  & $n_e$ & 0 & $1$ \tabularnewline
			$e_{\alpha}$ & $\left(\frac{1}{2},0\right)$ & $-\frac12$   & $\boldsymbol{1}$ & $-1$  & $n_e$ & 0 & $1$ \tabularnewline
			$e_{_\mathbb{C}\alpha}$ & $\left(\frac{1}{2},0\right)$ & $-\frac12$   & $\boldsymbol{1}$ & $1$  & $n_e$ & 0 & $-1$ \tabularnewline
			$u_{\alpha a}$     & $\left(\frac{1}{2},0\right)$ & $-\frac12$   & $\boldsymbol{3}$ & $\frac23$  & $n_u$ & $\frac13$ & 0 \tabularnewline
			$u_{_\mathbb{C}\alpha}^a$ & $\left(\frac{1}{2},0\right)$ & $-\frac12$   & $\overline{\boldsymbol{3}}$ & $-\frac23$  & $n_u$ & $-\frac13$ & 0 \tabularnewline
			$d_{\alpha a}$     & $\left(\frac{1}{2},0\right)$ & $-\frac12$   & $\boldsymbol{3}$ & $-\frac13$  & $n_d$ & $\frac13$ & 0 \tabularnewline
			$d_{_\mathbb{C}\alpha}^a$ & $\left(\frac{1}{2},0\right)$ & $-\frac12$   & $\overline{\boldsymbol{3}}$ & $\frac13$  & $n_d$ & $-\frac13$ & $0$ \tabularnewline
			\hline
		\end{tabular}
		\caption{\label{tab:GRLEFT-field-content}
			The field content of the GRLEFT, along with their representations under the Lorentz and gauge groups. The representation under Lorentz group is denoted by $(j_l,j_r)$, while the helicity of the field is given by $h = j_r-j_l$ .
			The numbers of lepton flavors, u-type quark flavors and d-type quark flavors are denoted as $n_e$, $n_u$ and $n_d$ respectively with $n_e=3$, $n_u=2$ and $n_d=3$ in the GRLEFT. All of the fields are accompanied with their Hermitian conjugates that are omitted, $(C_{\rm L \alpha\beta\gamma\delta})^\dagger = C_{\rm R \dot{\alpha}\dot{\beta}\dot{\gamma}\dot{\delta}}$ for gravitons, $(F_{\rm L \alpha\beta})^\dagger = F_{\rm R \dot\alpha\dot\beta}$ for gauge bosons and $(\psi_\alpha)^\dagger = (\psi^\dagger)_{\dot\alpha}$ for fermions, which are under the conjugate representations of all the groups. }
	\end{center}
\end{table}

\subsection{Dim-6 Operator and Amplitude Bases}
\underline{Class $C_{\rm{L}} F_{\rm{L}}{}^2      $}: 2 types

\begin{align}

&\begin{array}{c|l|l|}

\mathcal{O}_{C_{\rm{L}} F_{\rm{L}}{}^2      }

& C_{\rm L}^{\lambda\rho\mu\nu}F_{\rm L}{}_{\lambda\rho}F_{\rm L}{}_{\mu\nu}

&\langle 12\rangle ^2 \langle 13\rangle ^2

\end{array}\\

&\begin{array}{c|l|l|}

\mathcal{O}_{C_{\rm{L}} G_{\rm{L}}{}^2      }

& C_{\rm L}^{\lambda\rho\mu\nu}G_{\rm L}^A{}_{\lambda\rho}G_{\rm L}^A{}_{\mu\nu}

&\langle 12\rangle ^2 \langle 13\rangle ^2 \delta^{A_2A_3}

\end{array}

\end{align}

\underline{Class $C_{\rm{L}}{}^3        $}: 1 types

\begin{align}

&\begin{array}{c|l|l|}

\mathcal{O}_{C_{\rm{L}}{}^3        }

& C_{\rm L}{}_{\mu\nu\lambda\rho}C_{\rm L}^{\mu\nu\eta\xi}C_{\rm L}{}_{\eta\xi}{}^{\lambda\rho}

&\langle 12\rangle ^2 \langle 13\rangle ^2 \langle 23\rangle ^2

\end{array}

\end{align}

\subsection{Dim-7 Operator and Amplitude Bases}
\underline{Class $C_{\rm{L}} F_{\rm{L}} \psi {}^2    $}: 6 types

\begin{align}

&\begin{array}{c|l|l|}

\mathcal{O}_{C_{\rm{L}} F_{\rm{L}} \nu ^2    }

&\mathcal{Y}\left[\tiny{\young(p,r)}\right]C_{\rm L}{}_{\mu\nu\lambda\rho}F_{\rm L}^{\lambda\rho}\left(\nu_{\rm L}{}_p{}C\sigma^{\mu \nu}\nu_{\rm L}{}_r{}\right)

&\mc{C}_{f_3f_4}^{[2]} \langle 12\rangle ^2 \langle 13\rangle  \langle 14\rangle

\end{array}\\

&\begin{array}{c|l|l|}

\mathcal{O}_{C_{\rm{L}} F_{\rm{L}} \bar{e}_{\rm R} e_{\rm L}    }

& C_{\rm L}{}_{\mu\nu\lambda\rho}F_{\rm L}^{\lambda\rho}\left(\bar{e}_{\rm R}{}_{p}\sigma^{\mu \nu}e_{\rm L}{}_r{}\right)

&\mc{C}_{f_3,f_4} \langle 12\rangle ^2 \langle 13\rangle  \langle 14\rangle

\end{array}\\

&\begin{array}{c|l|l|}

\mathcal{O}_{C_{\rm{L}} F_{\rm{L}} \bar{d}_{\rm R} d_{\rm L}    }

& C_{\rm L}{}_{\mu\nu\lambda\rho}F_{\rm L}^{\lambda\rho}\left(\bar{d}_{\rm R}{}_{p}^{a}\sigma^{\mu \nu}dL_r{}_{a}\right)

&\mc{C}_{f_3,f_4} \langle 12\rangle ^2 \langle 13\rangle  \langle 14\rangle  \delta^{a_4}_{a_3}

\end{array}\\

&\begin{array}{c|l|l|}

\mathcal{O}_{C_{\rm{L}} F_{\rm{L}} \bar{u}_{\rm R} u_{\rm L}    }

& C_{\rm L}{}_{\mu\nu\lambda\rho}F_{\rm L}^{\lambda\rho}\left(\bar{u}_{\rm R}{}_{p}^{a}\sigma^{\mu \nu}uL_r{}_{a}\right)

&\mc{C}_{f_3,f_4} \langle 12\rangle ^2 \langle 13\rangle  \langle 14\rangle  \delta^{a_4}_{a_3}

\end{array}\\

&\begin{array}{c|l|l|}

\mathcal{O}_{C_{\rm{L}} G_{\rm{L}} \bar{d}_{\rm R} d_{\rm L}    }

& \lambda^A{}_{a}^{b}C_{\rm L}{}_{\mu\nu\lambda\rho}G_{\rm L}^A{}^{\lambda\rho}\left(\bar{d}_{\rm R}{}_{p}^{a}\sigma^{\mu \nu}dL_r{}_{b}\right)

&\mc{C}_{f_3,f_4} \langle 12\rangle ^2 \langle 13\rangle  \langle 14\rangle  \left(\lambda^{A_2}\right)^{a_4}_{a_3}

\end{array}\\

&\begin{array}{c|l|l|}

\mathcal{O}_{C_{\rm{L}} G_{\rm{L}} \bar{u}_{\rm R} u_{\rm L}    }

& \lambda^A{}_{a}^{b}C_{\rm L}{}_{\mu\nu\lambda\rho}G_{\rm L}^A{}^{\lambda\rho}\left(\bar{u}_{\rm R}{}_{p}^{a}\sigma^{\mu \nu}uL_r{}_{b}\right)

&\mc{C}_{f_3,f_4} \langle 12\rangle ^2 \langle 13\rangle  \langle 14\rangle  \left(\lambda^{A_2}\right)^{a_4}_{a_3}

\end{array}

\end{align}

\underline{Class $C_{\rm{L}}{}^2   \psi {}^2    $}: 4 types

\begin{align}

&\begin{array}{c|l|l|}

\mathcal{O}_{C_{\rm{L}}{}^2   \nu ^2    }

&\mathcal{Y}\left[\tiny{\young(pr)}\right]C_{\rm L}{}_{\mu\nu\lambda\rho}C_{\rm L}^{\mu\nu\lambda\rho}\nu_{\rm L}{}_p{}C\nu_{\rm L}{}_r{}

&\mc{C}_{f_3f_4}^{[2]} \langle 12\rangle ^4 \langle 34\rangle

\end{array}\\

&\begin{array}{c|l|l|}

\mathcal{O}_{C_{\rm{L}}{}^2   \bar{e}_{\rm R} e_{\rm L}    }

& \left(\bar{e}_{\rm R}{}_{p}e_{\rm L}{}_r{}\right)C_{\rm L}{}_{\mu\nu\lambda\rho}C_{\rm L}^{\mu\nu\lambda\rho}

&\mc{C}_{f_3,f_4} \langle 12\rangle ^4 \langle 34\rangle

\end{array}\\

&\begin{array}{c|l|l|}

\mathcal{O}_{C_{\rm{L}}{}^2   \bar{d}_{\rm R} d_{\rm L}    }

& \left(\bar{d}_{\rm R}{}_{p}^{a}dL_r{}_{a}\right)C_{\rm L}{}_{\mu\nu\lambda\rho}C_{\rm L}^{\mu\nu\lambda\rho}

&\mc{C}_{f_3,f_4} \langle 12\rangle ^4 \langle 34\rangle  \delta^{a_4}_{a_3}

\end{array}\\

&\begin{array}{c|l|l|}

\mathcal{O}_{C_{\rm{L}}{}^2   \bar{u}_{\rm R} u_{\rm L}    }

& \left(\bar{u}_{\rm R}{}_{p}^{a}uL_r{}_{a}\right)C_{\rm L}{}_{\mu\nu\lambda\rho}C_{\rm L}^{\mu\nu\lambda\rho}

&\mc{C}_{f_3,f_4} \langle 12\rangle ^4 \langle 34\rangle  \delta^{a_4}_{a_3}

\end{array}

\end{align}

\underline{Class $C_{\rm{L}}{}^2   \bar{\psi }{}^2    $}: 4 types

\begin{align}

&\begin{array}{c|l|l|}

\mathcal{O}_{C_{\rm{L}}{}^2   \bar{\nu} ^2    }

&\mathcal{Y}\left[\tiny{\young(pr)}\right]C_{\rm L}{}_{\mu\nu\lambda\rho}C_{\rm L}^{\mu\nu\lambda\rho}\bar{\nu}_{\rm L}{}_{p}{}C\bar{\nu}_{\rm L}{}_{r}{}

&\mc{C}_{f_3f_4}^{[2]} [34]  \langle 12\rangle ^4

\end{array}\\

&\begin{array}{c|l|l|}

\mathcal{O}_{C_{\rm{L}}{}^2   e_{\rm R}  \bar{e}_{\rm L}     }

& \left(\bar{e}_{\rm L}{}_{r}{}e_{\rm R}{}_p{}\right)C_{\rm L}{}_{\mu\nu\lambda\rho}C_{\rm L}^{\mu\nu\lambda\rho}

&\mc{C}_{f_3,f_4} [34]  \langle 12\rangle ^4

\end{array}\\

&\begin{array}{c|l|l|}

\mathcal{O}_{C_{\rm{L}}{}^2   d_{\rm R}  \bar{d}_{\rm L}     }

& \left(\bar{d}_{\rm L}{}_{r}{}^{a}d_{\rm R}{}_p{}_{a}\right)C_{\rm L}{}_{\mu\nu\lambda\rho}C_{\rm L}^{\mu\nu\lambda\rho}

&\mc{C}_{f_3,f_4} [34]  \langle 12\rangle ^4 \delta^{a_3}_{a_4}

\end{array}\\

&\begin{array}{c|l|l|}

\mathcal{O}_{C_{\rm{L}}{}^2   u_{\rm R}  \bar{u}_{\rm L}     }

& \left(\bar{u}_{\rm L}{}_{r}{}^{a}u_{\rm R}{}_p{}_{a}\right)C_{\rm L}{}_{\mu\nu\lambda\rho}C_{\rm L}^{\mu\nu\lambda\rho}

&\mc{C}_{f_3,f_4} [34]  \langle 12\rangle ^4 \delta^{a_3}_{a_4}

\end{array}

\end{align}

\subsection{Dim-8 Operator and Amplitude Bases}
\underline{Class $C_{\rm{L}} F_{\rm{L}}{}^3      $}: 1 types

\begin{align}

&\begin{array}{c|l|l|}

\mathcal{O}_{C_{\rm{L}} F_{\rm{L}} G_{\rm{L}}{}^2      }

& C_{\rm L}{}_{\mu\nu\lambda\rho}G_{\rm L}^A{}^{\lambda\eta}G_{\rm L}^A{}^{\mu\nu}F_{\rm L}{}_{\eta}{}^{\rho}

&\langle 12\rangle  \langle 13\rangle ^2 \langle 14\rangle  \langle 24\rangle  \delta^{A_3A_4}

\end{array}

\end{align}

\underline{Class $C_{\rm{L}}{}^2 F_{\rm{L}}{}^2      $}: 2 types

\begin{align}

&\begin{array}{c|l|l|}

\multirow{2}*{$\mathcal{O}_{C_{\rm{L}}{}^2 F_{\rm{L}}{}^2      }^{(1,2)}$}

& C_{\rm L}{}_{\mu\nu\lambda\rho}C_{\rm L}^{\mu\nu\lambda\rho}F_{\rm L}{}_{\eta\xi}F_{\rm L}^{\eta\xi}

&\langle 12\rangle ^4 \langle 34\rangle ^2

\\& C_{\rm L}{}_{\mu\nu\lambda\rho}F_{\rm L}^{\eta\xi}F_{\rm L}^{\mu\nu}C_{\rm L}{}_{\eta\xi}{}^{\lambda\rho}

&\langle 12\rangle ^2 \langle 13\rangle ^2 \langle 24\rangle ^2

\end{array}\\

&\begin{array}{c|l|l|}

\multirow{2}*{$\mathcal{O}_{C_{\rm{L}}{}^2 G_{\rm{L}}{}^2      }^{(1,2)}$}

& C_{\rm L}{}_{\mu\nu\lambda\rho}C_{\rm L}^{\mu\nu\lambda\rho}G_{\rm L}^A{}_{\eta\xi}G_{\rm L}^A{}^{\eta\xi}

&\langle 12\rangle ^4 \langle 34\rangle ^2 \delta^{A_3A_4}

\\& C_{\rm L}{}_{\mu\nu\lambda\rho}G_{\rm L}^A{}^{\eta\xi}G_{\rm L}^A{}^{\mu\nu}C_{\rm L}{}_{\eta\xi}{}^{\lambda\rho}

&\langle 12\rangle ^2 \langle 13\rangle ^2 \langle 24\rangle ^2 \delta^{A_3A_4}

\end{array}

\end{align}

\underline{Class $C_{\rm{L}}{}^4        $}: 1 types

\begin{align}

&\begin{array}{c|l|l|}

\mathcal{O}_{C_{\rm{L}}{}^4        }

& C_{\rm L}{}_{\eta\xi\tau\upsilon}C_{\rm L}{}_{\mu\nu\lambda\rho}C_{\rm L}^{\eta\xi\tau\upsilon}C_{\rm L}^{\mu\nu\lambda\rho}

&\langle 12\rangle ^4 \langle 34\rangle ^4

\end{array}

\end{align}

\underline{Class $C_{\rm{L}} F_{\rm{L}} \psi  \bar{\psi }   D$}: 11 types

\begin{align}

&\begin{array}{c|l|l|}

\mathcal{O}_{C_{\rm{L}} F_{\rm{L}} \nu  \bar{\nu}   D}

& C_{\rm L}{}_{\nu\lambda\rho}{}^{\mu}\left(D_{\mu}F_{\rm L}^{\nu\lambda}\right)\left(\bar{\nu}_{\rm L}{}_{r}{}\gamma^{\rho}\nu_{\rm L}{}_p{}\right)

&\mc{C}_{f_3,f_4} [34]  \langle 12\rangle ^2 \langle 13\rangle ^2

\end{array}\\

&\begin{array}{c|l|l|}

\mathcal{O}_{C_{\rm{L}} F_{\rm{L}} \bar{e}_{\rm R} e_{\rm R}    D}

& C_{\rm L}{}_{\nu\lambda\rho}{}^{\mu}\left(D_{\mu}F_{\rm L}^{\nu\lambda}\right)\left(\bar{e}_{\rm R}{}_{p}\gamma^{\rho}e_{\rm R}{}_r{}\right)

&\mc{C}_{f_3,f_4} [34]  \langle 12\rangle ^2 \langle 13\rangle ^2

\end{array}\\

&\begin{array}{c|l|l|}

\mathcal{O}_{C_{\rm{L}} F_{\rm{L}} e_{\rm L} \bar{e}_{\rm L}    D}

& C_{\rm L}{}_{\nu\lambda\rho}{}^{\mu}\left(D_{\mu}F_{\rm L}^{\nu\lambda}\right)\left(\bar{e}_{\rm L}{}_{r}{}\gamma^{\rho}e_{\rm L}{}_p{}\right)

&\mc{C}_{f_3,f_4} [34]  \langle 12\rangle ^2 \langle 13\rangle ^2

\end{array}\\

&\begin{array}{c|l|l|}

\mathcal{O}_{C_{\rm{L}} F_{\rm{L}} \bar{d}_{\rm R} d_{\rm R}    D}

& C_{\rm L}{}_{\nu\lambda\rho}{}^{\mu}\left(D_{\mu}F_{\rm L}^{\nu\lambda}\right)\left(\bar{d}_{\rm R}{}_{p}^{a}\gamma^{\rho}d_{\rm R}{}_r{}_{a}\right)

&\mc{C}_{f_3,f_4} [34]  \langle 12\rangle ^2 \langle 13\rangle ^2 \delta^{a_4}_{a_3}

\end{array}\\

&\begin{array}{c|l|l|}

\mathcal{O}_{C_{\rm{L}} F_{\rm{L}} d_{\rm L} \bar{d}_{\rm L}    D}

& C_{\rm L}{}_{\nu\lambda\rho}{}^{\mu}\left(D_{\mu}F_{\rm L}^{\nu\lambda}\right)\left(\bar{d}_{\rm L}{}_{r}{}^{a}\gamma^{\rho}dL_p{}_{a}\right)

&\mc{C}_{f_3,f_4} [34]  \langle 12\rangle ^2 \langle 13\rangle ^2 \delta^{a_3}_{a_4}

\end{array}\\

&\begin{array}{c|l|l|}

\mathcal{O}_{C_{\rm{L}} F_{\rm{L}} \bar{u}_{\rm R} u_{\rm R}    D}

& C_{\rm L}{}_{\nu\lambda\rho}{}^{\mu}\left(D_{\mu}F_{\rm L}^{\nu\lambda}\right)\left(\bar{u}_{\rm R}{}_{p}^{a}\gamma^{\rho}u_{\rm R}{}_r{}_{a}\right)

&\mc{C}_{f_3,f_4} [34]  \langle 12\rangle ^2 \langle 13\rangle ^2 \delta^{a_4}_{a_3}

\end{array}\\

&\begin{array}{c|l|l|}

\mathcal{O}_{C_{\rm{L}} F_{\rm{L}} u_{\rm L} \bar{u}_{\rm L}    D}

& C_{\rm L}{}_{\nu\lambda\rho}{}^{\mu}\left(D_{\mu}F_{\rm L}^{\nu\lambda}\right)\left(\bar{u}_{\rm L}{}_{r}{}^{a}\gamma^{\rho}uL_p{}_{a}\right)

&\mc{C}_{f_3,f_4} [34]  \langle 12\rangle ^2 \langle 13\rangle ^2 \delta^{a_3}_{a_4}

\end{array}\\

&\begin{array}{c|l|l|}

\mathcal{O}_{C_{\rm{L}} G_{\rm{L}} \bar{d}_{\rm R} d_{\rm R}    D}

& \lambda^A{}_{a}^{b}C_{\rm L}{}_{\nu\lambda\rho}{}^{\mu}\left(D_{\mu}G_{\rm L}^A{}^{\nu\lambda}\right)\left(\bar{d}_{\rm R}{}_{p}^{a}\gamma^{\rho}d_{\rm R}{}_r{}_{b}\right)

&\mc{C}_{f_3,f_4} [34]  \langle 12\rangle ^2 \langle 13\rangle ^2 \left(\lambda^{A_2}\right)^{a_4}_{a_3}

\end{array}\\

&\begin{array}{c|l|l|}

\mathcal{O}_{C_{\rm{L}} G_{\rm{L}} d_{\rm L} \bar{d}_{\rm L}    D}

& \lambda^A{}_{b}^{a}C_{\rm L}{}_{\nu\lambda\rho}{}^{\mu}\left(D_{\mu}G_{\rm L}^A{}^{\nu\lambda}\right)\left(\bar{d}_{\rm L}{}_{r}{}^{b}\gamma^{\rho}dL_p{}_{a}\right)

&\mc{C}_{f_3,f_4} [34]  \langle 12\rangle ^2 \langle 13\rangle ^2 \left(\lambda^{A_2}\right)^{a_3}_{a_4}

\end{array}\\

&\begin{array}{c|l|l|}

\mathcal{O}_{C_{\rm{L}} G_{\rm{L}} \bar{u}_{\rm R} u_{\rm R}    D}

& \lambda^A{}_{a}^{b}C_{\rm L}{}_{\nu\lambda\rho}{}^{\mu}\left(D_{\mu}G_{\rm L}^A{}^{\nu\lambda}\right)\left(\bar{u}_{\rm R}{}_{p}^{a}\gamma^{\rho}u_{\rm R}{}_r{}_{b}\right)

&\mc{C}_{f_3,f_4} [34]  \langle 12\rangle ^2 \langle 13\rangle ^2 \left(\lambda^{A_2}\right)^{a_4}_{a_3}

\end{array}\\

&\begin{array}{c|l|l|}

\mathcal{O}_{C_{\rm{L}} G_{\rm{L}} u_{\rm L} \bar{u}_{\rm L}    D}

& \lambda^A{}_{b}^{a}C_{\rm L}{}_{\nu\lambda\rho}{}^{\mu}\left(D_{\mu}G_{\rm L}^A{}^{\nu\lambda}\right)\left(\bar{u}_{\rm L}{}_{r}{}^{b}\gamma^{\rho}uL_p{}_{a}\right)

&\mc{C}_{f_3,f_4} [34]  \langle 12\rangle ^2 \langle 13\rangle ^2 \left(\lambda^{A_2}\right)^{a_3}_{a_4}

\end{array}

\end{align}

\underline{Class $C_{\rm{R}}{}^2 F_{\rm{L}}{}^2      $}: 2 types

\begin{align}

&\begin{array}{c|l|l|}

\mathcal{O}_{C_{\rm{R}}{}^2 F_{\rm{L}}{}^2      }

& C_{\rm R}{}_{\lambda\rho\eta\xi}C_{\rm R}^{\lambda\rho\eta\xi}F_{\rm L}{}_{\mu\nu}F_{\rm L}^{\mu\nu}

&[34]^4 \langle 12\rangle ^2

\end{array}\\

&\begin{array}{c|l|l|}

\mathcal{O}_{C_{\rm{R}}{}^2 G_{\rm{L}}{}^2      }

& C_{\rm R}{}_{\lambda\rho\eta\xi}C_{\rm R}^{\lambda\rho\eta\xi}G_{\rm L}^A{}_{\mu\nu}G_{\rm L}^A{}^{\mu\nu}

&[34]^4 \langle 12\rangle ^2 \delta^{A_1A_2}

\end{array}

\end{align}

\underline{Class $C_{\rm{L}}{}^2 C_{\rm{R}}{}^2        $}: 1 types

\begin{align}

&\begin{array}{c|l|l|}

\mathcal{O}_{C_{\rm{L}}{}^2 C_{\rm{R}}{}^2        }

& C_{\rm L}{}_{\mu\nu\lambda\rho}C_{\rm L}^{\mu\nu\lambda\rho}C_{\rm R}{}_{\eta\xi\tau\upsilon}C_{\rm R}^{\eta\xi\tau\upsilon}

&[34]^4 \langle 12\rangle ^4

\end{array}

\end{align}

\underline{Class $C_{\rm{L}}   \psi {}^4    $}: 17 types

\begin{align}

&\begin{array}{c|l|l|}

\mathcal{O}_{C_{\rm{L}}   \bar{e}_{\rm R} e_{\rm L} \nu ^2    }

&\mathcal{Y}\left[\tiny{\young(s,t)}\right]C_{\rm L}{}_{\mu\nu\lambda\rho}\left(\bar{e}_{\rm R}{}_{p}\sigma^{\lambda \rho}e_{\rm L}{}_r{}\right)\left(\nu_{\rm L}{}_s{}C\sigma^{\mu \nu}\nu_{\rm L}{}_t{}\right)

&\mc{C}_{f_2,f_3,f_4f_5}^{[2]} \langle 12\rangle  \langle 13\rangle  \langle 14\rangle  \langle 15\rangle

\end{array}\\

&\begin{array}{c|l|l|}

\mathcal{O}_{C_{\rm{L}}   \bar{e}_{\rm R}^2 e_{\rm L}^2    }

&\mathcal{Y}\left[\tiny{\young(p,r)},\tiny{\young(s,t)}\right]C_{\rm L}{}_{\mu\nu\lambda\rho}\left(e_{\rm L}{}_s{}C\sigma^{\mu \nu}e_{\rm L}{}_t{}\right)\left(\bar{e}_{\rm R}{}_{p}\sigma^{\lambda \rho}C\bar{e}_{\rm R}{}_{r}\right)

&\mc{C}_{f_2f_3,f_4f_5}^{[2],[2]} \langle 12\rangle  \langle 13\rangle  \langle 14\rangle  \langle 15\rangle

\end{array}\\

&\begin{array}{c|l|l|}

\mathcal{O}_{C_{\rm{L}}   \bar{d}_{\rm R}^3 e_{\rm L}    }

&\mathcal{Y}\left[\tiny{\young(prs)}\right]\epsilon_{abc}C_{\rm L}{}_{\mu\nu\lambda\rho}\left(\bar{d}_{\rm R}{}_{s}^{c}\sigma^{\mu \nu}e_{\rm L}{}_t{}\right)\left(\bar{d}_{\rm R}{}_{p}^{a}\sigma^{\lambda \rho}C\bar{d}_{\rm R}{}_{r}^{b}\right)

&\mc{C}_{f_2f_3f_4,f_5}^{[3]} \langle 12\rangle  \langle 13\rangle  \langle 14\rangle  \langle 15\rangle  \epsilon_{a_2a_3a_4}

\end{array}\\

&\begin{array}{c|l|l|}

\mathcal{O}_{C_{\rm{L}}   \bar{d}_{\rm R}^2 \nu  \bar{u}_{\rm R}    }

&\mathcal{Y}\left[\tiny{\young(pr)}\right]\epsilon_{abc}C_{\rm L}{}_{\mu\nu\lambda\rho}\left(\bar{u}_{\rm R}{}_{s}^{c}\sigma^{\mu \nu}\nu_{\rm L}{}_t{}\right)\left(\bar{d}_{\rm R}{}_{p}^{a}\sigma^{\lambda \rho}C\bar{d}_{\rm R}{}_{r}^{b}\right)

&\mc{C}_{f_2f_3,f_4,f_5}^{[2]} \langle 12\rangle  \langle 13\rangle  \langle 14\rangle  \langle 15\rangle  \epsilon_{a_2a_3a_4}

\end{array}\\

&\begin{array}{c|l|l|}

\mathcal{O}_{C_{\rm{L}}   \bar{d}_{\rm R} \bar{e}_{\rm R} \bar{u}_{\rm R}^2    }

&\mathcal{Y}\left[\tiny{\young(st)}\right]\epsilon_{abc}C_{\rm L}{}_{\mu\nu\lambda\rho}\left(\bar{d}_{\rm R}{}_{p}^{a}\sigma^{\lambda \rho}C\bar{e}_{\rm R}{}_{r}\right)\left(\bar{u}_{\rm R}{}_{s}^{b}\sigma^{\mu \nu}C\bar{u}_{\rm R}{}_{t}^{c}\right)

&\mc{C}_{f_2,f_3,f_4f_5}^{[2]} \langle 12\rangle  \langle 13\rangle  \langle 14\rangle  \langle 15\rangle  \epsilon_{a_2a_4a_5}

\end{array}\\

&\begin{array}{c|l|l|}

\mathcal{O}_{C_{\rm{L}}   \bar{d}_{\rm R} d_{\rm L} \nu ^2    }

&\mathcal{Y}\left[\tiny{\young(s,t)}\right]C_{\rm L}{}_{\mu\nu\lambda\rho}\left(\bar{d}_{\rm R}{}_{p}^{a}\sigma^{\lambda \rho}dL_r{}_{a}\right)\left(\nu_{\rm L}{}_s{}C\sigma^{\mu \nu}\nu_{\rm L}{}_t{}\right)

&\mc{C}_{f_2,f_3,f_4f_5}^{[2]} \langle 12\rangle  \langle 13\rangle  \langle 14\rangle  \langle 15\rangle  \delta^{a_3}_{a_2}

\end{array}\\

&\begin{array}{c|l|l|}

\mathcal{O}_{C_{\rm{L}}   \bar{d}_{\rm R} d_{\rm L} \bar{e}_{\rm R} e_{\rm L}    }

& C_{\rm L}{}_{\mu\nu\lambda\rho}\left(\bar{d}_{\rm R}{}_{p}^{a}\sigma^{\lambda \rho}dL_r{}_{a}\right)\left(\bar{e}_{\rm R}{}_{s}\sigma^{\mu \nu}e_{\rm L}{}_t{}\right)

&\mc{C}_{f_2,f_3,f_4,f_5} \langle 12\rangle  \langle 13\rangle  \langle 14\rangle  \langle 15\rangle  \delta^{a_3}_{a_2}

\end{array}\\

&\begin{array}{c|l|l|}

\mathcal{O}_{C_{\rm{L}}   d_{\rm L} \bar{e}_{\rm R} \nu  \bar{u}_{\rm R}    }

& C_{\rm L}{}_{\mu\nu\lambda\rho}\left(\bar{e}_{\rm R}{}_{r}\sigma^{\lambda \rho}dL_p{}_{a}\right)\left(\bar{u}_{\rm R}{}_{s}^{a}\sigma^{\mu \nu}\nu_{\rm L}{}_t{}\right)

&\mc{C}_{f_2,f_3,f_4,f_5} \langle 12\rangle  \langle 13\rangle  \langle 14\rangle  \langle 15\rangle  \delta^{a_2}_{a_4}

\end{array}\\

&\begin{array}{c|l|l|}

\mathcal{O}_{C_{\rm{L}}   \bar{d}_{\rm R} e_{\rm L} \nu  u_{\rm L}    }

& C_{\rm L}{}_{\mu\nu\lambda\rho}\left(\bar{d}_{\rm R}{}_{p}^{a}\sigma^{\lambda \rho}e_{\rm L}{}_r{}\right)\left(uL_s{}_{a}C\sigma^{\mu \nu}\nu_{\rm L}{}_t{}\right)

&\mc{C}_{f_2,f_3,f_4,f_5} \langle 12\rangle  \langle 13\rangle  \langle 14\rangle  \langle 15\rangle  \delta^{a_4}_{a_2}

\end{array}\\

&\begin{array}{c|l|l|}

\mathcal{O}_{C_{\rm{L}}   \nu ^2 \bar{u}_{\rm R} u_{\rm L}    }

&\mathcal{Y}\left[\tiny{\young(s,t)}\right]C_{\rm L}{}_{\mu\nu\lambda\rho}\left(\bar{u}_{\rm R}{}_{p}^{a}\sigma^{\lambda \rho}uL_r{}_{a}\right)\left(\nu_{\rm L}{}_s{}C\sigma^{\mu \nu}\nu_{\rm L}{}_t{}\right)

&\mc{C}_{f_2,f_3,f_4f_5}^{[2]} \langle 12\rangle  \langle 13\rangle  \langle 14\rangle  \langle 15\rangle  \delta^{a_3}_{a_2}

\end{array}\\

&\begin{array}{c|l|l|}

\mathcal{O}_{C_{\rm{L}}   \bar{e}_{\rm R} e_{\rm L} \bar{u}_{\rm R} u_{\rm L}    }

& C_{\rm L}{}_{\mu\nu\lambda\rho}\left(\bar{e}_{\rm R}{}_{p}\sigma^{\lambda \rho}e_{\rm L}{}_r{}\right)\left(\bar{u}_{\rm R}{}_{s}^{a}\sigma^{\mu \nu}uL_t{}_{a}\right)

&\mc{C}_{f_2,f_3,f_4,f_5} \langle 12\rangle  \langle 13\rangle  \langle 14\rangle  \langle 15\rangle  \delta^{a_5}_{a_4}

\end{array}\\

&\begin{array}{c|l|l|}

\multirow{2}*{$\mathcal{O}_{C_{\rm{L}}   \bar{d}_{\rm R}^2 d_{\rm L}^2    }^{(1,2)}$}

&\mathcal{Y}\left[\tiny{\young(pr)},\tiny{\young(st)}\right]C_{\rm L}{}_{\mu\nu\lambda\rho}\left(dL_s{}_{a}C\sigma^{\mu \nu}dL_t{}_{b}\right)\left(\bar{d}_{\rm R}{}_{p}^{a}\sigma^{\lambda \rho}C\bar{d}_{\rm R}{}_{r}^{b}\right)

&\mc{C}_{f_2f_3,f_4f_5}^{[2],[2]} \langle 12\rangle  \langle 13\rangle  \langle 14\rangle  \langle 15\rangle  \delta^{a_4}_{a_2} \delta^{a_5}_{a_3}

\\&\mathcal{Y}\left[\tiny{\young(p,r)},\tiny{\young(s,t)}\right]C_{\rm L}{}_{\mu\nu\lambda\rho}\left(dL_s{}_{a}C\sigma^{\mu \nu}dL_t{}_{b}\right)\left(\bar{d}_{\rm R}{}_{p}^{a}\sigma^{\lambda \rho}C\bar{d}_{\rm R}{}_{r}^{b}\right)

&\mc{C}_{f_2f_3,f_4f_5}^{[1,1],[1,1]} \langle 12\rangle  \langle 13\rangle  \langle 14\rangle  \langle 15\rangle  \delta^{a_4}_{a_2} \delta^{a_5}_{a_3}

\end{array}\\

&\begin{array}{c|l|l|}

\multirow{2}*{$\mathcal{O}_{C_{\rm{L}}   \bar{d}_{\rm R} d_{\rm L} \bar{u}_{\rm R} u_{\rm L}    }^{(1,2)}$}

& C_{\rm L}{}_{\mu\nu\lambda\rho}\left(\bar{d}_{\rm R}{}_{p}^{a}\sigma^{\lambda \rho}dL_r{}_{a}\right)\left(\bar{u}_{\rm R}{}_{s}^{c}\sigma^{\mu \nu}uL_t{}_{c}\right)

&\mc{C}_{f_2,f_3,f_4,f_5} \langle 12\rangle  \langle 13\rangle  \langle 14\rangle  \langle 15\rangle  \delta^{a_3}_{a_2} \delta^{a_5}_{a_4}

\\& C_{\rm L}{}_{\mu\nu\lambda\rho}\left(\bar{d}_{\rm R}{}_{p}^{a}\sigma^{\lambda \rho}dL_r{}_{b}\right)\left(\bar{u}_{\rm R}{}_{s}^{b}\sigma^{\mu \nu}uL_t{}_{a}\right)

&\mc{C}_{f_2,f_3,f_4,f_5} \langle 12\rangle  \langle 13\rangle  \langle 14\rangle  \langle 15\rangle  \delta^{a_5}_{a_2} \delta^{a_3}_{a_4}

\end{array}\\

&\begin{array}{c|l|l|}

\multirow{2}*{$\mathcal{O}_{C_{\rm{L}}   \bar{u}_{\rm R}^2 u_{\rm L}^2    }^{(1,2)}$}

&\mathcal{Y}\left[\tiny{\young(pr)},\tiny{\young(st)}\right]C_{\rm L}{}_{\mu\nu\lambda\rho}\left(\bar{u}_{\rm R}{}_{p}^{a}\sigma^{\lambda \rho}C\bar{u}_{\rm R}{}_{r}^{b}\right)\left(uL_s{}_{a}C\sigma^{\mu \nu}uL_t{}_{b}\right)

&\mc{C}_{f_2f_3,f_4f_5}^{[2],[2]} \langle 12\rangle  \langle 13\rangle  \langle 14\rangle  \langle 15\rangle  \delta^{a_4}_{a_2} \delta^{a_5}_{a_3}

\\&\mathcal{Y}\left[\tiny{\young(p,r)},\tiny{\young(s,t)}\right]C_{\rm L}{}_{\mu\nu\lambda\rho}\left(\bar{u}_{\rm R}{}_{p}^{a}\sigma^{\lambda \rho}C\bar{u}_{\rm R}{}_{r}^{b}\right)\left(uL_s{}_{a}C\sigma^{\mu \nu}uL_t{}_{b}\right)

&\mc{C}_{f_2f_3,f_4f_5}^{[1,1],[1,1]} \langle 12\rangle  \langle 13\rangle  \langle 14\rangle  \langle 15\rangle  \delta^{a_4}_{a_2} \delta^{a_5}_{a_3}

\end{array}\\

&\begin{array}{c|l|l|}

\mathcal{O}_{C_{\rm{L}}   d_{\rm L}^3 \bar{e}_{\rm R}    }

&\mathcal{Y}\left[\tiny{\young(prs)}\right]\epsilon^{abc}C_{\rm L}{}_{\mu\nu\lambda\rho}\left(\bar{e}_{\rm R}{}_{t}\sigma^{\mu \nu}dL_s{}_{c}\right)\left(dL_p{}_{a}C\sigma^{\lambda \rho}dL_r{}_{b}\right)

&\mc{C}_{f_2f_3f_4,f_5}^{[3]} \langle 12\rangle  \langle 13\rangle  \langle 14\rangle  \langle 15\rangle  \epsilon^{a_2a_3a_4}

\end{array}\\

&\begin{array}{c|l|l|}

\mathcal{O}_{C_{\rm{L}}   d_{\rm L}^2 \nu  u_{\rm L}    }

&\mathcal{Y}\left[\tiny{\young(pr)}\right]\epsilon^{abc}C_{\rm L}{}_{\mu\nu\lambda\rho}\left(dL_p{}_{a}C\sigma^{\lambda \rho}dL_r{}_{b}\right)\left(uL_s{}_{c}C\sigma^{\mu \nu}\nu_{\rm L}{}_t{}\right)

&\mc{C}_{f_2f_3,f_4,f_5}^{[2]} \langle 12\rangle  \langle 13\rangle  \langle 14\rangle  \langle 15\rangle  \epsilon^{a_2a_3a_4}

\end{array}\\

&\begin{array}{c|l|l|}

\mathcal{O}_{C_{\rm{L}}   d_{\rm L} e_{\rm L} u_{\rm L}^2    }

&\mathcal{Y}\left[\tiny{\young(st)}\right]\epsilon^{abc}C_{\rm L}{}_{\mu\nu\lambda\rho}\left(dL_p{}_{a}C\sigma^{\lambda \rho}e_{\rm L}{}_r{}\right)\left(uL_s{}_{b}C\sigma^{\mu \nu}uL_t{}_{c}\right)

&\mc{C}_{f_2,f_3,f_4f_5}^{[2]} \langle 12\rangle  \langle 13\rangle  \langle 14\rangle  \langle 15\rangle  \epsilon^{a_2a_4a_5}

\end{array}

\end{align}

\subsection{Dim-9 Operator and Amplitude Bases}
\underline{Class $C_{\rm{L}} F_{\rm{L}} \psi {}^2   D^2$}: 6 types

\begin{align}

&
\begin{array}{c|l|l|}

\multirow{2}*{$\mathcal{O}_{C_{\rm{L}} F_{\rm{L}} \nu ^2   D^2}^{(1,2)}$}

&\mathcal{Y}\left[\tiny{\young(pr)}\right]F_{\rm L}^{\rho\eta}\left(D_{\mu}C_{\rm L}{}_{\nu\lambda\rho\eta}\right)\left(\left(D^{\mu}\nu_{\rm L}{}_p{}\right)C\sigma^{\nu \lambda}\nu_{\rm L}{}_r{}\right)

&\mc{C}_{f_3f_4}^{[2]} [34]  \langle 12\rangle  \langle 13\rangle ^2 \langle 14\rangle  \langle 24\rangle

\\&\mathcal{Y}\left[\tiny{\young(p,r)}\right]C_{\rm L}{}_{\lambda\rho}{}^{\nu\mu}\left(D_{\mu}F_{\rm L}^{\lambda\rho}\right)\left(\left(D_{\nu}\nu_{\rm L}{}_p{}\right)C\nu_{\rm L}{}_r{}\right)

&\mc{C}_{f_3f_4}^{[1,1]} s_{34} \langle 12\rangle ^2 \langle 13\rangle  \langle 14\rangle

\end{array}\\

&\begin{array}{c|l|l|}

\multirow{2}*{$\mathcal{O}_{C_{\rm{L}} F_{\rm{L}} \bar{e}_{\rm R} e_{\rm L}   D^2}^{(1,2)}$}

& C_{\rm L}{}_{\lambda\rho}{}^{\nu\mu}\left(D_{\mu}F_{\rm L}^{\lambda\rho}\right)\left(\left(D_{\nu}\bar{e}_{\rm R}{}_{p}\right)e_{\rm L}{}_r{}\right)

&\mc{C}_{f_3,f_4} s_{34} \langle 12\rangle ^2 \langle 13\rangle  \langle 14\rangle

\\& F_{\rm L}^{\rho\eta}\left(D_{\mu}C_{\rm L}{}_{\nu\lambda\rho\eta}\right)\left(\left(D^{\mu}\bar{e}_{\rm R}{}_{p}\right)\sigma^{\nu \lambda}e_{\rm L}{}_r{}\right)

&\mc{C}_{f_3,f_4} [34]  \langle 12\rangle  \langle 13\rangle ^2 \langle 14\rangle  \langle 24\rangle

\end{array}\\

&\begin{array}{c|l|l|}

\multirow{2}*{$\mathcal{O}_{C_{\rm{L}} F_{\rm{L}} \bar{d}_{\rm R} d_{\rm L}   D^2}^{(1,2)}$}

& C_{\rm L}{}_{\lambda\rho}{}^{\nu\mu}\left(D_{\mu}F_{\rm L}^{\lambda\rho}\right)\left(\left(D_{\nu}\bar{d}_{\rm R}{}_{p}^{a}\right)dL_r{}_{a}\right)

&\mc{C}_{f_3,f_4} s_{34} \langle 12\rangle ^2 \langle 13\rangle  \langle 14\rangle  \delta^{a_4}_{a_3}

\\& F_{\rm L}^{\rho\eta}\left(D_{\mu}C_{\rm L}{}_{\nu\lambda\rho\eta}\right)\left(\left(D^{\mu}\bar{d}_{\rm R}{}_{p}^{a}\right)\sigma^{\nu \lambda}dL_r{}_{a}\right)

&\mc{C}_{f_3,f_4} [34]  \langle 12\rangle  \langle 13\rangle ^2 \langle 14\rangle  \langle 24\rangle  \delta^{a_4}_{a_3}

\end{array}\\

&\begin{array}{c|l|l|}

\multirow{2}*{$\mathcal{O}_{C_{\rm{L}} F_{\rm{L}} \bar{u}_{\rm R} u_{\rm L}   D^2}^{(1,2)}$}

& C_{\rm L}{}_{\lambda\rho}{}^{\nu\mu}\left(D_{\mu}F_{\rm L}^{\lambda\rho}\right)\left(\left(D_{\nu}\bar{u}_{\rm R}{}_{p}^{a}\right)uL_r{}_{a}\right)

&\mc{C}_{f_3,f_4} s_{34} \langle 12\rangle ^2 \langle 13\rangle  \langle 14\rangle  \delta^{a_4}_{a_3}

\\& F_{\rm L}^{\rho\eta}\left(D_{\mu}C_{\rm L}{}_{\nu\lambda\rho\eta}\right)\left(\left(D^{\mu}\bar{u}_{\rm R}{}_{p}^{a}\right)\sigma^{\nu \lambda}uL_r{}_{a}\right)

&\mc{C}_{f_3,f_4} [34]  \langle 12\rangle  \langle 13\rangle ^2 \langle 14\rangle  \langle 24\rangle  \delta^{a_4}_{a_3}

\end{array}\\

&\begin{array}{c|l|l|}

\multirow{2}*{$\mathcal{O}_{C_{\rm{L}} G_{\rm{L}} \bar{d}_{\rm R} d_{\rm L}   D^2}^{(1,2)}$}

& \lambda^A{}_{a}^{b}C_{\rm L}{}_{\lambda\rho}{}^{\nu\mu}\left(D_{\mu}G_{\rm L}^A{}^{\lambda\rho}\right)\left(\left(D_{\nu}\bar{d}_{\rm R}{}_{p}^{a}\right)dL_r{}_{b}\right)

&\mc{C}_{f_3,f_4} s_{34} \langle 12\rangle ^2 \langle 13\rangle  \langle 14\rangle  \left(\lambda^{A_2}\right)^{a_4}_{a_3}

\\& \lambda^A{}_{a}^{b}G_{\rm L}^A{}^{\rho\eta}\left(D_{\mu}C_{\rm L}{}_{\nu\lambda\rho\eta}\right)\left(\left(D^{\mu}\bar{d}_{\rm R}{}_{p}^{a}\right)\sigma^{\nu \lambda}dL_r{}_{b}\right)

&\mc{C}_{f_3,f_4} [34]  \langle 12\rangle  \langle 13\rangle ^2 \langle 14\rangle  \langle 24\rangle  \left(\lambda^{A_2}\right)^{a_4}_{a_3}

\end{array}\\

&\begin{array}{c|l|l|}

\multirow{2}*{$\mathcal{O}_{C_{\rm{L}} G_{\rm{L}} \bar{u}_{\rm R} u_{\rm L}   D^2}^{(1,2)}$}

& \lambda^A{}_{a}^{b}C_{\rm L}{}_{\lambda\rho}{}^{\nu\mu}\left(D_{\mu}G_{\rm L}^A{}^{\lambda\rho}\right)\left(\left(D_{\nu}\bar{u}_{\rm R}{}_{p}^{a}\right)uL_r{}_{b}\right)

&\mc{C}_{f_3,f_4} s_{34} \langle 12\rangle ^2 \langle 13\rangle  \langle 14\rangle  \left(\lambda^{A_2}\right)^{a_4}_{a_3}

\\& \lambda^A{}_{a}^{b}G_{\rm L}^A{}^{\rho\eta}\left(D_{\mu}C_{\rm L}{}_{\nu\lambda\rho\eta}\right)\left(\left(D^{\mu}\bar{u}_{\rm R}{}_{p}^{a}\right)\sigma^{\nu \lambda}uL_r{}_{b}\right)

&\mc{C}_{f_3,f_4} [34]  \langle 12\rangle  \langle 13\rangle ^2 \langle 14\rangle  \langle 24\rangle  \left(\lambda^{A_2}\right)^{a_4}_{a_3}

\end{array}

\end{align}

\underline{Class $C_{\rm{L}}{}^2   \psi {}^2   D^2$}: 4 types

\begin{align}

&\begin{array}{c|l|l|}

\multirow{2}*{$\mathcal{O}_{C_{\rm{L}}{}^2   \nu ^2   D^2}^{(1,2)}$}

&\mathcal{Y}\left[\tiny{\young(pr)}\right]\nu_{\rm L}{}_p{}C\nu_{\rm L}{}_r{}\left(D_{\mu}C_{\rm L}{}_{\nu\lambda\rho\eta}\right)\left(D^{\mu}C_{\rm L}^{\nu\lambda\rho\eta}\right)

&\mc{C}_{f_3f_4}^{[2]} s_{34} \langle 12\rangle ^4 \langle 34\rangle

\\&\mathcal{Y}\left[\tiny{\young(pr)}\right]C_{\rm L}{}_{\xi}{}^{\lambda\rho\eta}\left(D_{\mu}C_{\rm L}{}_{\nu\lambda\rho\eta}\right)\left(\left(D^{\mu}\nu_{\rm L}{}_p{}\right)C\sigma^{\xi \nu}\nu_{\rm L}{}_r{}\right)

&\mc{C}_{f_3f_4}^{[2]} [34]  \langle 12\rangle ^2 \langle 13\rangle ^2 \langle 24\rangle ^2

\end{array}\\

&\begin{array}{c|l|l|}

\multirow{2}*{$\mathcal{O}_{C_{\rm{L}}{}^2   \bar{e}_{\rm R} e_{\rm L}   D^2}^{(1,2)}$}

& \left(\bar{e}_{\rm R}{}_{p}e_{\rm L}{}_r{}\right)\left(D_{\mu}C_{\rm L}{}_{\nu\lambda\rho\eta}\right)\left(D^{\mu}C_{\rm L}^{\nu\lambda\rho\eta}\right)

&\mc{C}_{f_3,f_4} s_{34} \langle 12\rangle ^4 \langle 34\rangle

\\& C_{\rm L}{}_{\xi}{}^{\lambda\rho\eta}\left(D_{\mu}C_{\rm L}{}_{\nu\lambda\rho\eta}\right)\left(\left(D^{\mu}\bar{e}_{\rm R}{}_{p}\right)\sigma^{\xi \nu}e_{\rm L}{}_r{}\right)

&\mc{C}_{f_3,f_4} [34]  \langle 12\rangle ^2 \langle 13\rangle ^2 \langle 24\rangle ^2

\end{array}\\

&\begin{array}{c|l|l|}

\multirow{2}*{$\mathcal{O}_{C_{\rm{L}}{}^2   \bar{d}_{\rm R} d_{\rm L}   D^2}^{(1,2)}$}

& \left(\bar{d}_{\rm R}{}_{p}^{a}dL_r{}_{a}\right)\left(D_{\mu}C_{\rm L}{}_{\nu\lambda\rho\eta}\right)\left(D^{\mu}C_{\rm L}^{\nu\lambda\rho\eta}\right)

&\mc{C}_{f_3,f_4} s_{34} \langle 12\rangle ^4 \langle 34\rangle  \delta^{a_4}_{a_3}

\\& C_{\rm L}{}_{\xi}{}^{\lambda\rho\eta}\left(D_{\mu}C_{\rm L}{}_{\nu\lambda\rho\eta}\right)\left(\left(D^{\mu}\bar{d}_{\rm R}{}_{p}^{a}\right)\sigma^{\xi \nu}dL_r{}_{a}\right)

&\mc{C}_{f_3,f_4} [34]  \langle 12\rangle ^2 \langle 13\rangle ^2 \langle 24\rangle ^2 \delta^{a_4}_{a_3}

\end{array}\\

&\begin{array}{c|l|l|}

\multirow{2}*{$\mathcal{O}_{C_{\rm{L}}{}^2   \bar{u}_{\rm R} u_{\rm L}   D^2}^{(1,2)}$}

& \left(\bar{u}_{\rm R}{}_{p}^{a}uL_r{}_{a}\right)\left(D_{\mu}C_{\rm L}{}_{\nu\lambda\rho\eta}\right)\left(D^{\mu}C_{\rm L}^{\nu\lambda\rho\eta}\right)

&\mc{C}_{f_3,f_4} s_{34} \langle 12\rangle ^4 \langle 34\rangle  \delta^{a_4}_{a_3}

\\& C_{\rm L}{}_{\xi}{}^{\lambda\rho\eta}\left(D_{\mu}C_{\rm L}{}_{\nu\lambda\rho\eta}\right)\left(\left(D^{\mu}\bar{u}_{\rm R}{}_{p}^{a}\right)\sigma^{\xi \nu}uL_r{}_{a}\right)

&\mc{C}_{f_3,f_4} [34]  \langle 12\rangle ^2 \langle 13\rangle ^2 \langle 24\rangle ^2 \delta^{a_4}_{a_3}

\end{array}

\end{align}

\underline{Class $C_{\rm{L}} F_{\rm{R}} \psi {}^2   D^2$}: 6 types

\begin{align}

&\begin{array}{c|l|l|}

\mathcal{O}_{C_{\rm{L}} F_{\rm{R}} \nu ^2   D^2}

&\mathcal{Y}\left[\tiny{\young(p,r)}\right]F_{\rm R}{}_{\eta}{}^{\rho}C_{\rm L}{}_{\lambda}{}^{\nu}{}_{\rho}{}^{\mu}\left(\left(D_{\mu}D_{\nu}\nu_{\rm L}{}_p{}\right)C\sigma^{\eta \lambda}\nu_{\rm L}{}_r{}\right)

&\mc{C}_{f_2f_3}^{[2]} [34]^2  \langle 12\rangle  \langle 13\rangle ^3

\end{array}\\

&\begin{array}{c|l|l|}

\mathcal{O}_{C_{\rm{L}} F_{\rm{R}} \bar{e}_{\rm R} e_{\rm L}   D^2}

& F_{\rm R}{}_{\eta}{}^{\rho}C_{\rm L}{}_{\lambda}{}^{\nu}{}_{\rho}{}^{\mu}\left(\left(D_{\mu}D_{\nu}\bar{e}_{\rm R}{}_{p}\right)\sigma^{\eta \lambda}e_{\rm L}{}_r{}\right)

&\mc{C}_{f_2,f_3} [34]^2  \langle 12\rangle  \langle 13\rangle ^3

\end{array}\\

&\begin{array}{c|l|l|}

\mathcal{O}_{C_{\rm{L}} F_{\rm{R}} \bar{d}_{\rm R} d_{\rm L}   D^2}

& F_{\rm R}{}_{\eta}{}^{\rho}C_{\rm L}{}_{\lambda}{}^{\nu}{}_{\rho}{}^{\mu}\left(\left(D_{\mu}D_{\nu}\bar{d}_{\rm R}{}_{p}^{a}\right)\sigma^{\eta \lambda}dL_r{}_{a}\right)

&\mc{C}_{f_2,f_3} [34]^2  \langle 12\rangle  \langle 13\rangle ^3 \delta^{a_3}_{a_2}

\end{array}\\

&\begin{array}{c|l|l|}

\mathcal{O}_{C_{\rm{L}} F_{\rm{R}} \bar{u}_{\rm R} u_{\rm L}   D^2}

& F_{\rm R}{}_{\eta}{}^{\rho}C_{\rm L}{}_{\lambda}{}^{\nu}{}_{\rho}{}^{\mu}\left(\left(D_{\mu}D_{\nu}\bar{u}_{\rm R}{}_{p}^{a}\right)\sigma^{\eta \lambda}uL_r{}_{a}\right)

&\mc{C}_{f_2,f_3} [34]^2  \langle 12\rangle  \langle 13\rangle ^3 \delta^{a_3}_{a_2}

\end{array}\\

&\begin{array}{c|l|l|}

\mathcal{O}_{C_{\rm{L}} G_{\rm{R}} \bar{d}_{\rm R} d_{\rm L}   D^2}

& \lambda^A{}_{a}^{b}G_{\rm R}^A{}_{\eta}{}^{\rho}C_{\rm L}{}_{\lambda}{}^{\nu}{}_{\rho}{}^{\mu}\left(\left(D_{\mu}D_{\nu}\bar{d}_{\rm R}{}_{p}^{a}\right)\sigma^{\eta \lambda}dL_r{}_{b}\right)

&\mc{C}_{f_2,f_3} [34]^2  \langle 12\rangle  \langle 13\rangle ^3 \left(\lambda^{A_4}\right)^{a_3}_{a_2}

\end{array}\\

&\begin{array}{c|l|l|}

\mathcal{O}_{C_{\rm{L}} G_{\rm{R}} \bar{u}_{\rm R} u_{\rm L}   D^2}

& \lambda^A{}_{a}^{b}G_{\rm R}^A{}_{\eta}{}^{\rho}C_{\rm L}{}_{\lambda}{}^{\nu}{}_{\rho}{}^{\mu}\left(\left(D_{\mu}D_{\nu}\bar{u}_{\rm R}{}_{p}^{a}\right)\sigma^{\eta \lambda}uL_r{}_{b}\right)

&\mc{C}_{f_2,f_3} [34]^2  \langle 12\rangle  \langle 13\rangle ^3 \left(\lambda^{A_4}\right)^{a_3}_{a_2}

\end{array}

\end{align}

\underline{Class $C_{\rm{L}} F_{\rm{L}} \bar{\psi }{}^2   D^2$}: 6 types

\begin{align}

&\begin{array}{c|l|l|}

\mathcal{O}_{C_{\rm{L}} F_{\rm{L}} \bar{\nu} ^2   D^2}

&\mathcal{Y}\left[\tiny{\young(p,r)}\right]C_{\rm L}{}_{\lambda\rho}{}^{\nu\mu}\left(D_{\mu}F_{\rm L}^{\lambda\rho}\right)\left(\left(D_{\nu}\bar{\nu}_{\rm L}{}_{p}{}\right)C\bar{\nu}_{\rm L}{}_{r}{}\right)

&\mc{C}_{f_3f_4}^{[2]} [34]^2  \langle 12\rangle ^2 \langle 13\rangle  \langle 14\rangle

\end{array}\\

&\begin{array}{c|l|l|}

\mathcal{O}_{C_{\rm{L}} F_{\rm{L}} e_{\rm R}  \bar{e}_{\rm L}    D^2}

& C_{\rm L}{}_{\lambda\rho}{}^{\nu\mu}\left(D_{\mu}F_{\rm L}^{\lambda\rho}\right)\left(\bar{e}_{\rm L}{}_{r}{}\left(D_{\nu}e_{\rm R}{}_p{}\right)\right)

&\mc{C}_{f_3,f_4} [34]^2  \langle 12\rangle ^2 \langle 13\rangle  \langle 14\rangle

\end{array}\\

&\begin{array}{c|l|l|}

\mathcal{O}_{C_{\rm{L}} F_{\rm{L}} d_{\rm R}  \bar{d}_{\rm L}    D^2}

& C_{\rm L}{}_{\lambda\rho}{}^{\nu\mu}\left(D_{\mu}F_{\rm L}^{\lambda\rho}\right)\left(\bar{d}_{\rm L}{}_{r}{}^{a}\left(D_{\nu}d_{\rm R}{}_p{}_{a}\right)\right)

&\mc{C}_{f_3,f_4} [34]^2  \langle 12\rangle ^2 \langle 13\rangle  \langle 14\rangle  \delta^{a_3}_{a_4}

\end{array}\\

&\begin{array}{c|l|l|}

\mathcal{O}_{C_{\rm{L}} F_{\rm{L}} u_{\rm R}  \bar{u}_{\rm L}    D^2}

& C_{\rm L}{}_{\lambda\rho}{}^{\nu\mu}\left(D_{\mu}F_{\rm L}^{\lambda\rho}\right)\left(\bar{u}_{\rm L}{}_{r}{}^{a}\left(D_{\nu}u_{\rm R}{}_p{}_{a}\right)\right)

&\mc{C}_{f_3,f_4} [34]^2  \langle 12\rangle ^2 \langle 13\rangle  \langle 14\rangle  \delta^{a_3}_{a_4}

\end{array}\\

&\begin{array}{c|l|l|}

\mathcal{O}_{C_{\rm{L}} G_{\rm{L}} d_{\rm R}  \bar{d}_{\rm L}    D^2}

& \lambda^A{}_{b}^{a}C_{\rm L}{}_{\lambda\rho}{}^{\nu\mu}\left(D_{\mu}G_{\rm L}^A{}^{\lambda\rho}\right)\left(\bar{d}_{\rm L}{}_{r}{}^{b}\left(D_{\nu}d_{\rm R}{}_p{}_{a}\right)\right)

&\mc{C}_{f_3,f_4} [34]^2  \langle 12\rangle ^2 \langle 13\rangle  \langle 14\rangle  \left(\lambda^{A_2}\right)^{a_3}_{a_4}

\end{array}\\

&\begin{array}{c|l|l|}

\mathcal{O}_{C_{\rm{L}} G_{\rm{L}} u_{\rm R}  \bar{u}_{\rm L}    D^2}

& \lambda^A{}_{b}^{a}C_{\rm L}{}_{\lambda\rho}{}^{\nu\mu}\left(D_{\mu}G_{\rm L}^A{}^{\lambda\rho}\right)\left(\bar{u}_{\rm L}{}_{r}{}^{b}\left(D_{\nu}u_{\rm R}{}_p{}_{a}\right)\right)

&\mc{C}_{f_3,f_4} [34]^2  \langle 12\rangle ^2 \langle 13\rangle  \langle 14\rangle  \left(\lambda^{A_2}\right)^{a_3}_{a_4}

\end{array}

\end{align}

\underline{Class $C_{\rm{L}}{}^2   \bar{\psi }{}^2   D^2$}: 4 types

\begin{align}

&\begin{array}{c|l|l|}

\mathcal{O}_{C_{\rm{L}}{}^2   \bar{\nu} ^2   D^2}

&\mathcal{Y}\left[\tiny{\young(pr)}\right]\bar{\nu}_{\rm L}{}_{p}{}C\bar{\nu}_{\rm L}{}_{r}{}\left(D_{\mu}C_{\rm L}{}_{\nu\lambda\rho\eta}\right)\left(D^{\mu}C_{\rm L}^{\nu\lambda\rho\eta}\right)

&\mc{C}_{f_3f_4}^{[2]} [34]  s_{34} \langle 12\rangle ^4

\end{array}\\

&\begin{array}{c|l|l|}

\mathcal{O}_{C_{\rm{L}}{}^2   e_{\rm R}  \bar{e}_{\rm L}    D^2}

& \left(\bar{e}_{\rm L}{}_{r}{}e_{\rm R}{}_p{}\right)\left(D_{\mu}C_{\rm L}{}_{\nu\lambda\rho\eta}\right)\left(D^{\mu}C_{\rm L}^{\nu\lambda\rho\eta}\right)

&\mc{C}_{f_3,f_4} [34]  s_{34} \langle 12\rangle ^4

\end{array}\\

&\begin{array}{c|l|l|}

\mathcal{O}_{C_{\rm{L}}{}^2   d_{\rm R}  \bar{d}_{\rm L}    D^2}

& \left(\bar{d}_{\rm L}{}_{r}{}^{a}d_{\rm R}{}_p{}_{a}\right)\left(D_{\mu}C_{\rm L}{}_{\nu\lambda\rho\eta}\right)\left(D^{\mu}C_{\rm L}^{\nu\lambda\rho\eta}\right)

&\mc{C}_{f_3,f_4} [34]  s_{34} \langle 12\rangle ^4 \delta^{a_3}_{a_4}

\end{array}\\

&\begin{array}{c|l|l|}

\mathcal{O}_{C_{\rm{L}}{}^2   u_{\rm R}  \bar{u}_{\rm L}    D^2}

& \left(\bar{u}_{\rm L}{}_{r}{}^{a}u_{\rm R}{}_p{}_{a}\right)\left(D_{\mu}C_{\rm L}{}_{\nu\lambda\rho\eta}\right)\left(D^{\mu}C_{\rm L}^{\nu\lambda\rho\eta}\right)

&\mc{C}_{f_3,f_4} [34]  s_{34} \langle 12\rangle ^4 \delta^{a_3}_{a_4}

\end{array}

\end{align}

\underline{Class $C_{\rm{L}} F_{\rm{L}}{}^2 \psi {}^2    $}: 10 types

\begin{align}

&\begin{array}{c|l|l|}

\multirow{2}*{$\mathcal{O}_{C_{\rm{L}} F_{\rm{L}}{}^2 \nu ^2    }^{(1,2)}$}

&\mathcal{Y}\left[\tiny{\young(pr)}\right]C_{\rm L}{}_{\mu\nu\lambda\rho}F_{\rm L}^{\lambda\rho}F_{\rm L}^{\mu\nu}\nu_{\rm L}{}_p{}C\nu_{\rm L}{}_r{}

&\mc{C}_{f_4f_5}^{[2]} \langle 12\rangle ^2 \langle 13\rangle ^2 \langle 45\rangle

\\&\mathcal{Y}\left[\tiny{\young(p,r)}\right]C_{\rm L}{}_{\mu\nu\lambda\rho}F_{\rm L}^{\lambda\rho}F_{\rm L}{}_{\eta}{}^{\nu}\left(\nu_{\rm L}{}_p{}C\sigma^{\eta \mu}\nu_{\rm L}{}_r{}\right)

&\mc{C}_{f_4f_5}^{[1,1]} \langle 12\rangle ^2 \langle 13\rangle  \langle 14\rangle  \langle 35\rangle

\end{array}\\

&\begin{array}{c|l|l|}

\multirow{2}*{$\mathcal{O}_{C_{\rm{L}} F_{\rm{L}}{}^2 \bar{e}_{\rm R} e_{\rm L}    }^{(1,2)}$}

& \left(\bar{e}_{\rm R}{}_{p}e_{\rm L}{}_r{}\right)C_{\rm L}{}_{\mu\nu\lambda\rho}F_{\rm L}^{\lambda\rho}F_{\rm L}^{\mu\nu}

&\mc{C}_{f_4,f_5} \langle 12\rangle ^2 \langle 13\rangle ^2 \langle 45\rangle

\\& C_{\rm L}{}_{\mu\nu\lambda\rho}F_{\rm L}^{\lambda\rho}F_{\rm L}{}_{\eta}{}^{\nu}\left(\bar{e}_{\rm R}{}_{p}\sigma^{\eta \mu}e_{\rm L}{}_r{}\right)

&\mc{C}_{f_4,f_5} \langle 12\rangle ^2 \langle 13\rangle  \langle 14\rangle  \langle 35\rangle

\end{array}\\

&\begin{array}{c|l|l|}

\multirow{2}*{$\mathcal{O}_{C_{\rm{L}} F_{\rm{L}}{}^2 \bar{d}_{\rm R} d_{\rm L}    }^{(1,2)}$}

& \left(\bar{d}_{\rm R}{}_{p}^{a}dL_r{}_{a}\right)C_{\rm L}{}_{\mu\nu\lambda\rho}F_{\rm L}^{\lambda\rho}F_{\rm L}^{\mu\nu}

&\mc{C}_{f_4,f_5} \langle 12\rangle ^2 \langle 13\rangle ^2 \langle 45\rangle  \delta^{a_5}_{a_4}

\\& C_{\rm L}{}_{\mu\nu\lambda\rho}F_{\rm L}^{\lambda\rho}F_{\rm L}{}_{\eta}{}^{\nu}\left(\bar{d}_{\rm R}{}_{p}^{a}\sigma^{\eta \mu}dL_r{}_{a}\right)

&\mc{C}_{f_4,f_5} \langle 12\rangle ^2 \langle 13\rangle  \langle 14\rangle  \langle 35\rangle  \delta^{a_5}_{a_4}

\end{array}\\

&\begin{array}{c|l|l|}

\multirow{2}*{$\mathcal{O}_{C_{\rm{L}} F_{\rm{L}}{}^2 \bar{u}_{\rm R} u_{\rm L}    }^{(1,2)}$}

& \left(\bar{u}_{\rm R}{}_{p}^{a}uL_r{}_{a}\right)C_{\rm L}{}_{\mu\nu\lambda\rho}F_{\rm L}^{\lambda\rho}F_{\rm L}^{\mu\nu}

&\mc{C}_{f_4,f_5} \langle 12\rangle ^2 \langle 13\rangle ^2 \langle 45\rangle  \delta^{a_5}_{a_4}

\\& C_{\rm L}{}_{\mu\nu\lambda\rho}F_{\rm L}^{\lambda\rho}F_{\rm L}{}_{\eta}{}^{\nu}\left(\bar{u}_{\rm R}{}_{p}^{a}\sigma^{\eta \mu}uL_r{}_{a}\right)

&\mc{C}_{f_4,f_5} \langle 12\rangle ^2 \langle 13\rangle  \langle 14\rangle  \langle 35\rangle  \delta^{a_5}_{a_4}

\end{array}\\

&\begin{array}{c|l|l|}

\multirow{3}*{$\mathcal{O}_{C_{\rm{L}} F_{\rm{L}} G_{\rm{L}} \bar{d}_{\rm R} d_{\rm L}    }^{(1\sim 3)}$}

& \left(\bar{d}_{\rm R}{}_{p}^{a}dL_r{}_{b}\right)\lambda^A{}_{a}^{b}C_{\rm L}{}_{\mu\nu\lambda\rho}F_{\rm L}^{\lambda\rho}G_{\rm L}^A{}^{\mu\nu}

&\mc{C}_{f_4,f_5} \langle 12\rangle ^2 \langle 13\rangle ^2 \langle 45\rangle  \left(\lambda^{A_3}\right)^{a_5}_{a_4}

\\& \lambda^A{}_{a}^{b}C_{\rm L}{}_{\mu\nu\lambda\rho}F_{\rm L}^{\lambda\rho}G_{\rm L}^A{}_{\eta}{}^{\nu}\left(\bar{d}_{\rm R}{}_{p}^{a}\sigma^{\eta \mu}dL_r{}_{b}\right)

&\mc{C}_{f_4,f_5} \langle 12\rangle ^2 \langle 13\rangle  \langle 14\rangle  \langle 35\rangle  \left(\lambda^{A_3}\right)^{a_5}_{a_4}

\\& \lambda^A{}_{a}^{b}C_{\rm L}{}_{\mu\nu\lambda\rho}G_{\rm L}^A{}^{\lambda\eta}F_{\rm L}{}_{\eta}{}^{\rho}\left(\bar{d}_{\rm R}{}_{p}^{a}\sigma^{\mu \nu}dL_r{}_{b}\right)

&\mc{C}_{f_4,f_5} \langle 12\rangle  \langle 13\rangle ^2 \langle 14\rangle  \langle 25\rangle  \left(\lambda^{A_3}\right)^{a_5}_{a_4}

\end{array}\\

&\begin{array}{c|l|l|}

\multirow{3}*{$\mathcal{O}_{C_{\rm{L}} F_{\rm{L}} G_{\rm{L}} \bar{u}_{\rm R} u_{\rm L}    }^{(1\sim 3)}$}

& \left(\bar{u}_{\rm R}{}_{p}^{a}uL_r{}_{b}\right)\lambda^A{}_{a}^{b}C_{\rm L}{}_{\mu\nu\lambda\rho}F_{\rm L}^{\lambda\rho}G_{\rm L}^A{}^{\mu\nu}

&\mc{C}_{f_4,f_5} \langle 12\rangle ^2 \langle 13\rangle ^2 \langle 45\rangle  \left(\lambda^{A_3}\right)^{a_5}_{a_4}

\\& \lambda^A{}_{a}^{b}C_{\rm L}{}_{\mu\nu\lambda\rho}F_{\rm L}^{\lambda\rho}G_{\rm L}^A{}_{\eta}{}^{\nu}\left(\bar{u}_{\rm R}{}_{p}^{a}\sigma^{\eta \mu}uL_r{}_{b}\right)

&\mc{C}_{f_4,f_5} \langle 12\rangle ^2 \langle 13\rangle  \langle 14\rangle  \langle 35\rangle  \left(\lambda^{A_3}\right)^{a_5}_{a_4}

\\& \lambda^A{}_{a}^{b}C_{\rm L}{}_{\mu\nu\lambda\rho}G_{\rm L}^A{}^{\lambda\eta}F_{\rm L}{}_{\eta}{}^{\rho}\left(\bar{u}_{\rm R}{}_{p}^{a}\sigma^{\mu \nu}uL_r{}_{b}\right)

&\mc{C}_{f_4,f_5} \langle 12\rangle  \langle 13\rangle ^2 \langle 14\rangle  \langle 25\rangle  \left(\lambda^{A_3}\right)^{a_5}_{a_4}

\end{array}\\

&\begin{array}{c|l|l|}

\multirow{2}*{$\mathcal{O}_{C_{\rm{L}} G_{\rm{L}}{}^2 \nu ^2    }^{(1,2)}$}

&\mathcal{Y}\left[\tiny{\young(pr)}\right]C_{\rm L}{}_{\mu\nu\lambda\rho}G_{\rm L}^A{}^{\lambda\rho}G_{\rm L}^A{}^{\mu\nu}\nu_{\rm L}{}_p{}C\nu_{\rm L}{}_r{}

&\mc{C}_{f_4f_5}^{[2]} \langle 12\rangle ^2 \langle 13\rangle ^2 \langle 45\rangle  \delta^{A_2A_3}

\\&\mathcal{Y}\left[\tiny{\young(p,r)}\right]C_{\rm L}{}_{\mu\nu\lambda\rho}G_{\rm L}^A{}^{\lambda\rho}G_{\rm L}^A{}_{\eta}{}^{\nu}\left(\nu_{\rm L}{}_p{}C\sigma^{\eta \mu}\nu_{\rm L}{}_r{}\right)

&\mc{C}_{f_4f_5}^{[1,1]} \langle 12\rangle ^2 \langle 13\rangle  \langle 14\rangle  \langle 35\rangle  \delta^{A_2A_3}

\end{array}\\

&\begin{array}{c|l|l|}

\multirow{2}*{$\mathcal{O}_{C_{\rm{L}} G_{\rm{L}}{}^2 \bar{e}_{\rm R} e_{\rm L}    }^{(1,2)}$}

& \left(\bar{e}_{\rm R}{}_{p}e_{\rm L}{}_r{}\right)C_{\rm L}{}_{\mu\nu\lambda\rho}G_{\rm L}^A{}^{\lambda\rho}G_{\rm L}^A{}^{\mu\nu}

&\mc{C}_{f_4,f_5} \langle 12\rangle ^2 \langle 13\rangle ^2 \langle 45\rangle  \delta^{A_2A_3}

\\& C_{\rm L}{}_{\mu\nu\lambda\rho}G_{\rm L}^A{}^{\lambda\rho}G_{\rm L}^A{}_{\eta}{}^{\nu}\left(\bar{e}_{\rm R}{}_{p}\sigma^{\eta \mu}e_{\rm L}{}_r{}\right)

&\mc{C}_{f_4,f_5} \langle 12\rangle ^2 \langle 13\rangle  \langle 14\rangle  \langle 35\rangle  \delta^{A_2A_3}

\end{array}\\

&\begin{array}{c|l|l|}

\multirow{5}*{$\mathcal{O}_{C_{\rm{L}} G_{\rm{L}}{}^2 \bar{d}_{\rm R} d_{\rm L}    }^{(1\sim 5)}$}

& \left(\bar{d}_{\rm R}{}_{p}^{a}dL_r{}_{a}\right)C_{\rm L}{}_{\mu\nu\lambda\rho}G_{\rm L}^A{}^{\lambda\rho}G_{\rm L}^A{}^{\mu\nu}

&\mc{C}_{f_4,f_5} \langle 12\rangle ^2 \langle 13\rangle ^2 \langle 45\rangle  \delta^{a_5}_{a_4} \delta^{A_2A_3}

\\& C_{\rm L}{}_{\mu\nu\lambda\rho}G_{\rm L}^A{}^{\lambda\rho}G_{\rm L}^A{}_{\eta}{}^{\nu}\left(\bar{d}_{\rm R}{}_{p}^{a}\sigma^{\eta \mu}dL_r{}_{a}\right)

&\mc{C}_{f_4,f_5} \langle 12\rangle ^2 \langle 13\rangle  \langle 14\rangle  \langle 35\rangle  \delta^{a_5}_{a_4} \delta^{A_2A_3}

\\& \left(\bar{d}_{\rm R}{}_{p}^{a}dL_r{}_{b}\right)d^{ABC}\lambda^C{}_{a}^{b}C_{\rm L}{}_{\mu\nu\lambda\rho}G_{\rm L}^A{}^{\lambda\rho}G_{\rm L}^B{}^{\mu\nu}

&\mc{C}_{f_4,f_5} \langle 12\rangle ^2 \langle 13\rangle ^2 \langle 45\rangle  d^{\text{C}A_2A_3} \left(\lambda^{\text{C}}\right)^{a_5}_{a_4}

\\& d^{ABC}\lambda^C{}_{a}^{b}C_{\rm L}{}_{\mu\nu\lambda\rho}G_{\rm L}^A{}^{\lambda\rho}G_{\rm L}^B{}_{\eta}{}^{\nu}\left(\bar{d}_{\rm R}{}_{p}^{a}\sigma^{\eta \mu}dL_r{}_{b}\right)

&\mc{C}_{f_4,f_5} \langle 12\rangle ^2 \langle 13\rangle  \langle 14\rangle  \langle 35\rangle  d^{\text{C}A_2A_3} \left(\lambda^{\text{C}}\right)^{a_5}_{a_4}

\\& f^{ABC}\lambda^C{}_{a}^{b}C_{\rm L}{}_{\mu\nu\lambda\rho}G_{\rm L}^A{}^{\lambda\rho}G_{\rm L}^B{}_{\eta}{}^{\nu}\left(\bar{d}_{\rm R}{}_{p}^{a}\sigma^{\eta \mu}dL_r{}_{b}\right)

&\mc{C}_{f_4,f_5} \langle 12\rangle ^2 \langle 13\rangle  \langle 14\rangle  \langle 35\rangle  f^{\text{C}A_2A_3} \left(\lambda^{\text{C}}\right)^{a_5}_{a_4}

\end{array}\\

&\begin{array}{c|l|l|}

\multirow{5}*{$\mathcal{O}_{C_{\rm{L}} G_{\rm{L}}{}^2 \bar{u}_{\rm R} u_{\rm L}    }^{(1\sim 5)}$}

& \left(\bar{u}_{\rm R}{}_{p}^{a}uL_r{}_{a}\right)C_{\rm L}{}_{\mu\nu\lambda\rho}G_{\rm L}^A{}^{\lambda\rho}G_{\rm L}^A{}^{\mu\nu}

&\mc{C}_{f_4,f_5} \langle 12\rangle ^2 \langle 13\rangle ^2 \langle 45\rangle  \delta^{a_5}_{a_4} \delta^{A_2A_3}

\\& C_{\rm L}{}_{\mu\nu\lambda\rho}G_{\rm L}^A{}^{\lambda\rho}G_{\rm L}^A{}_{\eta}{}^{\nu}\left(\bar{u}_{\rm R}{}_{p}^{a}\sigma^{\eta \mu}uL_r{}_{a}\right)

&\mc{C}_{f_4,f_5} \langle 12\rangle ^2 \langle 13\rangle  \langle 14\rangle  \langle 35\rangle  \delta^{a_5}_{a_4} \delta^{A_2A_3}

\\& \left(\bar{u}_{\rm R}{}_{p}^{a}uL_r{}_{b}\right)d^{ABC}\lambda^C{}_{a}^{b}C_{\rm L}{}_{\mu\nu\lambda\rho}G_{\rm L}^A{}^{\lambda\rho}G_{\rm L}^B{}^{\mu\nu}

&\mc{C}_{f_4,f_5} \langle 12\rangle ^2 \langle 13\rangle ^2 \langle 45\rangle  d^{\text{C}A_2A_3} \left(\lambda^{\text{C}}\right)^{a_5}_{a_4}

\\& d^{ABC}\lambda^C{}_{a}^{b}C_{\rm L}{}_{\mu\nu\lambda\rho}G_{\rm L}^A{}^{\lambda\rho}G_{\rm L}^B{}_{\eta}{}^{\nu}\left(\bar{u}_{\rm R}{}_{p}^{a}\sigma^{\eta \mu}uL_r{}_{b}\right)

&\mc{C}_{f_4,f_5} \langle 12\rangle ^2 \langle 13\rangle  \langle 14\rangle  \langle 35\rangle  d^{\text{C}A_2A_3} \left(\lambda^{\text{C}}\right)^{a_5}_{a_4}

\\& f^{ABC}\lambda^C{}_{a}^{b}C_{\rm L}{}_{\mu\nu\lambda\rho}G_{\rm L}^A{}^{\lambda\rho}G_{\rm L}^B{}_{\eta}{}^{\nu}\left(\bar{u}_{\rm R}{}_{p}^{a}\sigma^{\eta \mu}uL_r{}_{b}\right)

&\mc{C}_{f_4,f_5} \langle 12\rangle ^2 \langle 13\rangle  \langle 14\rangle  \langle 35\rangle  f^{\text{C}A_2A_3} \left(\lambda^{\text{C}}\right)^{a_5}_{a_4}

\end{array}

\end{align}

\underline{Class $C_{\rm{L}}{}^2 F_{\rm{L}} \psi {}^2    $}: 6 types

\begin{align}

&\begin{array}{c|l|l|}

\multirow{2}*{$\mathcal{O}_{C_{\rm{L}}{}^2 F_{\rm{L}} \nu ^2    }^{(1,2)}$}

&\mathcal{Y}\left[\tiny{\young(p,r)}\right]C_{\rm L}{}_{\mu\nu\lambda\rho}C_{\rm L}^{\mu\nu\lambda\rho}F_{\rm L}{}_{\eta\xi}\left(\nu_{\rm L}{}_p{}C\sigma^{\eta \xi}\nu_{\rm L}{}_r{}\right)

&\mc{C}_{f_4f_5}^{[1,1]} \langle 12\rangle ^4 \langle 34\rangle  \langle 35\rangle

\\&\mathcal{Y}\left[\tiny{\young(p,r)}\right]C_{\rm L}{}_{\mu\nu\lambda\rho}F_{\rm L}^{\mu\nu}C_{\rm L}{}_{\eta\xi}{}^{\lambda\rho}\left(\nu_{\rm L}{}_p{}C\sigma^{\eta \xi}\nu_{\rm L}{}_r{}\right)

&\mc{C}_{f_4f_5}^{[1,1]} \langle 12\rangle ^2 \langle 13\rangle ^2 \langle 24\rangle  \langle 25\rangle

\end{array}\\

&\begin{array}{c|l|l|}

\multirow{2}*{$\mathcal{O}_{C_{\rm{L}}{}^2 F_{\rm{L}} \bar{e}_{\rm R} e_{\rm L}    }^{(1,2)}$}

& C_{\rm L}{}_{\mu\nu\lambda\rho}C_{\rm L}^{\mu\nu\lambda\rho}F_{\rm L}{}_{\eta\xi}\left(\bar{e}_{\rm R}{}_{p}\sigma^{\eta \xi}e_{\rm L}{}_r{}\right)

&\mc{C}_{f_4,f_5} \langle 12\rangle ^4 \langle 34\rangle  \langle 35\rangle

\\& C_{\rm L}{}_{\mu\nu\lambda\rho}F_{\rm L}^{\mu\nu}C_{\rm L}{}_{\eta\xi}{}^{\lambda\rho}\left(\bar{e}_{\rm R}{}_{p}\sigma^{\eta \xi}e_{\rm L}{}_r{}\right)

&\mc{C}_{f_4,f_5} \langle 12\rangle ^2 \langle 13\rangle ^2 \langle 24\rangle  \langle 25\rangle

\end{array}\\

&\begin{array}{c|l|l|}

\multirow{2}*{$\mathcal{O}_{C_{\rm{L}}{}^2 F_{\rm{L}} \bar{d}_{\rm R} d_{\rm L}    }^{(1,2)}$}

& C_{\rm L}{}_{\mu\nu\lambda\rho}C_{\rm L}^{\mu\nu\lambda\rho}F_{\rm L}{}_{\eta\xi}\left(\bar{d}_{\rm R}{}_{p}^{a}\sigma^{\eta \xi}dL_r{}_{a}\right)

&\mc{C}_{f_4,f_5} \langle 12\rangle ^4 \langle 34\rangle  \langle 35\rangle  \delta^{a_5}_{a_4}

\\& C_{\rm L}{}_{\mu\nu\lambda\rho}F_{\rm L}^{\mu\nu}C_{\rm L}{}_{\eta\xi}{}^{\lambda\rho}\left(\bar{d}_{\rm R}{}_{p}^{a}\sigma^{\eta \xi}dL_r{}_{a}\right)

&\mc{C}_{f_4,f_5} \langle 12\rangle ^2 \langle 13\rangle ^2 \langle 24\rangle  \langle 25\rangle  \delta^{a_5}_{a_4}

\end{array}\\

&\begin{array}{c|l|l|}

\multirow{2}*{$\mathcal{O}_{C_{\rm{L}}{}^2 F_{\rm{L}} \bar{u}_{\rm R} u_{\rm L}    }^{(1,2)}$}

& C_{\rm L}{}_{\mu\nu\lambda\rho}C_{\rm L}^{\mu\nu\lambda\rho}F_{\rm L}{}_{\eta\xi}\left(\bar{u}_{\rm R}{}_{p}^{a}\sigma^{\eta \xi}uL_r{}_{a}\right)

&\mc{C}_{f_4,f_5} \langle 12\rangle ^4 \langle 34\rangle  \langle 35\rangle  \delta^{a_5}_{a_4}

\\& C_{\rm L}{}_{\mu\nu\lambda\rho}F_{\rm L}^{\mu\nu}C_{\rm L}{}_{\eta\xi}{}^{\lambda\rho}\left(\bar{u}_{\rm R}{}_{p}^{a}\sigma^{\eta \xi}uL_r{}_{a}\right)

&\mc{C}_{f_4,f_5} \langle 12\rangle ^2 \langle 13\rangle ^2 \langle 24\rangle  \langle 25\rangle  \delta^{a_5}_{a_4}

\end{array}\\

&\begin{array}{c|l|l|}

\multirow{2}*{$\mathcal{O}_{C_{\rm{L}}{}^2 G_{\rm{L}} \bar{d}_{\rm R} d_{\rm L}    }^{(1,2)}$}

& \lambda^A{}_{a}^{b}C_{\rm L}{}_{\mu\nu\lambda\rho}C_{\rm L}^{\mu\nu\lambda\rho}G_{\rm L}^A{}_{\eta\xi}\left(\bar{d}_{\rm R}{}_{p}^{a}\sigma^{\eta \xi}dL_r{}_{b}\right)

&\mc{C}_{f_4,f_5} \langle 12\rangle ^4 \langle 34\rangle  \langle 35\rangle  \left(\lambda^{A_3}\right)^{a_5}_{a_4}

\\& \lambda^A{}_{a}^{b}C_{\rm L}{}_{\mu\nu\lambda\rho}G_{\rm L}^A{}^{\mu\nu}C_{\rm L}{}_{\eta\xi}{}^{\lambda\rho}\left(\bar{d}_{\rm R}{}_{p}^{a}\sigma^{\eta \xi}dL_r{}_{b}\right)

&\mc{C}_{f_4,f_5} \langle 12\rangle ^2 \langle 13\rangle ^2 \langle 24\rangle  \langle 25\rangle  \left(\lambda^{A_3}\right)^{a_5}_{a_4}

\end{array}\\

&\begin{array}{c|l|l|}

\multirow{2}*{$\mathcal{O}_{C_{\rm{L}}{}^2 G_{\rm{L}} \bar{u}_{\rm R} u_{\rm L}    }^{(1,2)}$}

& \lambda^A{}_{a}^{b}C_{\rm L}{}_{\mu\nu\lambda\rho}C_{\rm L}^{\mu\nu\lambda\rho}G_{\rm L}^A{}_{\eta\xi}\left(\bar{u}_{\rm R}{}_{p}^{a}\sigma^{\eta \xi}uL_r{}_{b}\right)

&\mc{C}_{f_4,f_5} \langle 12\rangle ^4 \langle 34\rangle  \langle 35\rangle  \left(\lambda^{A_3}\right)^{a_5}_{a_4}

\\& \lambda^A{}_{a}^{b}C_{\rm L}{}_{\mu\nu\lambda\rho}G_{\rm L}^A{}^{\mu\nu}C_{\rm L}{}_{\eta\xi}{}^{\lambda\rho}\left(\bar{u}_{\rm R}{}_{p}^{a}\sigma^{\eta \xi}uL_r{}_{b}\right)

&\mc{C}_{f_4,f_5} \langle 12\rangle ^2 \langle 13\rangle ^2 \langle 24\rangle  \langle 25\rangle  \left(\lambda^{A_3}\right)^{a_5}_{a_4}

\end{array}

\end{align}

\underline{Class $C_{\rm{L}}{}^3   \psi {}^2    $}: 4 types

\begin{align}

&\begin{array}{c|l|l|}

\mathcal{O}_{C_{\rm{L}}{}^3   \nu ^2    }

&\mathcal{Y}\left[\tiny{\young(pr)}\right]C_{\rm L}{}_{\mu\nu\lambda\rho}C_{\rm L}^{\mu\nu\eta\xi}C_{\rm L}{}_{\eta\xi}{}^{\lambda\rho}\nu_{\rm L}{}_p{}C\nu_{\rm L}{}_r{}

&\mc{C}_{f_4f_5}^{[2]} \langle 12\rangle ^2 \langle 13\rangle ^2 \langle 23\rangle ^2 \langle 45\rangle

\end{array}\\

&\begin{array}{c|l|l|}

\mathcal{O}_{C_{\rm{L}}{}^3   \bar{e}_{\rm R} e_{\rm L}    }

& \left(\bar{e}_{\rm R}{}_{p}e_{\rm L}{}_r{}\right)C_{\rm L}{}_{\mu\nu\lambda\rho}C_{\rm L}^{\mu\nu\eta\xi}C_{\rm L}{}_{\eta\xi}{}^{\lambda\rho}

&\mc{C}_{f_4,f_5} \langle 12\rangle ^2 \langle 13\rangle ^2 \langle 23\rangle ^2 \langle 45\rangle

\end{array}\\

&\begin{array}{c|l|l|}

\mathcal{O}_{C_{\rm{L}}{}^3   \bar{d}_{\rm R} d_{\rm L}    }

& \left(\bar{d}_{\rm R}{}_{p}^{a}dL_r{}_{a}\right)C_{\rm L}{}_{\mu\nu\lambda\rho}C_{\rm L}^{\mu\nu\eta\xi}C_{\rm L}{}_{\eta\xi}{}^{\lambda\rho}

&\mc{C}_{f_4,f_5} \langle 12\rangle ^2 \langle 13\rangle ^2 \langle 23\rangle ^2 \langle 45\rangle  \delta^{a_5}_{a_4}

\end{array}\\

&\begin{array}{c|l|l|}

\mathcal{O}_{C_{\rm{L}}{}^3   \bar{u}_{\rm R} u_{\rm L}    }

& \left(\bar{u}_{\rm R}{}_{p}^{a}uL_r{}_{a}\right)C_{\rm L}{}_{\mu\nu\lambda\rho}C_{\rm L}^{\mu\nu\eta\xi}C_{\rm L}{}_{\eta\xi}{}^{\lambda\rho}

&\mc{C}_{f_4,f_5} \langle 12\rangle ^2 \langle 13\rangle ^2 \langle 23\rangle ^2 \langle 45\rangle  \delta^{a_5}_{a_4}

\end{array}

\end{align}

\underline{Class $C_{\rm{L}} F_{\rm{L}}{}^2 \bar{\psi }{}^2    $}: 10 types

\begin{align}

&\begin{array}{c|l|l|}

\mathcal{O}_{C_{\rm{L}} F_{\rm{L}}{}^2 \bar{\nu} ^2    }

&\mathcal{Y}\left[\tiny{\young(pr)}\right]C_{\rm L}{}_{\mu\nu\lambda\rho}F_{\rm L}^{\lambda\rho}F_{\rm L}^{\mu\nu}\bar{\nu}_{\rm L}{}_{p}{}C\bar{\nu}_{\rm L}{}_{r}{}

&\mc{C}_{f_4f_5}^{[2]} [45]  \langle 12\rangle ^2 \langle 13\rangle ^2

\end{array}\\

&\begin{array}{c|l|l|}

\mathcal{O}_{C_{\rm{L}} F_{\rm{L}}{}^2 e_{\rm R}  \bar{e}_{\rm L}     }

& \left(\bar{e}_{\rm L}{}_{r}{}e_{\rm R}{}_p{}\right)C_{\rm L}{}_{\mu\nu\lambda\rho}F_{\rm L}^{\lambda\rho}F_{\rm L}^{\mu\nu}

&\mc{C}_{f_4,f_5} [45]  \langle 12\rangle ^2 \langle 13\rangle ^2

\end{array}\\

&\begin{array}{c|l|l|}

\mathcal{O}_{C_{\rm{L}} F_{\rm{L}}{}^2 d_{\rm R}  \bar{d}_{\rm L}     }

& \left(\bar{d}_{\rm L}{}_{r}{}^{a}d_{\rm R}{}_p{}_{a}\right)C_{\rm L}{}_{\mu\nu\lambda\rho}F_{\rm L}^{\lambda\rho}F_{\rm L}^{\mu\nu}

&\mc{C}_{f_4,f_5} [45]  \langle 12\rangle ^2 \langle 13\rangle ^2 \delta^{a_4}_{a_5}

\end{array}%

\end{align}

\subsection{Dim-10 Operator and Amplitude Bases}
\input{GRLEFT10}

\section{Conclusion}

In this study, we step-by-step build the amplitude-operator correspondence for spin-2 particles and fields.
We first demonstrate that the Ricci tensor and Ricci scalar components of the Riemann Tensor can always be dropped by field redefinition in the basis construction.
Thus the only component that should be retained in the Lagrangian is the Weyl tensor $C_{\mu\nu\rho\sigma}$. 
Second, we prove with Bianchi identity that both 
$D^2 C_{\mu\nu\rho\sigma}$ and $D^\mu C_{\mu\nu\rho\sigma}$ are functional of Ricci tensor and scalar, thus both of the term can be set to zero in finding operator basis at a given dimension, which corresponds to setting EOM of Weyl tensor to zero and put the graviton in the amplitude on-shell. 
Finally, we illustrate that in the spinor notation, only the single particle module for the building block $D^{n} C_{L/R}$ with totally symmetrized spinor indices need to be retained.  Because any anti-symmetrized spinor indices would yield $D^2 C_{\mu\nu\rho\sigma}$, $D^\mu C_{\mu\nu\rho\sigma}$ or $[D_\mu, D_\nu]$ and can be set to zero according to the proceeding argument.
This is exactly the same as the amplitude-operator correspondence for the fields with helicity $|h|\leq 1$.
Therefore, we can use the Young tensor method to construct the complete and independent operator basis for the gravity EFTs.

One difference compared with our previous work is that, when obtaining the operator m-basis, to avoid the time-consuming process for simplifying the chain of $\sigma$ matrices, we directly construct over-complete Lorentz structures for an operator type by exhausting all the different contractions between Lorentz indices. By mapping candidates of these operators to the corresponding on-shell amplitudes,  one can obtain the coordinates on the y-basis amplitudes, which helps us to select an independent set of monomial Lorentz structures as the m-basis. 

Using this updated algorithm, we obtain the complete and independent on-shell Lorentz structures and amplitudes for the gravity EFTs involving (Goldstone) scalar, gauge boson in Yang-Mills theory, and spin-1/2 fermions. Finally, taking the gauge and flavor structures into account, the GRSMEFT and GRLEFT operator/amplitude basis are obtained up to mass dimension 10.

\section*{Acknowledgments}
J.H.Y. is supported by the National Science Foundation of China under Grants No. 12022514, No. 11875003 and No. 12047503, and National Key Research and Development Program of China Grant No. 2020YFC2201501, No. 2021YFA0718304, and CAS Project for Young Scientists in Basic Research YSBR-006, the Key Research Program of the CAS Grant No. XDPB15. 
M.-L.X. is supported by the U.S. Department of Energy under contracts No. DE-AC02-06CH11357 at Argonne.
H.-L.L is supported by the 4.4517.08 IISN-F.N.R.S convention.

\section*{Note Added}
During the preparation of this work, Ref.~\cite{Harlander:2023psl} appeared, which implements our algorithm for enumeration of the SMEFT operators up to dimension-12 and GRSMEFT up to dimension-11. However, in their work, they express the operators in terms of the fields with purely spinor indices for the Lorentz group, and with purely fundamental indices for the gauge group, basically the y-basis in our notation. In contrast, in our work, we convert the y-basis operator into the form of m-basis, where the field strength tensor and Weyl tensors are written in terms of ordinary four-component Lorentz indices, and the corresponding gauge indices are expressed in a more commonly used notation. This conversion is a non-trivial task and is one of our new updates to our previous algorithm. In addition, we also provide the independent amplitudes for each operator type in the bracket notation.

\bibliographystyle{JHEP}
\bibliography{GRSMEFTref}

\end{document}